\documentclass[onecolumn]{aastex631}
\usepackage{xspace}
\usepackage{longtable}
\usepackage{placeins}
\usepackage{xcolor}
\usepackage{xcolor}
\usepackage{graphicx}
\usepackage{rotating}
\usepackage{url}

\shorttitle{GASKAP Galactic Absorption}
\shortauthors{Dickey, J.M. et al.}

\begin{document}
\title{GASKAP Pilot Survey Science II: ASKAP Zoom Observations of Galactic 21-cm Absorption}
\author[0000-0002-6300-7459]{John M. Dickey}%\altaffiliation{1}
\affiliation{School of Natural Sciences, Private Bag 37, University of Tasmania, Hobart, TAS, 7001, Australia}
\author[0000-0002-4899-4169]{J.M. Dempsey}
\affiliation{Research School of Astronomy \& Astrophysics, Australian National University, Canberra, Australia 2611}
\affiliation{CSIRO Information Management and Technology, GPO Box 1700, Canberra, ACT 2601, Australia}
\author[0000-0001-9504-7386]{N.M. Pingel}
\affiliation{Research School of Astronomy \& Astrophysics, Australian National University, Canberra, Australia 2611}
\author[0000-0003-2730-957X]{N.M. McClure-Griffiths}
\affiliation{Research School of Astronomy \& Astrophysics, Australian National University, Canberra, Australia 2611}
\author[0000-0001-7105-0994]{K. Jameson}
\affiliation{Bolton Fellow, CSIRO Space and Astronomy, 26 Dick Perry Avenue, Kensington WA 6151, Australia}
\author[0000-0003-0235-3347]{J.R. Dawson}
\affiliation{Department of Physics and Astronomy and MQ Research Centre in Astronomy, Astrophysics and Astrophotonics, Macquarie University, NSW 2109, Australia}
\affiliation{CSIRO Space and Astronomy, Australia Telescope National Facility, PO Box 76, Epping, NSW 1710, Australia}
\author[0000-0002-9214-8613]{H. D\'{e}nes}
\affiliation{ASTRON, The Netherlands Institute for Radio Astronomy, Oude Hoogeveensedijk 4, 7991 PD, Dwingeloo, the Netherlands}
\author[0000-0002-7633-3376]{S.E. Clark}
\affiliation{Department of Physics, Stanford University, Stanford, CA 94305-4060} 
\affiliation{Kavli Institute for Particle Astrophysics \& Cosmology (KIPAC), Stanford University, Stanford, CA 94305}
\author[0000-0001-7462-4818]{G. Joncas}
\affiliation{D\'{e}partment de Physique, de G\'{e}nie Physique et d'Optique, Universit\'{e} Laval, Qu\'{e}bec, G1V 0A6, Canada}
\author[0000-0002-4814-958X]{D. Leahy}
\affiliation{Department of Physics and Astronomy, University of Calgary, Calgary, AB T2N 1N4, Canada}
\author[0000-0002-9888-0784]{Min-Young Lee}
\affiliation{Korea Astronomy and Space Science Institute, 776 Daedeokdae-ro, 34055, Daejeon, Republic of Korea}
\author[0000-0002-7351-6062]{M.-A. Miville-Desch\^{e}nes}
\affiliation{Laboratoire AIM, CEA/CNRS/Universit\'{e} Paris-Saclay, 91191 Gif-sur-Yvette, France}
%\affil{Institut d'Astrophysique Spatiale, CNRS, Univ. Paris-Sud, Universit\'{e} Paris-Saclay, F-91405 Orsay, France}
\author[0000-0002-3418-7817]{S. Stanimirovi\'{c}}
\affiliation{Department of Astronomy, University of Wisconsin, Madison, WI 53706}
\author[0000-0002-4409-3515]{C. D. Tremblay}
\affiliation{CSIRO Space and Astronomy, Australia Telescope National Facility, PO Box 1130, Bentley, WA 6102, Australia}
\author[0000-0002-1272-3017]{J. Th. van Loon}
\affiliation{Lennard-Jones Laboratories, Keele University, ST5 5BG, UK}
%\author{Hotan, McConnell, Whiting, GASKAP Builders, ASKAP builders, ASKAP ACES members}
%\affiliation{CSIRO Astronomy and Space Science, PO Box 1130, Bentley, WA 6102, Australia}
\correspondingauthor{John Dickey}

\begin{abstract}

%The HI disk of the Milky Way is much more extended than the disk of stars, molecular clouds, and other tracers of star formation activity.  Surveys of 21-cm emission show the atomic gas disk extending to 35 kpc or more from the Galactic center.  Conditions at the outer edge of the disk are important but not fully understood; the medium must be partially ionized by an external ultraviolet field, and heated in part by shocks due to converging flows of accreting circumgalactic clouds.  

Using the Australian Square Kilometre Array Pathfinder to measure 21-cm absorption spectra toward continuum background sources, we study the cool phase of the neutral atomic gas in the far outer disk, and in the inner Galaxy near the end of the Galactic bar at longitude 340$^o$.  In the inner Galaxy the cool atomic gas has a smaller scale height than in the solar neighborhood, similar to the molecular gas and the superthin stellar population in the bar.  In the outer Galaxy the cool atomic gas is mixed with the warm, neutral medium, with the cool fraction staying roughly constant with Galactic radius.  
\textcolor{black}{The ratio of the emission brightness temperature to the absorption, i.e. $1-e^{-\tau}$, is roughly constant for velocities corresponding to Galactic radius greater than about twice the solar circle radius.  The ratio has a value of about 300 K, but this does not correspond to a physical temperature in the gas. If the gas causing the absorption has kinetic temperature of about 100 K, as in the solar neighborhood, then the value 300 K indicates that the fraction of the gas mass in this phase is one-third of the total HI mass}.

%The mean spin temperature is consistent with $\sim$300 K from at least radius 12 to 35 kpc.  This demonstrates that the two phases of the atomic medium can coexist even in the extreme conditions of the outer edge of the disk, where the gas has recently accreted from the circumgalactic medium.

\end{abstract}

\section{Background}

A galaxy's lifeblood is its interstellar medium (ISM).  A spiral galaxy evolves 
through the cycle of star formation and stellar mass-loss leading to 
gradually increasing gas-phase metallicity.  Physical conditions in the
ISM cover the full range between the coldest molecular clouds, with temperatures
$T\simeq 10$ K, to the fully ionized interiors of \textcolor{black}{supernova remnant-blown bubbles and chimneys with $T>10^6$ K.}
The atomic hydrogen (HI) populates regions of intermediate density and temperature,
primarily in two phases, the cool neutral medium (CNM) with temperatures between
20 and 150 K and the warm neutral medium (WNM) with temperatures between about 
6,000 and 10,000 K \citep{Kulkarni_Heiles_1987, Ferriere_2001}.  Deep surveys of
21-cm absorption with the Very Large Array telescope show that there is a population
of atomic clouds with intermediate temperatures, from a few hundred to a few 
thousand K, called the unstable neutral medium \citep[UNM,][]{Murray_etal_2015} 
because it is thermally unstable in the classic two-phase theory
of \citet{Field_etal_1969}.  As the more abundant
phase, the WNM dominates 21-cm emission spectra, but in 21-cm absorption the WNM
disappears and only the  
CNM is visible, with low-level traces of the UNM, which explains
the dramatic difference in shape between Galactic
21-cm emission and absorption spectra \citep{Clark_1965, Murray_etal_2021}.
In spiral galaxy disks, conditions allow heating-cooling equilibrium for both
the WNM and CNM at a range of typical ISM pressures, thus both phases coexist
%throughout the Milky Way (MW) thin disk \citep{Wolfire_etal_2003,Bialy_Sternberg_2019}.  The CNM is
inside and outside the solar circle \citep{Wolfire_etal_2003,Bialy_Sternberg_2019}.  The CNM is
more difficult to image than the WNM, because sensitive absorption spectra require bright
continuum background sources, but to understand the thermal balance, and the 
interaction between the gas and the radiation field more generally, we need 
to study the CNM along with the WNM, i.e. the 21-cm absorption along with the emission.

To maintain the star formation rate observed in the Milky Way and other spiral galaxies,
and to understand the abundance of metals and their recycling, the circumgalactic medium
(CGM) is indispensible \citep{Putman_etal_2012, Tumlinson_etal_2017}.  Although the gas halo outside the Milky Way (MW)
disk is mostly warm to very hot and highly ionized \citep{Miller_Bregman_2015}
it can coexist with atomic hydrogen structures like the Magellanic Stream
\citep{Fox_etal_2014}.  At the outer edges of the Milky Way disk, with galactocentric
radius $R$ in the range 30 to 40 kpc \citep{Kalberla_Dedes_2008}, the disk
and halo must have an interface.  This may be the region where 
cold-mode accretion from the CGM to the ISM is underway
\citep{Bland-Hawthorn_etal_2017}

Surveys of the 21-cm line in emission show the structure and rotation of the HI gas
in the MW disk and halo \citep{Kalberla_Kerp_2009}.  The
thermal velocity of the WNM smooths emission
spectra so that lines are typically $\sim$10 km s$^{-1}$ wide.  In absorption, 
CNM features are typically much narrower than in emission. When it is possible to
translate absorption velocities to distances, the narrower lines provide better distance
resolution.  The filling factor of the CNM is lower than the WNM by an order of 
magnitude \citep[$\sim$5\% and $\sim$50\% respectively]{Dickey_Lockman_1990}; absorption
picks out the dense regions.  For these reasons, we can sometimes get better
resolution of Galactic structure with the CNM than with the WNM, e.g. small offsets 
of the HI disk from the Galactic midplane.

The molecular clouds, type II supernovae, and synchrotron
emissivity all drop off with $R$ following an exponential distribution with
scale length similar to that of most stellar populations, typically
\textcolor{black}{2.0} to 3.9 kpc \citep{Miville-Deschenes_etal_2017, Bland-Hawthorn_Gerhard_2016, Chrobakova_etal_2020}.
But the WNM has a much larger radial extent.  For $R < \ \sim$12 kpc the surface
density of HI, $\Sigma$, is roughly flat based on 21-cm emission surveys
\citep{Kalberla_Dedes_2008}, beyond 12 kpc, $\Sigma$ drops exponentially with scale
length 3.75 kpc, to $R\sim 35$ kpc.  Beyond 35 kpc the radial scale length increases,
the disk disappears and the trace HI remaining is distributed in a halo. 
Since the CNM phase is intermediate between the WNM and the molecular
clouds, its radial variation might follow that of the molecular emission
or that of the HI emission, or something in between.  Although the sampling
was sparse, surveys of absorption
that were available 15 years ago indicated that the CNM has the same radial
distribution as the WNM, i.e. the phase mixture is constant with
radius \citep{Dickey_etal_2009}.  That result came as a surprise, because
at the very low pressures in the outer disk the heating-cooling equilibrium curve
favors the WNM, with a minimum pressure below which the CNM cannot exist at all
\citep{Wolfire_etal_2003,Bialy_Sternberg_2019}.  The need for a better understanding of
the phase mixture in the outer MW disk 
motivates further observations, particularly in the fourth quadrant where the
Galactic warp does not cause the outer disk to depart significantly from
latitude $b=0^o$.

The recently commissioned Australian Square Kilometre Array Pathfinder
(ASKAP) is a powerful
survey instrument, described by \citet{Hotan_etal_2021}.
%\citet{McConnell_etal_2016}, and \citet{Johnston_etal_2007}. 
The telescope
is built around a
revolutionary phased-array-feed (PAF) technology \citep{Chippendale_etal_2010, Chippendale_etal_2016}.  With 36 antennas, ASKAP has a huge number of short baselines ($<$1 km) that 
give it superior brightness sensitivity that is crucial for Galactic and extragalactic
21-cm emission mapping, and it has a large number of longer baselines, out to 6 km,
that allow a synthesized beam size smaller than 10\arcsec.  This resolution is 
optimal 
for extragalactic continuum source counts, and it is just what is needed for
Galactic 21-cm absorption observations, where the line emission must be resolved out by
the spatial high-pass filter obtained by using only the longer baselines.  

In this paper we consider 21-cm absorption in a single GASKAP pilot field centered at
($\ell,b$)=(340$^o,0^o$) that covers an area of $\sim$25 square degrees with 
resolution 10\arcsec.  Imaging a spectral cube of such immense proportions is
an intricate and long process \citep{Pingel_etal_2021}, but the absorption spectra
used in this paper can be obtained more quickly with a separate processing pipeline
\citep{Dempsey_etal_2021}.  The observations and spectra are described in 
section \ref{sec:obs}.  
%There are many Galactic continuum sources as well as extragalactic
%background sources, their spectra are segregated as described in section 3.
The inner Galaxy velocity range (negative velocities at this longitude) is rich
in absorption lines.  In section \ref{sec:termv} we discuss absorption spectral averages  
near the terminal velocity, corresponding to the sub-central point
where the line of sight (LoS) passes close to the far end of the MW bar
and the 3 kpc arm.  In section \ref{sec:outerD} we study the CNM in the outer disk, outside
the solar circle on the far side of the Galactic center.  There are many distinct
absorption lines that are detected in weighted averages of the spectra, with
kinematic distances out to $R\sim$40 kpc. We divide the
GASKAP field into six sub-fields and make weighted averages  of the spectra in
each area.
%order to confirm that the detections are robust.  
\textcolor{black}{Averaging spectra over the entire area we find little radial
variation of 
%the spin temperature, $T_{s}$, of the gas, and confirm that the mixture of 
the ratio of emission to absorption.  Thus the \textcolor{black}{ratio of} %keep 
 WNM and CNM mass densities is roughly constant
%, consistent with a value of $T_{s} \sim $300 K 
for 12 $< R <$ 40 kpc.}  Section \ref{sec:disc} discusses
the results in the context of cold-mode accretion and conditions in the outer disk,
with focus on the potential applications of the full GASKAP survey of the
Galactic plane that will begin in 2022.

%Measuring the scale height of the CNM disk at the terminal
%velocity gives about 50 pc, similar to the molecular cloud population and the
%``superthin'' stellar component of the bar.

\section{Observations \label{sec:obs}}

The observations were taken as part of the Pilot phase 1 survey science program on ASKAP on 26 and 27 April 2020 with total integration time of 16 hours (SBID 13531 and 13536).  The visibilities were calibrated 
with the standard ASKAPsoft \citep[version 1.0.14]{Guzman_etal_2019}  and ASKAP pipeline (version 1.0.15.1) steps.  The spectrometer used frequency zoom mode 5, that has channel spacing 1.1574 kHz over a total bandwidth of 18 MHz.  For the spectra used here there are 604 channels averaged to 1.0 km s$^{-1}$ spacing and bandwidth, running from $-268$ to $+335$ km s$^{-1}$ in the LSRK system \citep{Gordon_1976}.  As described by \citet{Dempsey_etal_2020,Dempsey_etal_2021}, the spectra are extracted
from small cubes covering just 50\arcsec x 50\arcsec centered on positions
of continuum sources selected by the {\small SELAVY} program \citep{Whiting_Humphreys_2012}
%\citep{Hancock_etal_2012} 
that is part of the ASKAP pipeline.
%in the field taken from the catalog of \citet{McConnell_etal_2020}.  

%For Galactic 21-cm absorption studies, the limiting factor has long been the 
%confusion due to variations in the 21-cm line emission that dominates the 
%brightness temperature for all but the strongest background continuum sources.  
\textcolor{black}{ For Galactic 21-cm absorption studies, the limiting factor has long been confusion due to small-scale variations in the 21-cm line emission that make it impossible to a ccurately interpolate the emission toward the continuum source.}
To eliminate this source of noise, we high-pass filter the spatial frequencies
of the images by deleting all data from baselines shorter than 1.5 k$\lambda$ 
= 317 m.
The clean beam size is 7.7\arcsec \xspace by 6.2\arcsec \xspace (position angle -89$^o$).
The resulting cubes include only continuum and line emission with
spatial scales significantly smaller than 3\arcmin.  This effectively removes 
all vestiges of the Galactic 21-cm line in emission, except in some high velocity clouds 
where the emission has extremely small-scale angular structure.  
The extended continuum brightness is also filtered out by this process, 
leaving only the point sources, and whatever fraction of the extended sources
may be concentrated on sizes smaller than about 20\arcsec.  

The noise in the absorption spectra is due to the system temperature of
the telescopes, primarily receiver noise, plus the sky brightness, which
is dominated by the HI emission line as seen by the individual dishes in the
telescope array.  We compute this by smoothing the Parkes Galactic All Sky
Survey \citep{McClure-Griffiths_etal_2009} to the 62\arcmin\xspace beam size of the
ASKAP dishes centered on the position of each beam formed from the PAF outputs.
Thus the noise is a function of velocity across the spectra, typically increasing
up to a factor of two compared to its off-line value.
%A total of 347 absorption spectra were generated, of which 323 have off-line rms noise
%in optical depth, $\sigma_{\tau} < 0.3$ and 186 have $\sigma_{\tau} < 0.1$.

The 21-cm emission cubes come from a very different and more involved processing
pipeline, described by \citet{Pingel_etal_2021}, that includes joint deconvolution
of all the fields together, followed by the delicately calibrated combination 
of single dish HI survey data from Parkes \citep{McClure-Griffiths_etal_2009}.
As this emission processing is on-going, we concentrate in this paper on 
the results from the absorption spectra, with comparison to the Parkes single dish
data taken from the HI4PI distribution \citep{Ben-Bekhti_etal_2016}.

\begin{figure}
\hspace{-.1in}\includegraphics[width=2.5in]{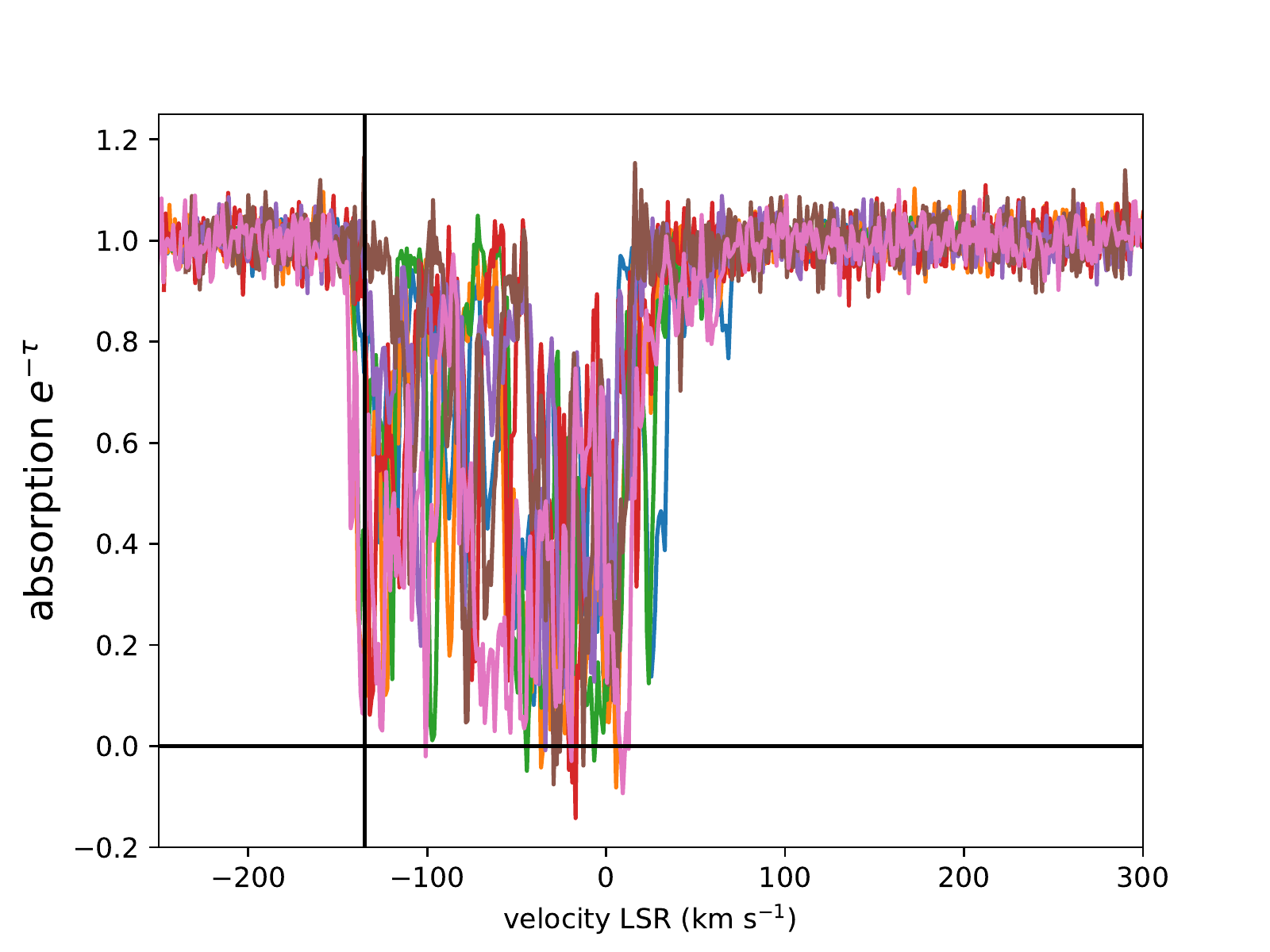}
\hspace{-.1in}\includegraphics[width=2.5in]{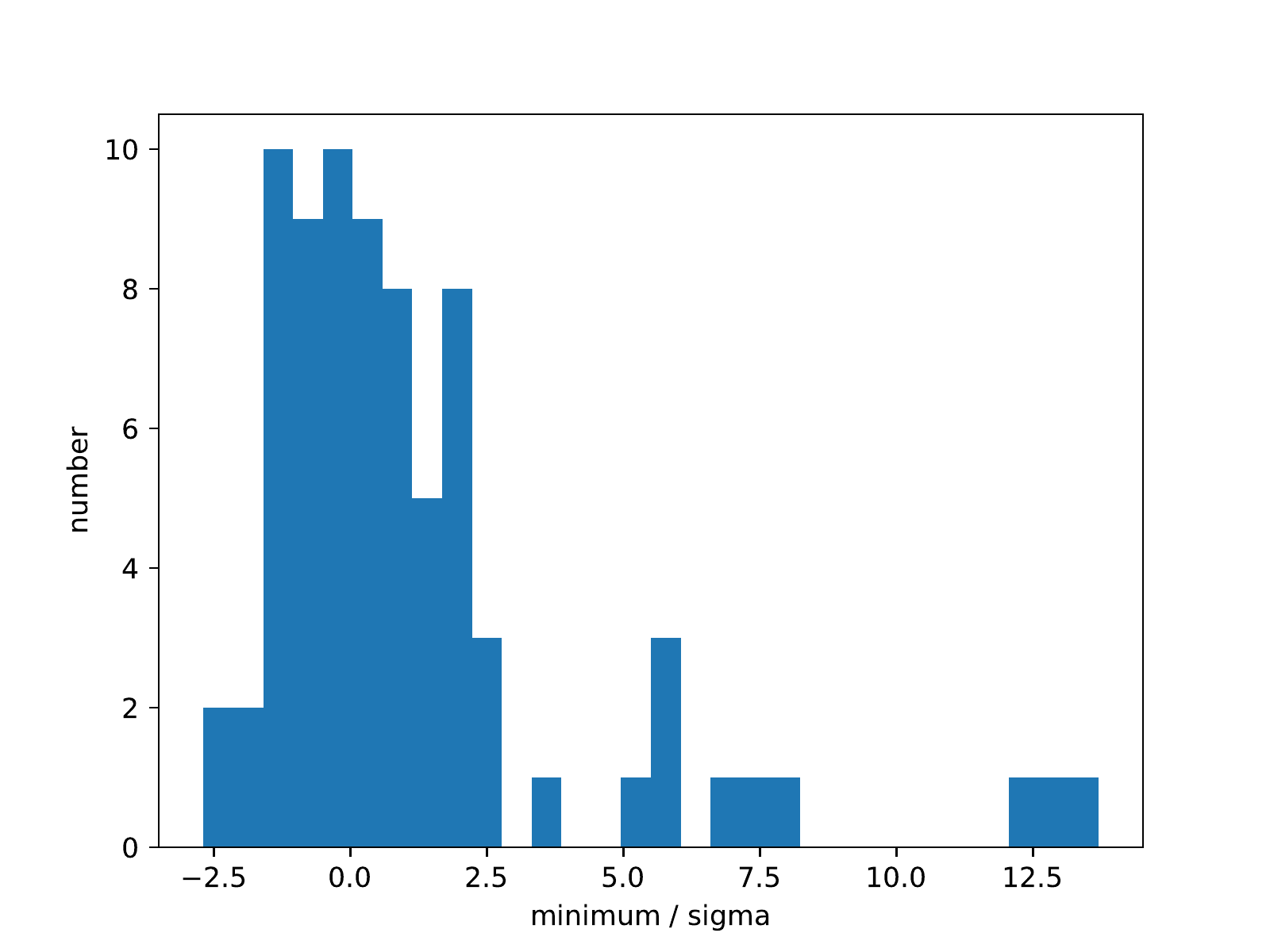}
\hspace{-.1in}\includegraphics[width=2.5in]{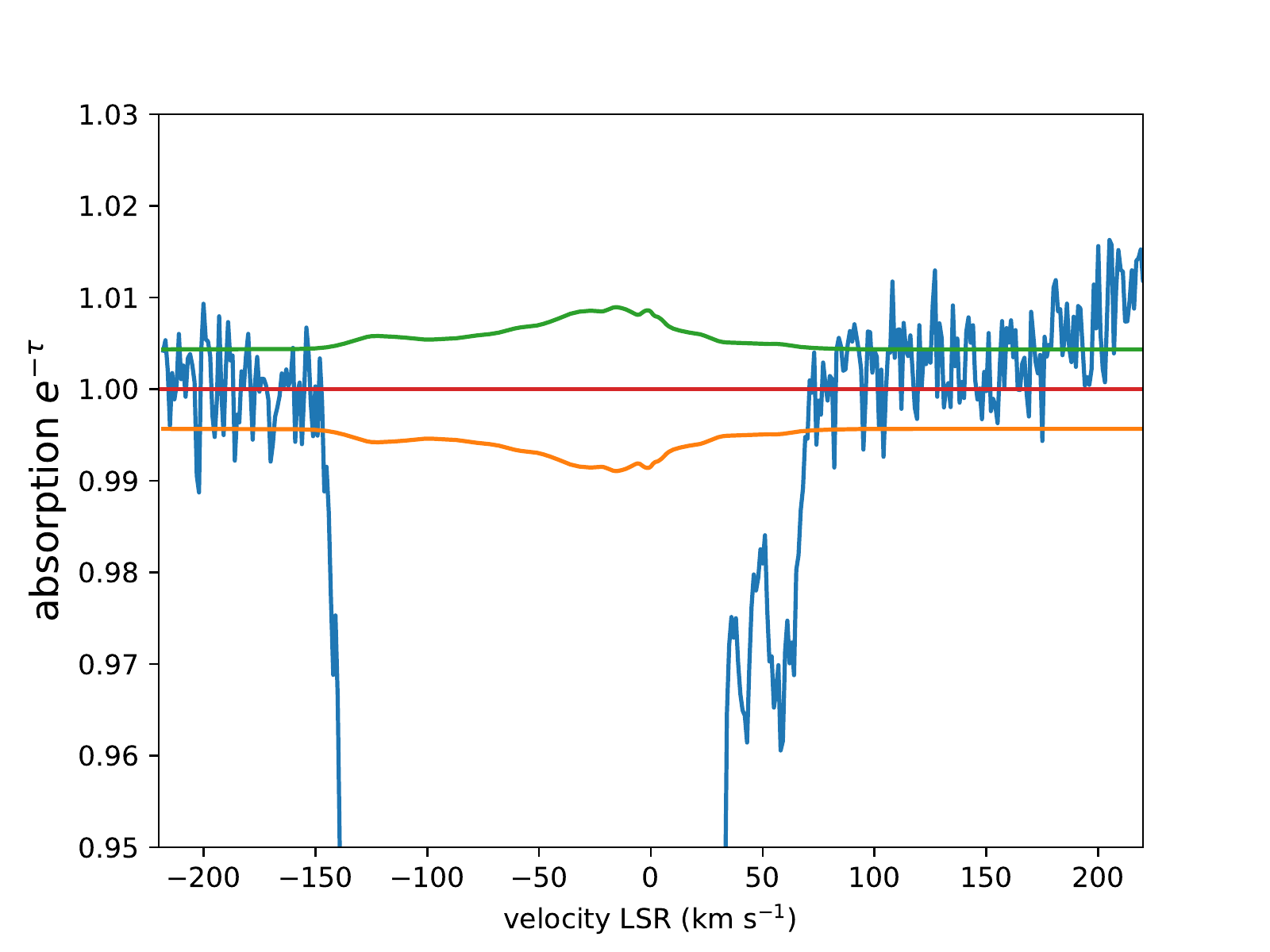}
\caption{Sample spectra and spectrum averages.  The left panel shows a superposition of seven extragalactic spectra at low latitudes ($|b|<0.5^o$).  In all cases,
the absorption spectra go close to zero at some velocities where the optical
depth is very high.  Most of the spectra go below zero, which is unphysical and often indicates problems with calibration of the bandpass shape or 
differences in processing of the continuum and spectral line channels.
The center panel scales this overshoot by the rms noise in $e^{-\tau}$.
The histogram shows the number of spectra vs. the ratio of the minimum
value of $e^{-\tau}$ divided by $\sigma_{\tau}$, for a sample of 77 extragalactic
sources.  The histogram shows
that the spectra overshoot by up to $\sim$2.5 times the rms
noise ($\sigma_{\tau}$), but mostly by about 1 $\sigma_{\tau}$.
A few spectra at higher latitudes ($|b|\sim 3^o$) have no deep absorption
lines, but most have at least one line that is nearly
saturated ($\tau \rightarrow \infty$).
The right hand panel shows an average of the 77 spectra, with the
scale expanded around $\tau = 0$ showing that the spectral baselines
are very flat.  Some curvature is noticeable at velocities v $>$
180 km s$^{-1}$.  The noise envelope broadens at velocities
where the 21-cm emission has increased the system temperature of the
receivers.
\label{fig:samples}}
\end{figure}

Some sample absorption spectra are shown in Figure \ref{fig:samples}.
The bandpass calibration is so precise that there was no need for
baseline fitting, the offline channels give a very flat baseline at
$e^{-\tau}$=1 precisely.  The deepest lines show the intensity going to zero, 
but not beyond zero by more than $\sim$2 times the rms noise.
This confirms that the continuum and spectral channels have no relative
calibration offsets.  Spectral line surveys can have optical depth scale
problems if the continuum is subtracted from the data as part of the
mapping process.  This can cause drastic errors in $\tau$ at high 
optical depths where the continuum is absorbed to almost zero. 

\subsection{Galactic vs. Extragalactic Background Sources}

The SELAVY tool used for continuum source-finding is powerful, but the results
need to be checked.  In this field there are many extended sources
that are heavily resolved by
the telescope and filtered into a few high points by the uv minimum.  
SELAVY catalogs these as separate sources, but by inspection and comparison
with lower resolution continuum maps
\citep[e.g.][]{Haverkorn_etal_2006}, we group together sources that are 
blended by the beam, or are clearly part of a single, larger continuum structure.
Many extragalactic sources are doubles.  These we leave as a single source
if the two components are blended by the beam, i.e. within about 15\arcsec,
if they are distinct points we treat them as seprate background sources.
After removing duplicates there are 295 distinct sources, whose
positions are determined by Gaussian fitting.  They are listed on 
tables \ref{tab:Galactic} and \ref{tab:exgal}.  

The density of spectra per square degree in this survey (8.2) is higher
than in the Southern Galactic Plane Survey \citep[SGPS]{McClure-Griffiths_etal_2005} by a
factor of six, and orders of magnitude higher than
in any other 21-cm absorption survey in the fourth quadrant.  This high density of
background sources makes it possible to trace the distribution of
CNM in the absorbing clouds, and to measure the covering factor of
absorption, i.e. the chance of a LoS {\bf not} showing
absorption at velocities where strong absorption is seen on nearby
sightlines.  But this analysis depends on separating the continuum
sources into two groups, Galactic and extragalactic.

\begin{figure}
\hspace{.6in}\includegraphics[width=5in]{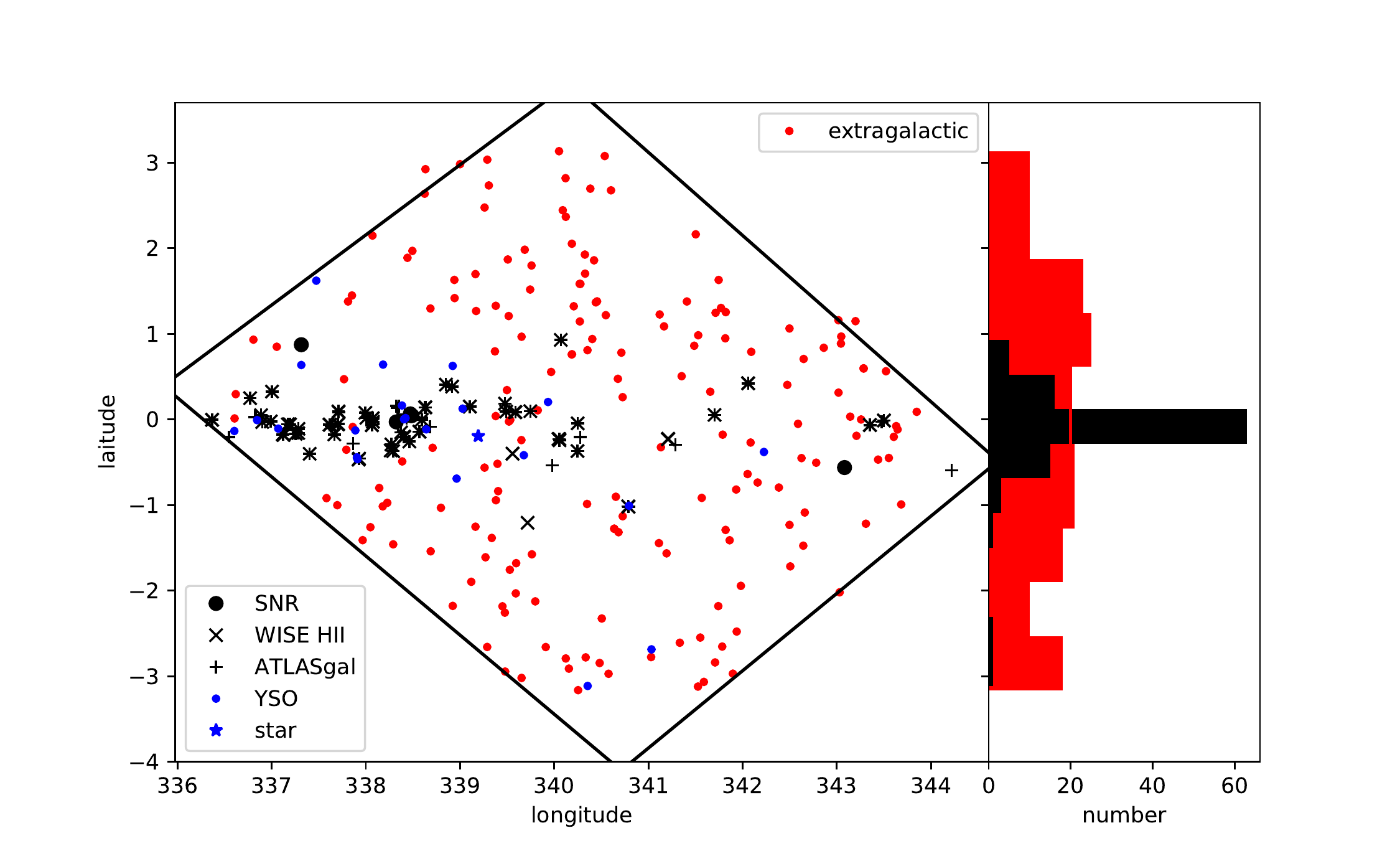}
%\vspace*{-.4in}

\caption{The distribution of continuum sources in the GASKAP field.
On the left is a map of the field showing all sources, extragalactic in red
and Galactic in black and blue, with different symbols for different types.
The right panel shows histograms of the latitude distribution of the
Galactic and extragalactic sources.  Since the longitude range decreases
with $|b|$, there are fewer extragalactic sources at high latitudes.
%The two lines indicate the width (in longitude), that
%would give the number vs. latitude as shown assuming source density
%of 8.5 per square degree.
The edges of the field are scalloped by the beam forming step,
but the four black lines show roughly the extent of the PAF footprint.
\label{fig:num_vs_lat}}
\end{figure}

Source identifications were made by matching catalogs from
VizieR\footnote{\url{https://vizier.u-strasbg.fr/viz-bin/VizieR}} using 
maximum separation
36\arcsec \xspace: from catalogs of supernova remnants \citep[][6 matches]{Green_2019},
%eDR3 \citep{Bailer_Jones_etal_2021},
submillimeter compact sources from ATLASGAL \citep[][86 matches]{Urquhart_etal_2014},
HII regions selected from the WISE survey \citep[][80 matches]{Anderson_etal_2015}, 14 young stellar objects (YSO), plus three planetary nebulae \citep{Riggi_etal_2021,Preite_Martinez_1988},
one pulsar
\citep{Manchester_etal_2005}\footnote{\url{https://www.atnf.csiro.au/research/pulsar/psrcat} v. 1.64}, and two sources on the rim of the S36 bubble
\citep{Churchwell_etal_2006}.
After this matching, we find 105 of the 295 continuum sources are Galactic,
the other 190 are assumed to be extragalactic.
%The Galactic sources are listed on Table \ref{tab:Galactic}, the
%extragalactic sources are listed on Table \ref{tab:exgal}.  
Figure
\ref{fig:num_vs_lat} shows the source positions and the latitude
distributions of the Galactic and extragalactic sources.

Many of the sources show interesting structure, in particular arcs or shells of
continuum are apparent in several Galactic sourcs.  
%Some of the more attractive examples are shown on Figure \ref{fig:contin_maps}.  
Because of the high-pass
filtering, absolute flux density values are not useful for comparison with other
surveys, and they are not tabulated.  The index numbers on tables \ref{tab:Galactic}
and \ref{tab:exgal} 
%and on Figure \ref{fig:contin_maps} 
are based on the original
SELAVY list of 347 sources, thus there are many gaps in the sequence.

%\begin{figure}
%\hspace{.5in}\includegraphics[width=5.5in]{f1.pdf}

%\caption{Continuum structure of some extended sources.  The index numbers correspond
%to the numbers on tables \ref{tab:Galactic} and \ref{tab:exgal}.  
%ICRF equatorial coordinates are around the outside, Galactic coordinates are
%on the inside edges of the figures in blue.
%\label{fig:contin_maps}}
%\end{figure}

%\begin{figure}
%\hspace{.5in}\includegraphics[width=5.5in]{f2.pdf}

%\figurenum{3}
%\caption{continued.}
%\end{figure}

%\begin{figure}
%\hspace{.5in}\includegraphics[width=5.5in]{f3.pdf}

%\figurenum{3} \caption{continued.}
%\end{figure}

%\begin{figure}
%\hspace{.5in}\includegraphics[width=5.5in]{f4.pdf}
%
%\figurenum{3} \caption{continued.}
%\end{figure}

%\begin{figure}
%\hspace{1.5in}\includegraphics[width=4.5in]{f6.pdf}

%\figurenum{3} \caption{continued.}
%\end{figure}

\subsection{Comparison with the Southern Galactic Plane Survey}

\begin{figure}
\hspace{-.1in}\includegraphics[width=2.5in]{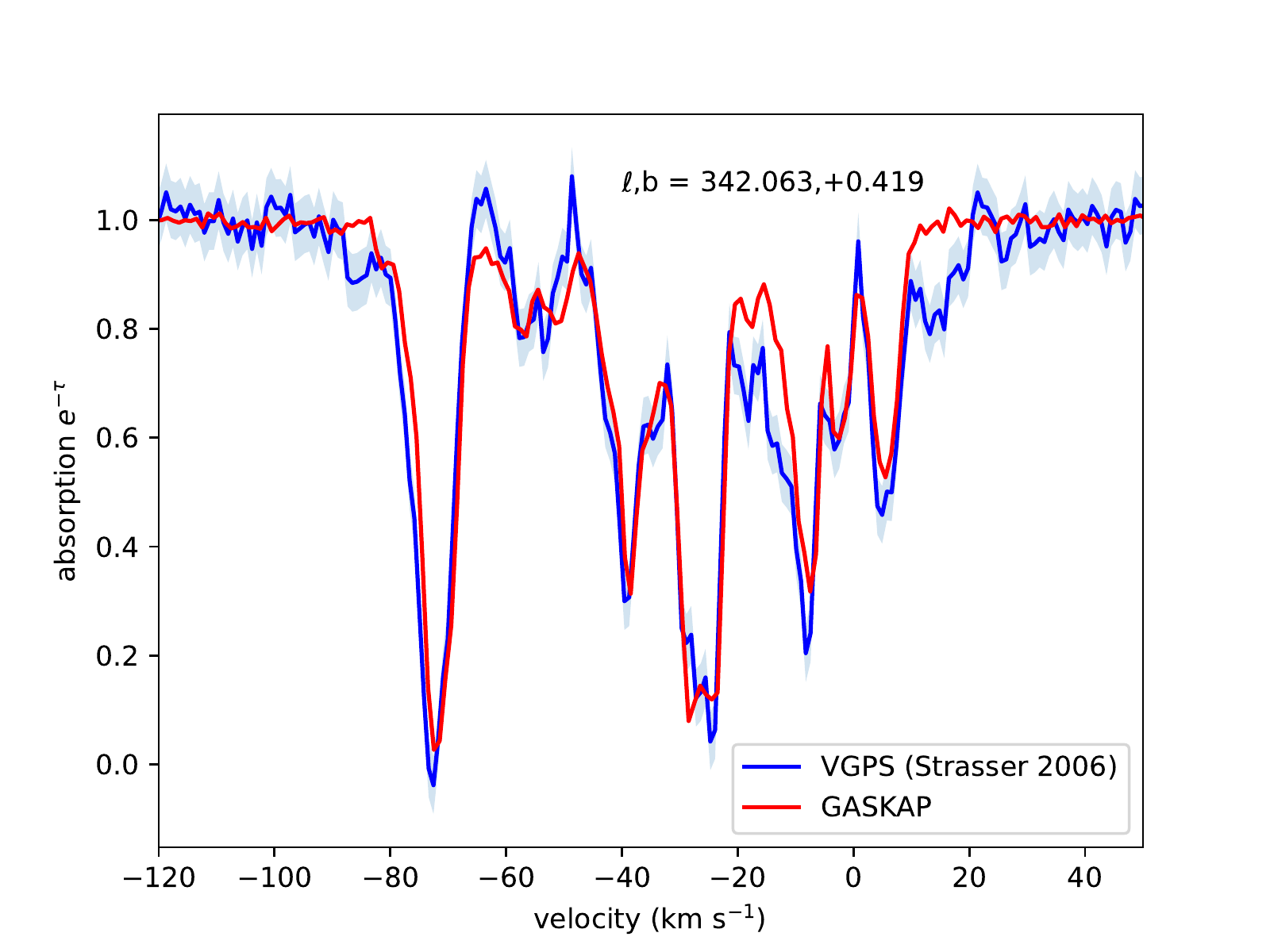}
\hspace{-.2in}\includegraphics[width=2.5in]{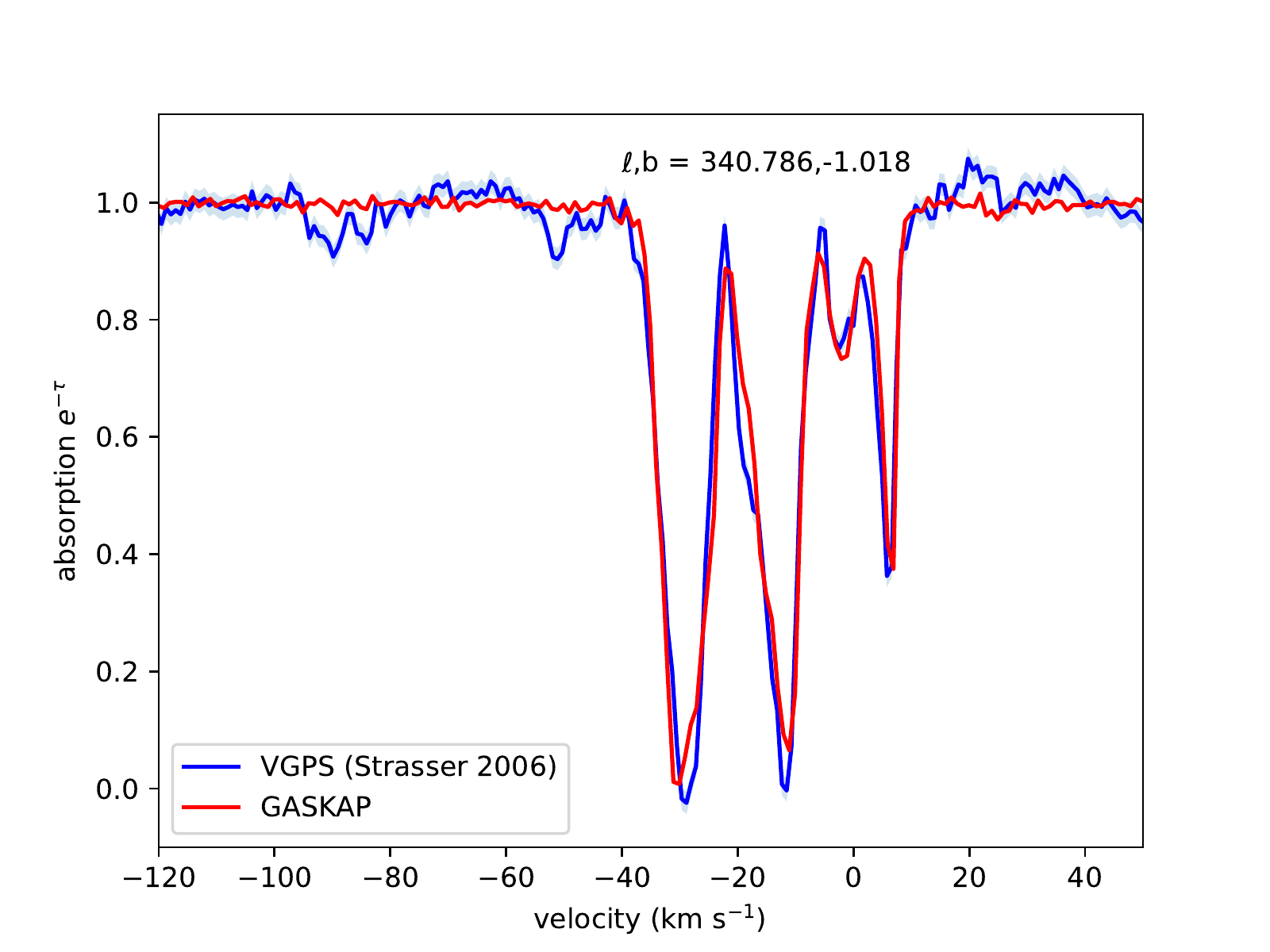}
\hspace{-.2in}\includegraphics[width=2.5in]{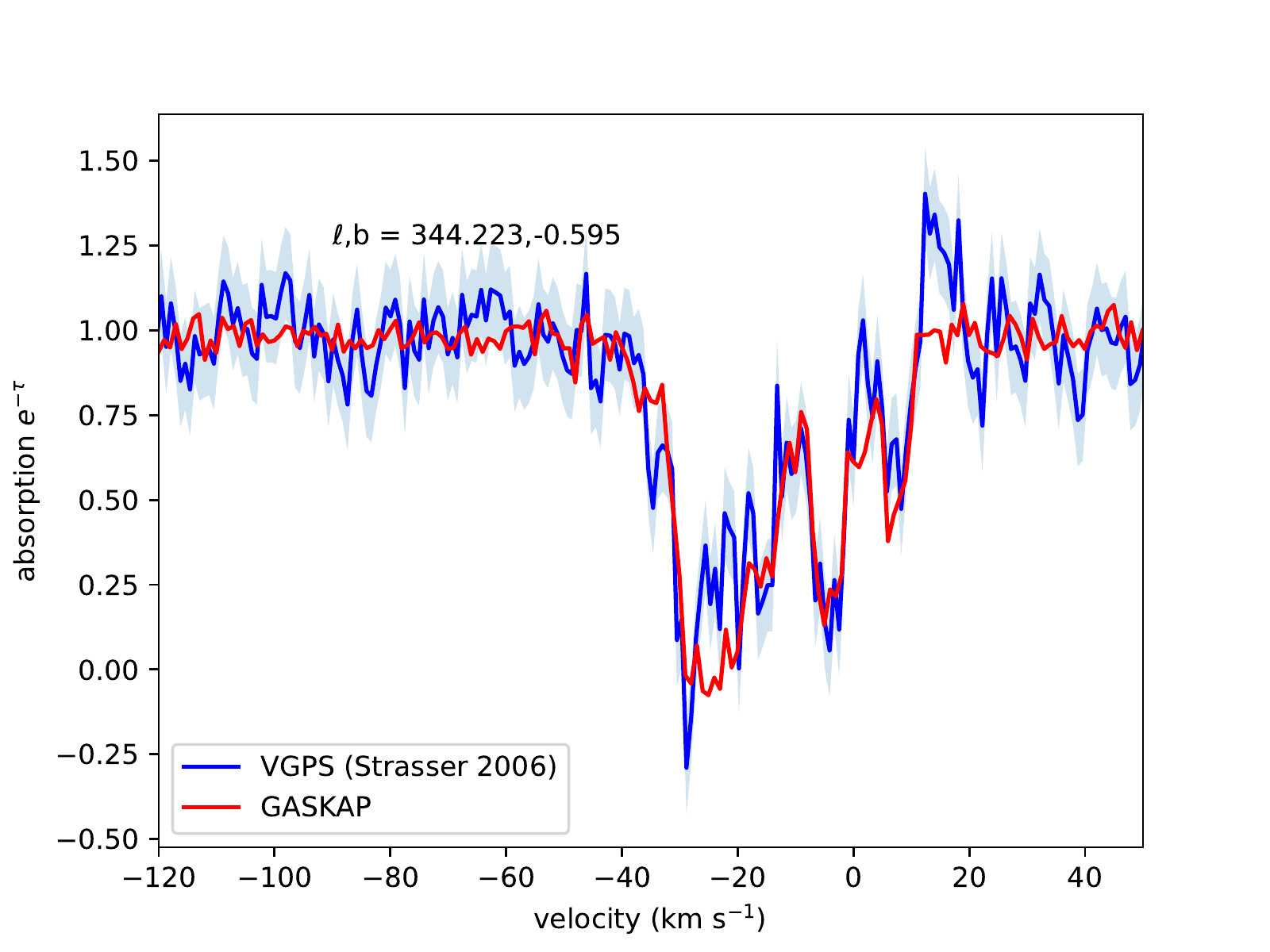}

\caption{Comparison with spectra from the SGPS \citep{Strasser_2006}.  The
blue shading indicates the 1$\sigma$ noise level in the SGPS data, which is typically
larger than that of the GASKAP spectra by a factor of three to ten.
The left and center panels are spectra toward bright continuum sources,
so that discrepencies between the two surveys are due to emission 
fluctuations causing confusion that manifests as positive or negative
``pseudo-absorption" \citep{Radhakrishnan_etal_1972}.  The right hand
panel shows spectra toward a weaker continuum source, so the radiometer
noise dominates at most velocities, but some discrepencies
above the noise are apparent.  
%The blue shading on the SGPS spectra
%indicate $\pm 1 \sigma$ with $\sigma$ computed from the off-line channels.
\label{fig:Strasser}}
\end{figure}

The field observed here was covered in the SGPS.  The strongest, most compact
continuum sources were studied for absorption in the 21-cm line by 
\citet{Strasser_2006}.  The rms noise was typically three to ten times larger
than in the GASKAP spectra, and there is confusion due to residual emission 
fluctuations.  Some comparisons are shown in Figure \ref{fig:Strasser}.
The spectra are generally consistent to within the noise, but the SGPS
spectra show occasional
spurious positive and negative fluctuations at extreme negative
and positive velocities corresponding to the more distant gas that are
not present in the GASKAP data.  The zero point ($\tau \rightarrow \infty$)
is also much better defined in the GASKAP spectra (Figure \ref{fig:samples}). 

\eject

\section{HI near the Terminal Velocity \label{sec:termv}}

\begin{figure}
\hspace{.5in}\includegraphics[width=5.9in]{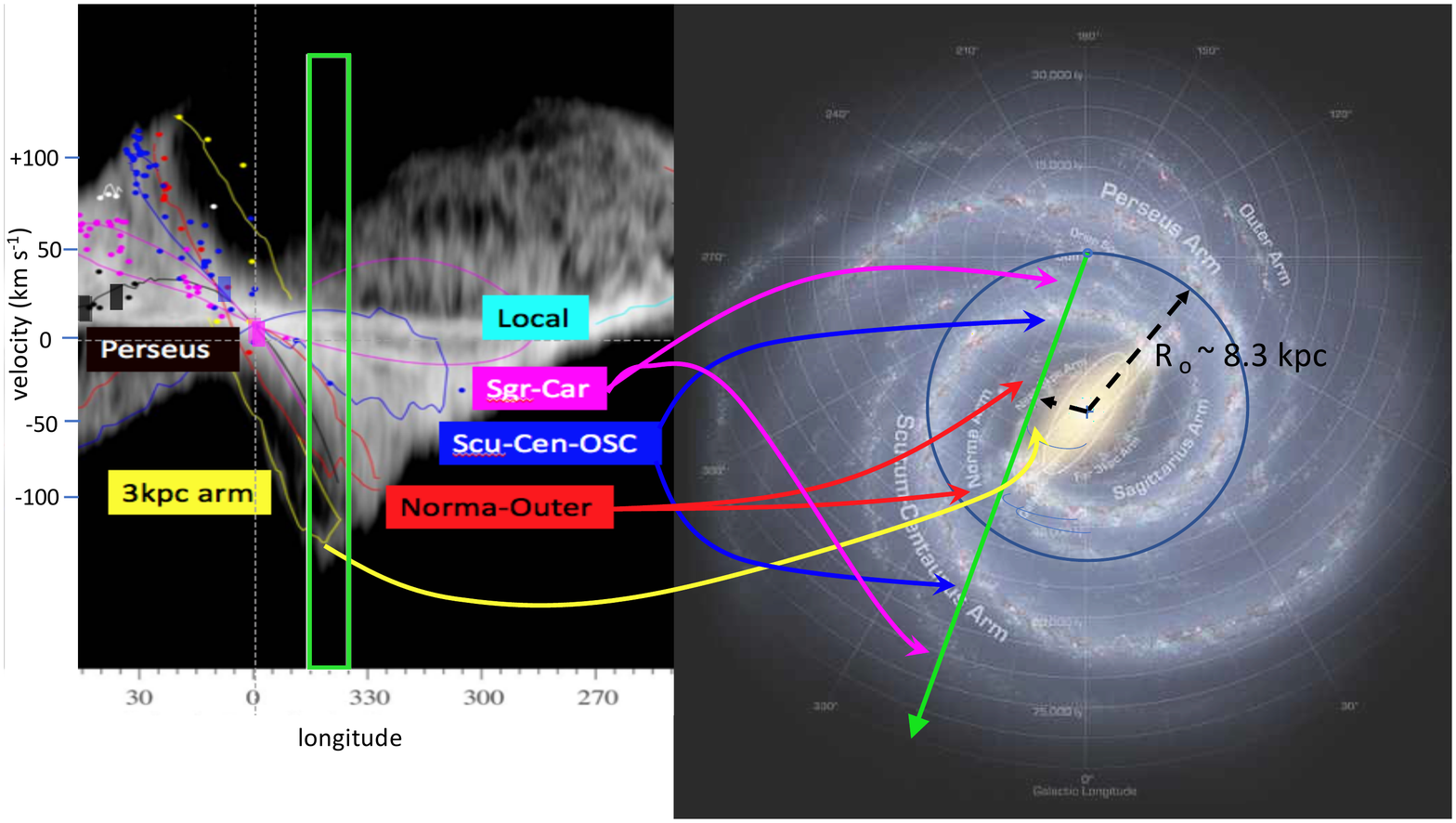}
\caption{Two views of the GASKAP Pilot field.  
%The left panel is a face-on view
%of the HI emission in the outer galaxy from \citet{Koo_etal_2017}, based
%on the LAB survey \citep{Kalberla_etal_2005} and using a flat rotation curve beyond the solar circle , assuming LSR velocity $\Theta_o$=240 km~s$^{-1}$.  
The left panel is a longitude velocity diagram of 21-cm emission taken from 
\citet[fig. 3]{Reid_etal_2019}, with grey scale showing the LAB survey emission
\citep{Kalberla_etal_2005}, and 
colored curves tracing spiral arms based on
VLBI parallax distances to masers in star formation regions.  
The white rectangle shows the longitude coverage of the GASKAP field.
The
right panel is the familiar Hurt/Benjamin/Churchwell figure \citep{Churchwell_etal_2009} with superposed the line of sight (LoS) at $340^o$ longitude in green.
Note that this LoS has tangent point in the midst of
the 3 kpc arm, it passes close to the far end of the Galactic bar (the ``long bar'' e.g. \citet{Benjamin_2008,Wegg_etal_2015}), and 
outside the solar circle it passes through the Scutum-Centaurus Arm 
at velocity roughly +5 km~s$^{-1}$,
then further out the Sagittarius-Carina Arm, at about +25 km~s$^{-1}$.
At larger Galactic radii the more distant features in the l-v diagram
are controversial, connecting either with the Perseus Arm and/or the 
Norma Arm and/or the Scutum-Outer Arm \citep[e.g.][]{Koo_etal_2017,Vallee_2020}, depending
on the pitch angles.
\label{fig:lvdiag}}
\end{figure}

Longitude 340$^o$ was chosen for the first GASKAP pilot field at $b = 0^o$
because this line of sight samples some of the most important features
in the inner and outer disk, including the 3-kpc arm and the far end of the
long bar, as well as the Scutum-Centaurus Arm both inside and again outside the
solar circle.  Figure \ref{fig:lvdiag} shows the Galactic structure 
visible through this small ``keyhole'' opened by the $\sim$5$^o$ 
square PAF field of view.  In the fourth quadrant, Doppler shifts due
to differential rotation give negative velocities inside the solar circle,
reaching to the terminal velocity at the sub-central or tangent point, where the
LoS comes closest to the Galactic center.  At longitude 340$^o$
the sub-central point is in the middle of the 3-kpc Arm, an anomalous 
structure that shows Galactocentric radial velocity of about 50 km~s$^{-1}$
at longitude 0$^o$.  At the subcentral point the Galactic radial velocity is perpendicular
to the LoS, so it does not affect the Doppler shift.

\begin{figure}
\hspace{-.2in}\includegraphics[width=3.8in]{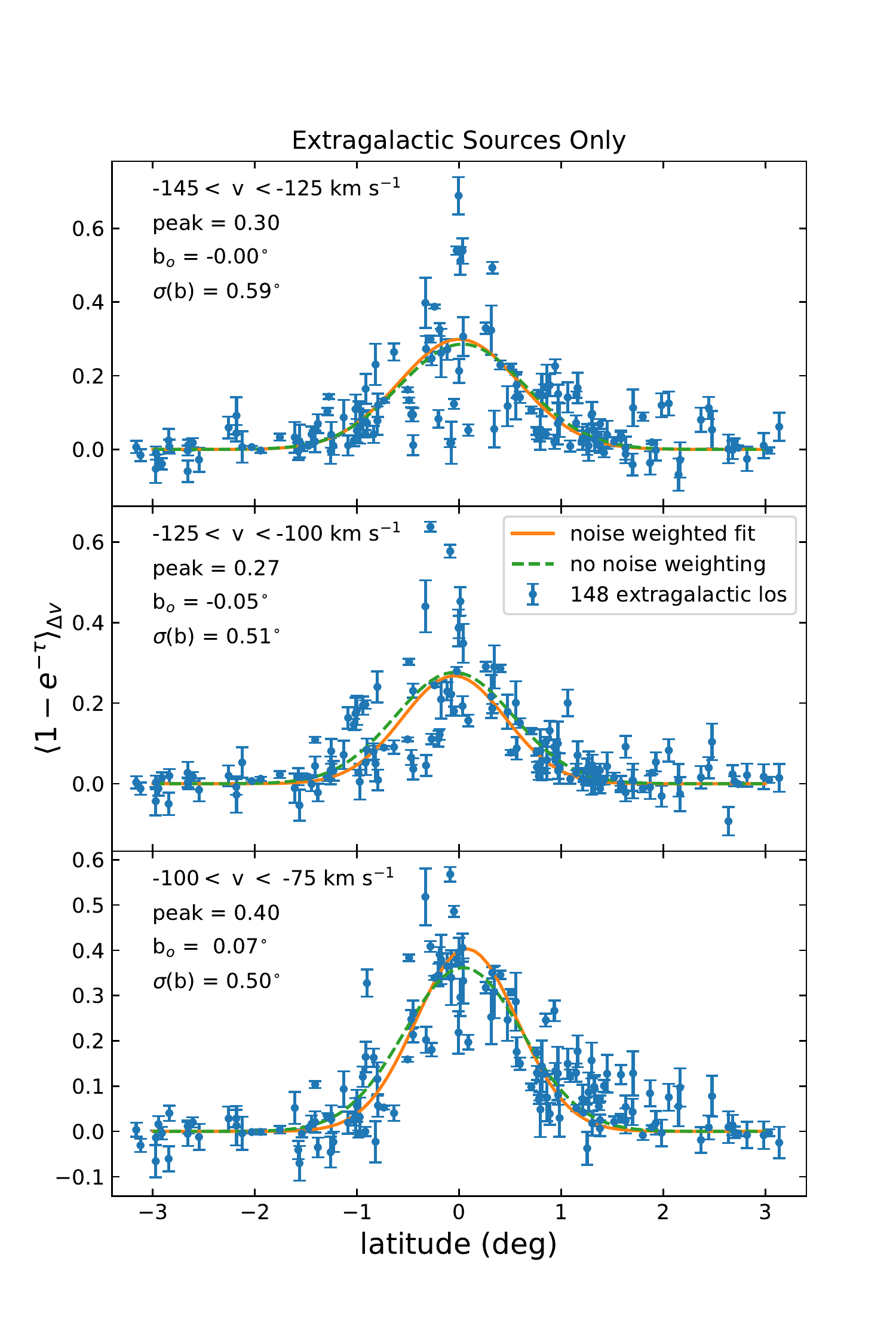}
\hspace{-.4in}\raisebox{4cm}{\includegraphics[width=4.4in]{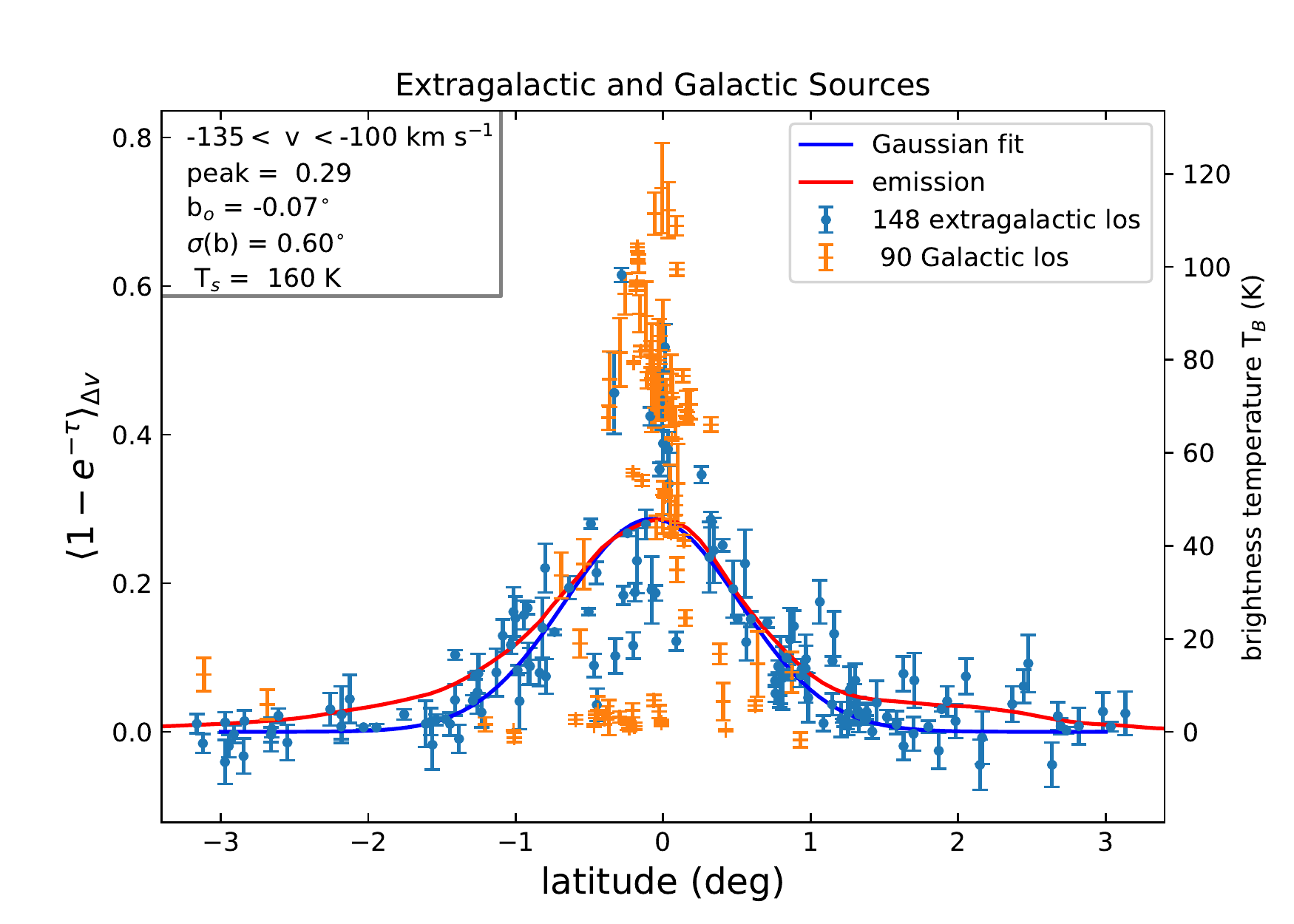}}
\caption{Mean absorption vs. latitude near the terminal velocity.  The left panel shows absorption,
$\langle 1-e^{-\tau} \rangle$ (eq. \ref{eq:equiv_w}), toward extragalactic sources only.
The right panel includes the Galactic source spectra, that show much larger 
scatter, and the emission brightness temperature.
\label{fig:termv}}
\end{figure}

The GASKAP spectra are rich with absorption features at
negative velocities, corresponding to CNM structures in the inner Galaxy.
Particularly prominent is the line at \textcolor{black}{v$_{LSR}\simeq -30$ km~s$^{-1}$
(figure \ref{fig:Strasser} middle and right panels)}
that is due to cold clouds in the Scutum-Centaurus Arm at distance
$d \sim 2.8$ kpc and Galactocentric radius R$ \simeq $ 5.8 kpc, 
\citep[using
kinematic distances from][based on $R_o$=8.31 kpc and $\Theta_o = 240$ km s$^{-1}$]{Wenger_etal_2018, Reid_etal_2019}.  Of particular interest is the gas near
the sub-central point, at a distance $d\sim$7.7 kpc and velocity
of about -135 km~s$^{-1}$.  The terminal velocity vs. longitude 
in the fourth quadrant has been shown to follow a nearly linear trend with $\sin{\ell}$:
\begin{equation}
 v_{term}(\ell) \ = \ -177.7 \cdot \left(1 - |\sin{(\ell)}|\right) \ - \ 18.4 \textrm{km\ s}^{-1} 
\label{eq:vlsr}
\end{equation}
\citep{McClure-Griffiths_jd_2016}.  
In the analysis shown on Figure \ref{fig:termv} we offset each absorption spectrum by the
difference between $v_{term}(\ell)$ and -135 km~s$^{-1}$ so that 
%the three velocity ranges represent approximately the same path length along the LoS.  
a given velocity interval near -135 km~s$^{-1}$ will include the same
line of sight interval near the sub-central point.

% Trey tool gives at l=340, v=-125 for R=3.12, d=6.86 or 9.12 (B+B "traditional" note offset from R etal))
% Monte Carlo:                     2.94+-0.2   6.94    8.54 +-0.6 (R etal 2014)
% Trey tool gives at l=340, v=-112.5 for R=2.99, d=6.92 or 8.75 (R et al)
% Monte Carlo:                     3.07+-0.2   6.44    9.03 +-0.6 (R etal 2014)
% Trey tool gives at l=340, v=-100 for R=3.58, d=5.89 or 10.08 (B&B)
% Monte Carlo:                     3.42+-0.3   5.98    9.57 +-0.5 (R etal 2014)
% Trey tool gives at l=340, v=-87.5 for R=3.68, d=5.52 or 10.16 (R et al)
%                                  3.70+-0.3   5.61    10.24 +-0.6 (R etal 2014)
% Trey tool gives at l=340, v=-75 for R=4.08, d=4.92 or 10.76 (R et al)
%                                  4.10+-0.3   4.87    10.80 +-0.4 (R etal 2014)
%
On Figure \ref{fig:termv} the left panel shows the mean absorption, 
as a function of latitude, where the mean absorption is defined as:  
\begin{equation}\label{eq:equivw}
\langle  1 - e^{-\tau} \rangle_{\Delta v} \ = \ 1 \ - \ \frac{\int_{\Delta v} \ e^{-\tau} \ dv}{\Delta v} \label{eq:equiv_w}
\end{equation}
i.e. one minus the equivalent width computed over a chosen velocity interval, $\Delta v$, divided by the velocity interval. 
Three different velocity intervals near the terminal velocity are shown on the 
three panels.  
\textcolor{black}{For the upper panel on the left of Figure \ref{fig:termv}, the velocity range, $\Delta v$,
is [-145,-125] km s$^{-1}$, which straddles the nominal terminal velocity (-135 km~s$^{-1}$ at \textcolor{black}{$\ell = 340^o$}) 
and thus is dominated by the 3-kpc Arm near the sub-central point.  The middle and lower panels step back in velocity from the
terminal velocity, and thus correspond to two regions along the LoS offset on either side of the
sub-central point, with $\Delta v$ edges at $v=-100$ km s$^{-1}$ ($R=3.6$ kpc) and $v=-75$ km s$^{-1}$ ($R=4.1$ kpc). 
The blue points are the values for each of the 151 brightest
extragalactic background sources that give the
most sensitive absorption spectra, having off-line
rms noise $\sigma_{\tau} < 0.3$.  Each point
is weighted by $\sigma_{\tau}^{-2}$ in the fitting
(using scipy.optimize routine curve\_fit) to fit the Gaussians 
shown as the solid curves on Figure \ref{fig:termv}.
We also compute the unweighted best fit Gaussians, i.e. weighting each point equally,
shown as the dashed curves. 
The resulting fit parameters for each velocity range are shown on the figures,
and tabulated on Table \ref{tab:Gaussians}.}

\textcolor{black}{
The right panel of Figure \ref{fig:termv} shows a
similar fit to the absorption toward the extragalactic sources, averaged over
a broader range, $\Delta v$=[-135,-100] km s$^{-1}$, with points for the Galactic
sources added to show the much larger scatter in absorption at the terminal velocity 
for this sample.  On the right panel, the 21-cm emission brightness temperature,
$T_b$, is the average of HI4PI spectra over the field area, integrated over the
same velocity range as for the absorption points. 
The $T_b$ scale on the right-hand axis is chosen so that the two curves peak
at the same height; there is a factor of 158 between the two scales.}

%The formal errors
%in the best-fit parameters for the unweighted fits (columns 6-8 on table \ref{tab:Gaussians}) are the square roots of the diagonal values of the covariance matrix.
%For the weighted fits
%the corresponding errors are unrealistically small, a more conservative estimate of the precision of the fitted values
%is the larger of the formal error of the unweighted fit parameters, and the difference between the
%weighted and unweighted results.  These are the errors given on table \ref{tab:Gaussians}, columns 2 - 4.

The value of $\sigma_b$ in the \textcolor{black}{Gaussian fits on Figure \ref{fig:termv}
and Table \ref{tab:Gaussians}} is particularly interesting; this gives an estimate for
the scale height ($\sigma_z$) of the CNM given the distance corresponding to the velocity
interval, $\Delta v$.  \textcolor{black}{These scale heights are given on Table \ref{tab:Scale_hts}.
At the terminal velocity the distance is
unique, while at the stepped-back velocity ranges of the lower panels on Figure \ref{fig:termv}
the two kinematic distances diverge, $d_1$ and $d_2$ on columns 4 and 5 \citep{Wenger_etal_2018}.  
These give two values for the scale height, $\sigma_z$ (columns 7 and 8).
In general the results for $\sigma_z$ are in the range 50 to 90 pc in the inner Galaxy.}
%Using $R_o = 8.31$ kpc,
%at $\ell = 340^o$ the sub-central point is at $R = 2.8$ kpc and the distance is 7.8 kpc, columns 3 and 4
%on Table \ref{tab:Scale_hts}.  Then the measured $\sigma_b$ from the Gaussian fits
%(column 3 Table \ref{tab:Gaussians} repeated on column 6, Table \ref{tab:Scale_hts}) gives the 
%scale height of the absorbing layer, $\sigma_z$=80 pc, column 7.
%Stepping back to $v_{LSR}$ = -112.5 and then to -87.5 km~s$^{-1}$, the two kinematic distances 
%\citep{Wenger_etal_2018} 
%lead to two values for $\sigma_z$ (columns 7 and 8).}

These values for $\sigma_z$, the scale height of the cool gas layer,
are considerably smaller than the estimate for the solar neighborhood,
$\sim$125 pc \citep{Lockman_1984, Lockman_Gehman_1991, Malhotra_1995, Ferriere_2001,  Su_etal_2021},
but this variation is to be expected given the gravitational effect of the superthin stellar
component in the bar, with scale height of just 45 pc, determined by \citet{Wegg_etal_2015}.
%The radial variation of the scale height is in good agreement with the values measured
%for the 21-cm emission in the first quadrant by \citet[][equation 21]{Celnik_etal_1979},
%although in their treatment they fit the self-consistent sech(z) functional form for a 
%self-gravitating disk.  Working with absorption instead of emission there is more 
%scatter, due to the low filling factor of CNM clouds.  
When a larger sample of absorption
spectra is available from the full GASKAP survey the improved precision
will \textcolor{black}{allow precision mapping of the scale height, warp and corrugations of the midplane,
and ultimately the filling factor of the CNM clouds as functions of $R$.}

The right hand panel of Figure \ref{fig:termv} includes the mean absorption
in the adjusted velocity range %$\Delta v $
$-135 < v < -100$ km~s$^{-1}$ for the
Galactic sources.  In this case there are many Galactic LoS that show no absorption at all
in this velocity interval, even though they are mostly all near latitude
$b=0^o$.  These are probably more nearby than the sub-central point.
There is a second group that shows about the same absorption as the
extragalactic sources, $(1-e^{-\tau}) \simeq 0.3$, and a third group
that shows much more absorption than most of the extragalactic sources,
from 0.4 to 0.65.  The latter group may correspond to continuum
sources near the end of the Galactic bar, where there is active star
formation and molecular gas.  The high optical depths would then be
due to the cold HI surrounding or mixed with the star forming clouds.

At the highest velocities (-125 to -145 km~s$^{-1}$, top left panel of
Figure \ref{fig:termv}), the extragalactic
absorption points resemble the Galactic sources on the right hand panel; 
the scatter around the fitted Gaussian is large, and there are some points
near latitude zero that show very little absorption, the
mean of $(1 - e^{-\tau})$ over this velocity range is below 0.1.
This suggests that absorption within 10 km~s$^{-1}$ of the terminal
velocity is spotty, sometimes there are CNM clouds very near the
sub-central point, sometimes not.  
So the CNM is not a \textcolor{black}{solid curtain, it has gaps that leave some LoS unabsorbed. } 
This should be kept in mind when assigning near-far
kinematic distances based on the presence or absence of absorption at
the terminal velocity.  When the absorption is present it is a good
indicator of the far side distance, but when absorption is not seen
in the last 10 km~s$^{-1}$ before the terminal velocity, this does not
prove that the source is at the near distance.  Averaging the absorption
spectrum over 25
km~s$^{-1}$ or more back from the terminal velocity, as in the lower two
panels on the left of Figure \ref{fig:termv}, is more conclusive.

%\hspace{.02in}\raisebox{2cm}{\includegraphics[width=3.6in]{Termv_1panel_22oct21_incl_Gal.pdf}
% updated 8 sept for wtmin=25, n=150  now 22 oct 21 wtmin=25 n=148 
%\FloatBarrier
%\startlongtable
%\begin{deluxetable}{|l|l|l|l|l|l|l|l|}
%\tablecaption{Gaussian Fits to the Absorption vs. Latitude near the Terminal Velocity
%\label{tab:Gaussians}}
%\tablehead{\colhead{velocity range} &
%\colhead{peak} & \colhead{b center} & \colhead{$\sigma{b}$ width} & \colhead{T${_s}$} & \colhead{peak}& \colhead{b center} & \colhead{$\sigma_b$b width}} \\ 
%\colhead{km s$^{-1}$} &
%\colhead{$1-e^{-\tau}$} & \colhead{(deg)} & \colhead{(deg)} & \colhead{(K)} &
%\colhead{$1-e^{-\tau}$} & \colhead{(deg)} & \colhead{(deg)}}\\
%
\begin{table} \caption{Gaussian Fits to Absorption vs. Latitude near the Terminal Velocity \label{tab:Gaussians}}
%\hspace{-2cm}\begin{tabular}{|l|lll|l|lll|}
\hspace{-2cm}\begin{tabular}{|c|ccc|c|ccc|}
\hline
 & \multicolumn{3}{c|}{weighted fit}& & \multicolumn{3}{c|}{unweighted fit}\\
velocity range &  peak & b center & $\sigma_b$ width &  T$_{s}$ &  peak & b center & $\sigma_b$ width  \\
km s$^{-1}$ &$1-e^{-\tau}$ & deg & deg & K & $1-e^{-\tau}$ & deg & deg \\ \hline
%\startdata
-145 $< v <$ -125 & 0.30$\pm$0.02 & 0.0$\pm$0.04 & 0.59$\pm$0.03  &   & 0.29$\pm$0.02 & +0.03$\pm$0.05 & 0.62$\pm$0.04 \\
\hline
-125 $<v<$ -100 & 0.27$\pm$0.02 & -0.05$\pm$0.03 & 0.51$\pm$0.03 &    & 0.28$\pm$0.02 & -0.05$\pm$0.04 & 0.58$\pm$0.04 \\
\hline
-100 $<v<$ -75  & 0.40$\pm$0.02 & +0.07$\pm$0.02 & 0.50$\pm$0.02 &    & 0.36$\pm$0.02 & +0.04$\pm$0.03 & 0.58$\pm$0.03  \\
\hline  
-135 $<v<$ -100 & 0.29$\pm$0.01 & -0.07$\pm$0.03 & 0.60$\pm$0.02 &  160 & 0.30$\pm$0.01 & -0.04$\pm$0.03 & 0.61$\pm$0.03  \\
\hline  
0 $<v<$ +8 & 0.37$\pm$0.03 & +0.11$\pm$0.05 & 0.59$\pm$0.05  & 123 & 0.37$\pm$0.03 & +0.08$\pm$0.05 & 0.47$\pm$0.06\\
$1-e^{-\tau}$ baseline & 0.34$\pm$0.02 &&&&  0.37$\pm$0.02&&\\
\hline 
+15 $<v<$ +30 & 0.39$\pm$0.02 & -0.62$\pm$0.03& 0.60$\pm$0.03 & 176 & 0.42$\pm$0.02 & -0.76$\pm$0.03 & 0.63$\pm$0.04 \\
\hline
+45 $<v<$+65 &0.05$\pm$0.01 & -0.27$\pm$0.052 & 1.20$\pm$0.06 & 325 & 0.05$\pm$0.01 & -0.15$\pm$0.20 & 1.43$\pm$0.25\\
\hline
\hline
\multicolumn{8}{|c|} {Two Phase Emission Fit} \\
\hline
& \multicolumn{3}{c|}{CNM} & T$_{CNM}$ & \multicolumn{3}{c|}{WNM}\\
-135 $<v<$ -100 & 35 K & -0.074 & 0.596 & 121 K & 11 K & +0.1 & 1.6 \\ 
\hline
\end{tabular}
\end{table}

\begin{table} \caption{\textcolor{black}{Scale Height of the Absorbing Layer \label{tab:Scale_hts}}}
\hspace{2cm}\begin{tabular}{|lrcccccc|}
\hline
$\Delta_v$ & $v_o$ & $R$ & $d_1$ & $d_2$ & $\sigma_b$ & $\sigma_{z1}$ & $\sigma_{z2}$ \\
 km s$^{-1}$ & km s$^{-1}$ & kpc & kpc & kpc &  & pc & pc \\
\hline
$\left[-145,-125\right]$ & -135 & 2.8 & 7.8 & \dots & 0.59$^{\circ}$ & 80 & \dots\\
$\left[-125, -100\right]$ & -112.5 & 3.0 & 6.5 & 9.3 & 0.51$^{\circ}$ & 58 & 83 \\
$\left[-100, -75\right]$ & -87.5 & 3.7 & 5.5 & 10.3 & 0.50$^{\circ}$ & 48 & 90 \\
$\left[0, +8\right]$ & +4  & 8.3 & 15.6 & \dots & 0.59$^{\circ}$ & 160 & \dots \\
$\left[+15, +30\right]$ & +22.5  & 12.5 & 20 & \dots & 0.60$^{\circ}$ & 210 & \dots \\
$\left[+45, +65\right]$ & +55  & 32 & 33 & \dots & 1.2$^{\circ}$ & 690 & \dots \\
\hline
\end{tabular}
\end{table}

\subsection{Comparison with Emission}

To measure the excitation temperature of the 21-cm line transition, i.e. the ``spin temperature", requires combination
of emission and absorption spectra.  The simplest, one-phase assumption gives
\begin{equation}
T_{s} \ = \ \frac{T_b}{1 - e^{-\tau}} \label{eq:tspin}
\end{equation}
where $T_b$ is the brightness temperature of the 21-cm emission corresponding to the
absorption, $e^{-\tau}$ \textcolor{black}{\citep[sec. 8.2]{Draine_2011}}.
\textcolor{black}{
Hereafter we will call $T_s$ the mean spin temperature, because it generally
represents an average of a mixture of different gas temperatures at the same velocity
along the line of sight.}
Since the spin temperature has a very wide range, from about 20 to 10$^4$ K \citep[e.g.][]{Murray_etal_2015,Murray_etal_2018}, values
derived from eq. \ref{eq:tspin} \textcolor{black}{can be quite different from either the CNM
or WNM temperatures.}  If two regions, $1$ and $2$, with low optical depths and
different spin temperatures, overlap in radial velocity, then the measured
value of $T_{s}$ is the harmonic mean:
\begin{equation}\label{eq:Tspin_2} T_{s} \ = \frac{N_1 + N_2}{\left(\frac{N_1}{T_1}\right)
+ \left(\frac{N_2}{T_2}\right)}  \ \ \ \ \ \ \ \ \left(\tau_1 , \tau_2 \ll 1\right) \end{equation}
where the column density, $N$, is the velocity integral of the emission brightness
temperature spectrum,
$T_b$, by
\begin{equation}\label{eq:N-T_b} \frac{N}{\rm cm^{-2}} \ = \ 1.82 \cdot 10^{18}\ \ \frac{\int_{\Delta v} T_b(v)\ dv}{\rm K\ \ km\ s^{-1}}
\hspace{.3in} \left( \tau \ll 1\right) \end{equation}
where the velocity range of the integral, $\Delta v$, corresponds to the velocity interval used in equation \ref{eq:equiv_w}. \textcolor{black}{We assume that the WNM contribution to $N$ is optically thin, so equation \ref{eq:N-T_b} gives the relationship between $N_w$ and brightness temperature for this phase in a two-phase approximation (Eq. \ref{eq:Tspin_2}).}  %\textcolor{black}{
The corresponding integral of the absorption spectrum gives:
\begin{equation}\label{eq:EW}
\frac{N}{T} \ \left(\frac{\rm K}{\rm cm^{-2}}\right) \ = \ 1.82 \cdot 10^{18}\ \ \frac{\int_{\Delta v} \tau(v)\ dv}{\rm  km\ s^{-1}}
\end{equation}
where $T$ on the left side would be the excitation temperature of the line if all the gas
were at the same temperature in the velocity range $\Delta v$.  
\textcolor{black}{In the two-phase approximation of Eq. \ref{eq:Tspin_2},  the left side of Eq. \ref{eq:EW}  becomes $\frac{N_{CNM}}{T_{CNM}}$, with $N_{CNM}$ the column density of cool phase gas, and
$T_{CNM}$ the cool phase temperature \citep[e.g.][]{Dickey_etal_2003}.}  Again, the atomic hydrogen shows variation
in temperature of at least two orders of magnitude, so $T_s$ does not represent a physical temperature,
but rather a harmonic mean as in equation \ref{eq:Tspin_2}.  
As such it is still a critically important quantity
for understanding the relative fractions of warm and cool phases.  A common use \citep[e.g.][]{Murray_etal_2020}
is to derive $f_c$, the fraction of cool phase HI by mass:
\begin{equation}\label{eq:fcool}
f_c \ \equiv \ \frac{N_{CNM}}{N_{CNM} + N_{WNM}}.  \end{equation}
\textcolor{black}{If the optical depth is small, then $f_c$ is simply:
\begin{equation}\label{eq:fcool2}
f_c \ = \ \frac{T_{CNM}}{T_s}  \ \ \ \ (\tau \ll 1)
\end{equation}
where $T_{CNM}$ is the cool phase excitation temperature, typically assumed to be 50 to 100 K based on results like
those of \citet{Heiles_Troland_2003} and \citet{Murray_etal_2018}.
If the optical depth is not small, then warm gas can be hidden behind the CNM.  
The two extremes are that all the WNM is behind all the CNM, and all the WNM is in front of the CNM.
They give lower and upper limits on $f_c$, respectively :
\begin{equation}\label{eq:fcool3}
\frac{\tau}{\tau \ + e^{\tau} \cdot \left(\frac{T_b}{T_{CNM}} - \frac{T_{b}}{T_s}\right)} \ \leq \ 
f_c \ \leq \ \frac{\tau}{\tau \ + \frac{T_b}{T_{CNM}} - \frac{T_{b}}{T_s}}
\end{equation} 
given the observed $T_b$ and $T_s$ and an assumed value of $T_{CNM}$.  For a 
typical case where half the WNM is in front of the CNM and half behind, then the simple formula
of Eq. \ref{eq:fcool2} is accurate to a few percent for $\tau \sim 1$.}
\textcolor{black}{For the analysis that follows we ignore self-absorption and compute $T_s$ from
Eq. \ref{eq:tspin} with the two-phase assumption of Eq. \ref{eq:Tspin_2} - \ref{eq:EW},  and leave more complex radiative transfer modelling to a future paper
when the GASKAP emission spectra are available.}

By comparing the emission as a function of
latitude, averaged over the area of the ASKAP field and over the same velocity 
intervals as for the absorption, we can compute the \textcolor{black}{mean} spin temperature of the 
gas, $T_s$, and estimate the cool phase temperature, $T_{CNM}$. 

The profile labelled ``emission" on the right hand panel of Figure \ref{fig:termv}
is an average of the emission spectra in the HI4PI survey \citep{Ben-Bekhti_etal_2016}.  
The HI4PI data in this area come from the Parkes GASS
survey \citep{McClure-Griffiths_etal_2009} and they have been carefully
corrected for stray radiation \citep{Kalberla_etal_2010}, thus the emission spectra
are trustworthy down to a 
small fraction of a K in brightness temperature.  For Figure \ref{fig:termv}
we average data from a rectangle defined by the black lines on Figure \ref{fig:num_vs_lat}, at each latitude we average the emission over the longitudes within
the PAF footprint.  For comparison with the absorption, we scale the
emission so that the peak coincides with the peak in the weighted Gaussian fit
to the measured absorption points.  
The scale factor is 1/158 K = 45.8/0.29, as the emission peak is 45.8 K
while the maxiumum absorption $(1 - e^{-\tau}) = 0.29$.    
If the HI gas were all at the same temperature, i.e. a
one-phase assumption, %e.g. \citet{Dickey_etal_2003},
then $T = T_{s} = 158$ K and the emission and absorption
profiles should look the same after this scaling.

At first glance the emission does look roughly the same as the absorption in
the right panel of
Figure \ref{fig:termv}, but the width (in latitude) of
the emission is broader than the Gaussian fitted to the absorption,
and at latitudes above $|b| \simeq 1^o$ the emission is well above the
absorption fit.  This would naturally result if the emission includes a contribution
from the WNM with a broader scale height than the CNM seen in absorption,
as
\textcolor{black}{
\begin{equation}\label{eq:2phase}
T_b \ = \ T_{b,WNM} \ + \ T_{CNM} \left( 1 \ - \ e^{-\tau} \right) 
\end{equation}
}
where $T_{b,WNM}$ is the emission due to the WNM, and the second term on the right
is the contribution of the CNM to the emission brightness
temperature, $T_b$.  Fitting the 
emission profile with two Gaussians shows how much WNM vs. CNM is needed.
We force the emission from the CNM component to have the same center and
width in latitude
as the absorption, and fit a Gaussian to the WNM component using only the 
data with $|b| > 1.2^o$. The result is shown on the last line of table \ref{tab:Gaussians}.
The Gaussian fitted to the emission wings has width $\sigma_b$ = 1.6$^o$ and peak
$T_b =$10.7 K centered at b=+0.1$^o$.  The remaining emission is due to the CNM, with peak 
45.8-10.7 = 35.1 K.  At midplane the CNM constitutes 35.1/45.8 = 77\% of the total
HI, but above $|b|= 1.5^o$ the HI is nearly all in the WNM phase.  We can estimate
the \textcolor{black}{mean} spin temperature of the CNM by dividing its contribution to the emission at 
midplane, 35.1 K, by the peak value of $\langle 1 - e^{-\tau}\rangle$, i.e. 35/0.29 = 121 K
(bottom line of Table \ref{tab:Gaussians}).

%Figure \ref{fig:two_phase} shows this in more detail.
%In the left panel is the same data as on the right panel of 
%Fig. \ref{fig:termv}, 
%with the emission decomposed into two components.
%One component of the emission is assumed to have the same center and width
%in latitude as the absorption Gaussian, fourth line on 
%Table \ref{tab:Gaussians}.  We identify this component with the CNM clouds.
%Using only data at the higher latitudes, $|b|>1^o$ or so, where the
%absorption profile has dropped almost to zero, we fit another Gaussian
%to the emission profile.  This is labelled ``Warm medium fit" on the panels
%of Fig. \ref{fig:two_phase}.
%
%Continuing with the assumption that the emission is the sum of CNM and
%WNM contributions, we add the fit to the warm medium to the fit to the
%absorption, with the latter scaled so that the sum of the two
%fits has the same peak
%value as the emission.  The scale factor needed is $T_{s}=$45 K.  The bottom line
%on Table \ref{tab:Gaussians} gives the values:  the peak of the emission
%is 18.3 K (plotted as 18.3/63 = 0.29 to match the absorption peak).
%This 18.3 K is the sum of 5.3 K in WNM plus 13 K in CNM, where the 
%13 K is 45$\times$0.29, i.e. T$_{s} \times (1 - e^{- \tau})$.

Even with this two-phase analysis, the 121 K result for the CNM \textcolor{black}{mean} spin
temperature represents a blend of HI temperatures.  The molecular
phase is so dominant in the inner Galaxy that a significant fraction
of the atomic hydrogen may be mixed in or between the molecular
clumps, with spin temperature in equilibrium with
the kinetic temperature of $\sim$50 K or less,
similar to $T_{s}$ values found in the Perseus molecular cloud by
\citet{Stanimirovic_etal_2014}.  The 121 K value would then
result from a blend of such very cold HI with some warmer gas, at
spin temperatures of 50 to several hundred K.  Careful analysis of individual
absorption features combined with the high resolution GASKAP emission
cubes may further clarify the
temperature distribution of the CNM in the 3 kpc Arm and near the
far end of the bar.  Study of the GASKAP emission cube will be the
topic of the next paper in this series.

\begin{figure}
\hspace{-.4in}\raisebox{2cm}{\includegraphics[width=4.0in]{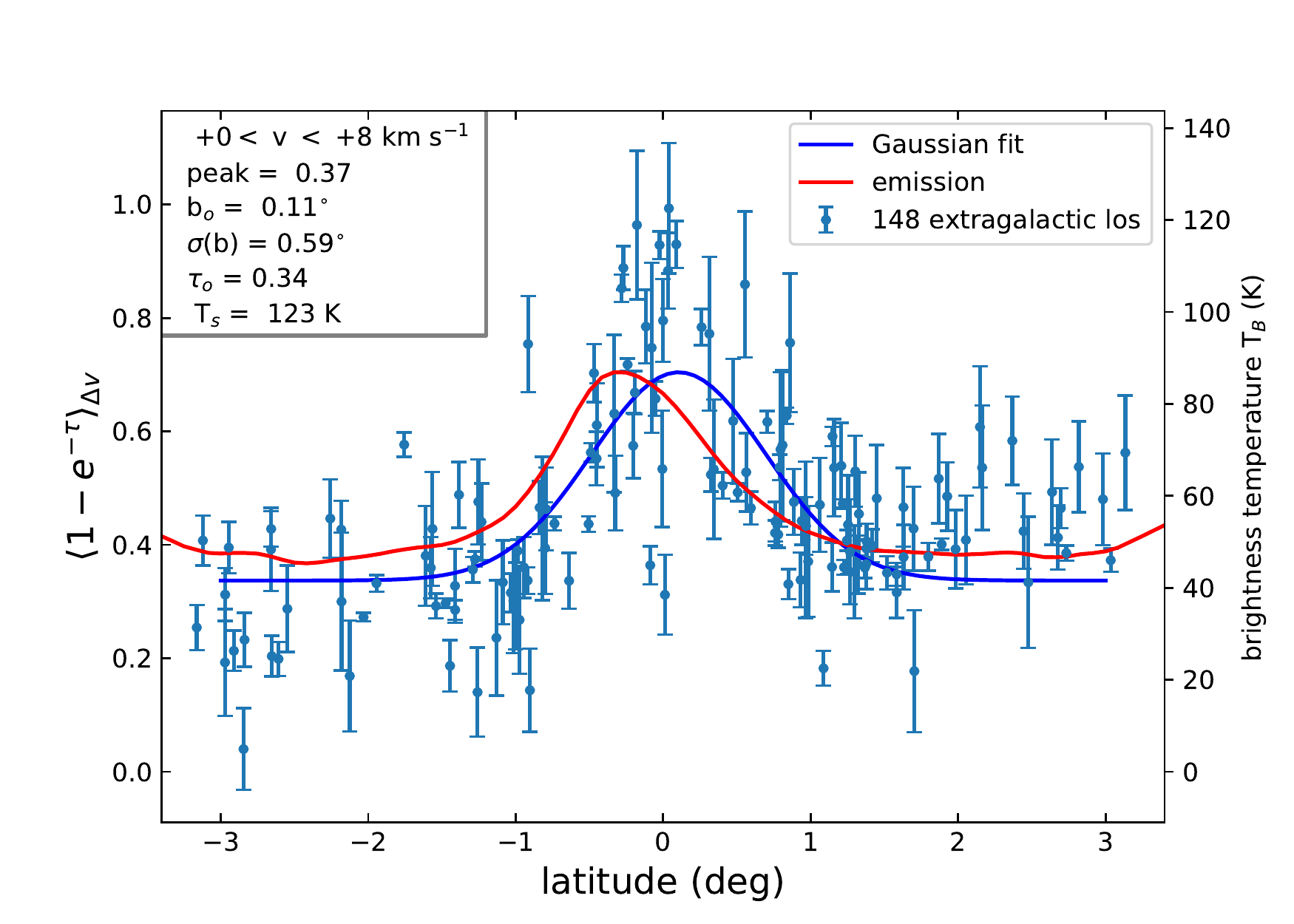}}
\hspace{-.2in}\includegraphics[width=3.7in]{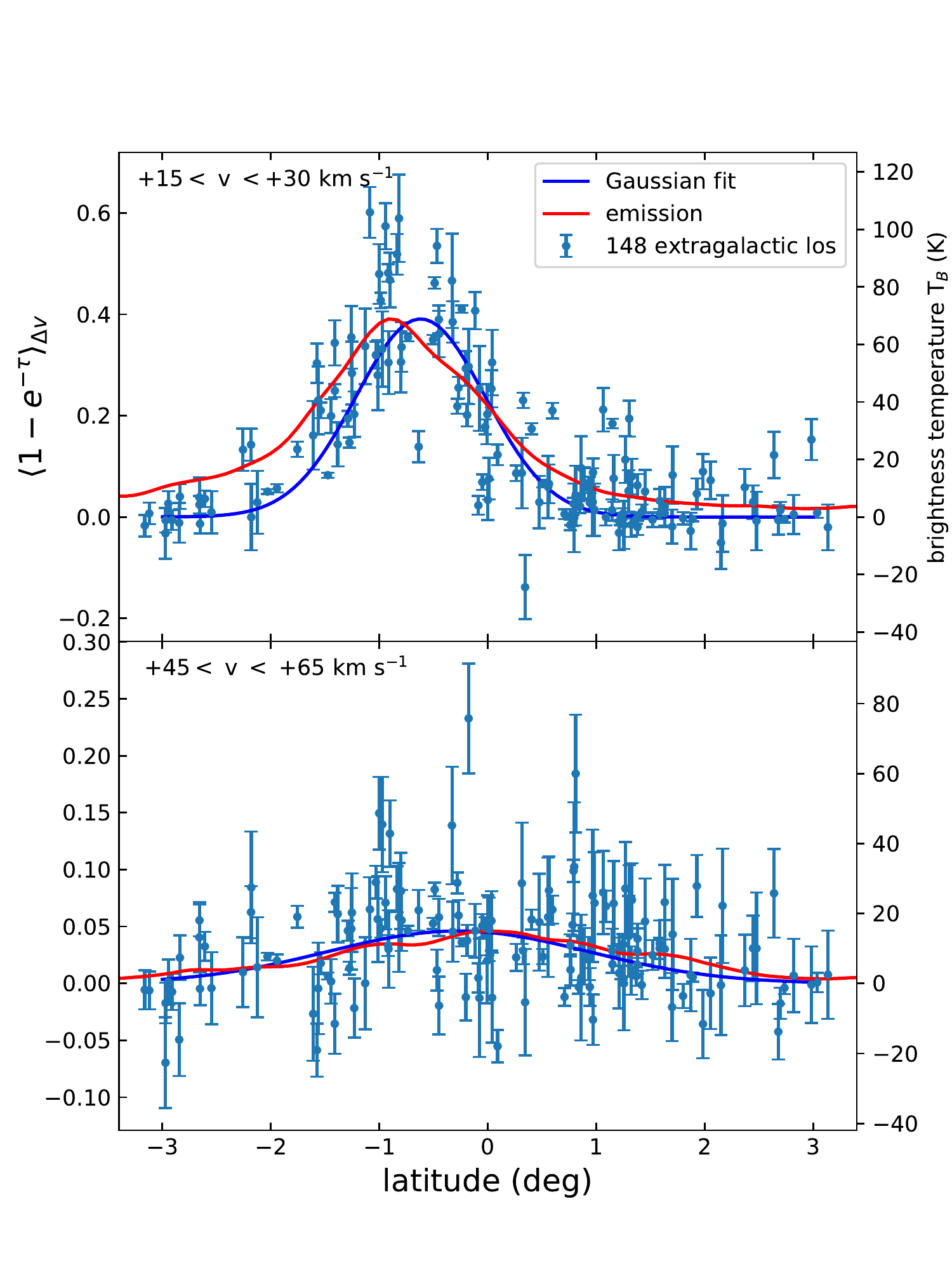}
\caption{Mean absorption vs. latitude at the solar circle and beyond.  
The left panel shows the latitude dependence of the absorption at velocities near
0 km s$^{-1}$, which selects gas at the solar circle both in the solar neighborhood and
on the far side of the Galaxy.  The two right panels show the latitude dependence
of the absorption beyond the solar circle on the far side of the Galaxy.  The 
upper panel displays the velocity range +15$<v<$+30 km s$^{-1}$, that includes the 
Sgr-Car Arm at velocity $\sim +25$ km s $^{-1}$ and distance about 20 kpc.
The lower panel shows the velocity range +45 to +65 km s$^{-1}$.  
\label{fig:Termv_posvel}}
\end{figure}

Figure \ref{fig:Termv_posvel} shows plots of $1 - e^{-\tau}$ vs. latitude
as on Figure \ref{fig:termv}, but now for $\Delta v$ intervals at
positive velocities, corresponding to gas near the solar circle, and then
progressively further outside the solar circle on the far side of the Galactic
center.  The
left panel shows absorption, $\langle 1-e^{-\tau}\rangle_{\Delta v}$, vs. latitude 
for velocity range $\Delta v$ = [0,+8] km s$^{-1}$.  The CNM that causes absorption
in this velocity range is partly in the solar neighborhood, and partly at or just
beyond the solar circle at a distance of 15.6 $\frac{R_o}{8.31 {\tiny \textrm{kpc}}}$
kpc.  The local gas is spread very widely in latitude, so widely that 
$\langle 1 - e^{-\tau}\rangle$ is roughly constant for -3$^o<b<+3^o$, while the distant CNM
has a Gaussian shape with width $\sigma_b=0.59^o$, similar to the width at the 
tangent point.  Since this gas is twice as far away, the physical scale height
of the CNM cloud population must be about 160 pc,
for a full width to half-maximum CNM cloud layer about 375 pc thick.
The peak of the Gaussian corresponding to the far side feature, $1 - e^{\tau} =0.37$,
is quite similar to the baseline level, $1 - e^{-\tau_o} = 0.34$, 
due to the local CNM (blue curve on Figure \ref{fig:Termv_posvel}), \textcolor{black}{
and similar to the results of surveys of absorption in nearby molecular clouds \citep{Nguyen_etal_2019}}.

The right panels on Figure \ref{fig:Termv_posvel} show profiles through the disk
at much greater Galactic radii, $R\sim 12$ kpc for the upper panel and 
20$<R<$40 kpc for the lower panel.  The upper panel shows that there is 
abundant CNM in the Sgr-Car arm at about 25 km s$^{-1}$.  The baseline
that was prominent due to local gas on the left panel, \textcolor{black}{($1 - e^{-\tau_o}) =  0.34$,}
has disappeared at these higher positive velocities, and the CNM layer is
continuing to flare, since the angular width, $\sigma_b \simeq 0.56^o$,
gives $\sigma_z =$200 pc for a distance of 20 kpc ($R = 1.5 R_o$).

The lower panel on the right side of Figure \ref{fig:Termv_posvel} shows that
at $R>3 R_o$ the situation changes considerably.  There is still a population
of CNM clouds, but their aggregate optical depth is much smaller, and 
their scale height is much greater.  Taking distance 33 kpc ($R \sim 4 R_o$)
and scale height given by the Gaussian fit, $\sigma_b \simeq 1.3^o$,
gives $\sigma_z$ about 750 pc.  The CNM traces the flaring of the disk just
as the HI emission does \citep[Fig. 7]{Kalberla_Dedes_2008}.
At this point the optical depth on each LoS is so small that we need to
stack (co-add) the spectra to get sufficient signal-to-noise to
study the relationship between the emission and absorption in this very
distant gas.

\section{The Outer Disk \label{sec:outerD}}

\textcolor{black}{
Moving radially across the MW outer disk, far outside the solar circle,
the HI is the dominant ISM constituent, and the 21-cm line is the only 
tracer of the structure and dynamics of the gas.  Using the emission and
absorption together, we can determine the conditions in the gas, at least
as far as the mass fractions of CNM and WNM.  A convenient quantity to use
is the ratio of emission to absorption, i.e. $T_s$ (Equation \ref{eq:tspin}).
Although it does not measure the kinetic temperature nor even the excitation temperature
of the gas, since it is always a weighted average of warm and cool phases
(Equation \ref{eq:Tspin_2}), $T_s$ is a convenient indicator of the presence or
absence of cool clouds in the warm, partially ionized surrounding gas
\citep{Bland-Hawthorn_etal_2017} that may include recently accreted 
circum-galactic material \citep{Stewart_etal_2011,Trapp_etal_2021}.
For a rough estimate of the mass ratio of the phases, $f_c$ (Equation \ref{eq:fcool}),
we only need to know $T_s$ and an estimate of $T_{CNM}$, the cool phase temperature,
predicted by e.g. \citet[][fig. 6]{Bialy_Sternberg_2019}.  Even without knowing
$T_{CNM}$ or $f_c$, we can use $T_s$ to convert between column density, Eq. \ref{eq:N-T_b},
and equivalent width, Eq. \ref{eq:EW}, for example to interpret the results of
extragalactic absorption line surveys \citep[e.g. FLASH][]{Allison_2021}, since
$T_s$ is defined as the ratio of emission to absorption (Eq. \ref{eq:tspin}).
}

In the outer Galaxy, far outside the solar circle, kinematic distances
depend hardly at all on the choice of rotation curve.  A simple, flat
rotation curve, with $\Theta(R_o) = R \ \Omega(R_o) = \Theta_o$ gives
very nearly the same distance vs. radial velocity, $v_r$, as the
\citet{Brand_Blitz_1993} or the
\citet{Reid_etal_2014, Reid_etal_2019} A5 model curves,  
using the same solar neighborhood parameters, 
$\Theta(R)=\Theta_o=240$ km s $^{-1}$, and $R_{o} = 8.31$ kpc.
At longitude $\ell=340^o$ the difference in velocity predicted by the 
A5 curve and a flat curve is only 0.3 km s$^{-1}$ at $R$=20 kpc,
increasing to
just 0.45 km s$^{-1}$ at $R=40$ kpc.  In this section we use the flat curve to
estimate distances for absorption features at positive velocities,
i.e. beyond the solar circle on the far side of the Galaxy.

\begin{figure}[h]
\hspace{2.2in}\includegraphics[width=3.5in]{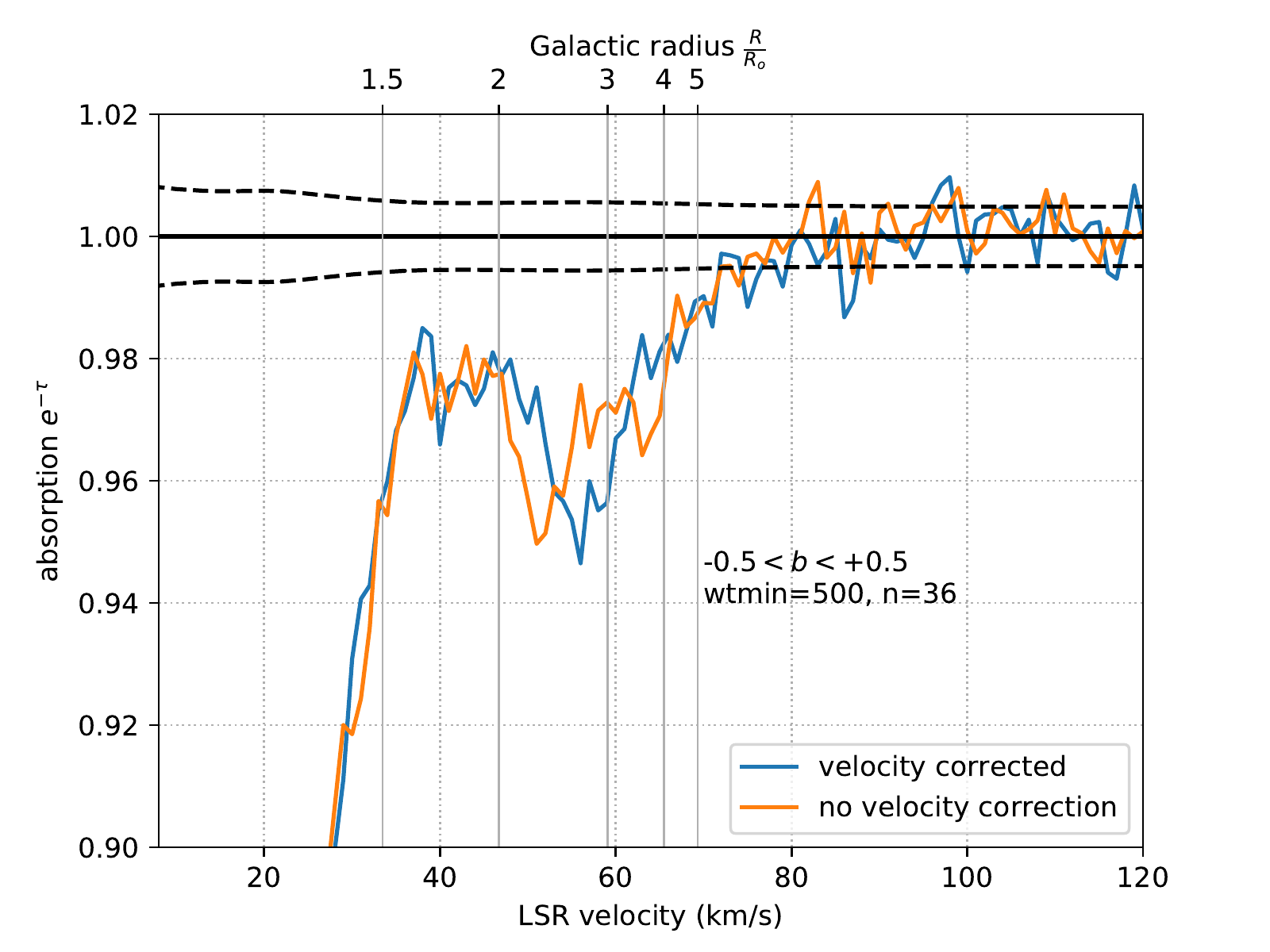}

\caption{The effect of aligning spectral features at the same distance
by shifting the velocity scales before averaging, using a factor of 
$\frac{\sin{(\ell)}}{\sin{340^o}}$ (equation \ref{eq:v_of_R}).  The spectra 
are weighted averages of the 36 absorption spectra with the lowest noise
($\sigma_{\tau} < 0.045$) at low latitudes 
($-0.5 < b < +0.5$).  The velocity corrected spectrum is the average
after adjusting the velocity scales using equation \ref{eq:v_of_R}, the other
is the average without adjusting.  The faint absorption line at $v \sim +55$ km s$^{-1}$
is narrower in the velocity corrected stack, indicating that many of the 36
LoS contributing to the average at different longitudes all show absorption due
to CNM at roughly the same distance. 
The upper scale on the x-axis shows the corresponding kinematic galactocentric radius, $R$.  \label{fig:vscales}}
\end{figure}

Using a flat rotation curve, kinematic distances vary as sin$(\ell)$ simply due to the change
in the projection of the LSR velocity along the LoS:
\begin{equation}
 v_r \ = \ \Theta_{o} \left[ \frac{R_{0}}{R} \ - \ 1\right]  \sin{(\ell)} \ \cos{(b)} \end{equation}
becomes 
\begin{equation}
\frac{v_r}{\Theta_{o} \ \sin{(\ell)}} \ = \ \left[ \frac{R_{0}}{R} \ - \ 1\right]
\label{eq:v_of_R}
\end{equation}
% def Rofv(v):
%   radeg=180/np.pi
%   f1=1.+(v/(240*np.sin(-20/radeg)))
%   R = 8.31/f1  
%   return R
%     gives Rofv(50)=21.3 vs. Wenger tool: 21.3 (+2.65,-0.74) kpc
at latitude $b$=0,
where $v_r$ is the observed LSR velocity of a spectral feature, and
$R$ is its galactocentric radius.  We adjust 
the velocity scales of all the spectra 
%to match the distance vs.  velocity law at $\ell=340^o$ 
by multiplying the velocities by 
$\frac{\sin{\ell}}{\sin{(340^o)}}$
to align features at the same distance but different longitudes to the
same standard velocity scale, i.e. equation \ref{eq:v_of_R} with
$\ell \equiv 340^o$.
Figure \ref{fig:vscales} shows the effect of this correction for the
average of the 36 highest sensitivity spectra toward extragalactic
sources with latitude -0.5$^o < b < +0.5^o$.  The result is that 
the broad, noisy feature between about 45 and 65 km s$^{-1}$
in the raw average narrows to a width of about 5 km s$^{-1}$
centered at about 55 km s$^{-1}$, a kinematic Galactic 
radius $R \simeq 3 R_0$ or $\sim$25 kpc.
Kinematic distances become quite uncertain at large $R$ because
the velocity gradient is small; $\frac{dv}{dR} \sim 0.5$ km s$^{-1}$ kpc$^{-1}$
at $\ell = 340^o$ and $R=4R_o$,
so that random motions of $\sim$ 5 km s$^{-1}$ can give errors in $R$ of 10 kpc.
In spite of this uncertainty, it is worthwhile to shift all spectra to a common
distance scale using equation \ref{eq:v_of_R} before stacking so that absorption
features at the same distance will align in velocity.

The spectra shown in Figure \ref{fig:vscales} are weighted averages 
\begin{equation}
\langle S(v)\rangle \ = \ \frac{\sum_i \ w_i \times S_i(v)}{\sum_i w_i}
\label{eq:average}
\end{equation}
where $S_i(v)$ are the absorption spectra ($e^{-\tau}(v)$) and
the weights $w_i$ are inversely proportional to the off-line rms noise, $\sigma_i$, squared:
\begin{equation}
w_i \ = \ \frac{1}{\sigma_i^2} 
\label{eq:wts}
\end{equation}
The error envelope of this average spectrum is 
the weighted average of the noise spectra, given by
\begin{equation}
\langle E(v)\rangle \ = \frac{\sqrt{\sum \ \left(w_i \times E_i\right)^2}}{\sum_i w_i}
\label{eq:weighted_error}
\end{equation}
where the individual spectra $S_i(v)$ and their respective
noise envelopes, $E_i(v)$, have been shifted by $\frac{\sin{(\ell)}}{\sin{340^o}}$ as discussed above.  For the averages analysed in this section, the maximum allowed weight is 900, 
%($\sigma_{\tau} = 0.032$)
 spectra with lower noise are weighted by 900, so that a small number do not dominate the average.  Spectra with weight less than 10 ($\sigma_{\tau} > 0.31$) are not used in
the averaging at all.

For comparison with 21-cm emission, we use the HI4PI compendium of surveys \citep{Ben-Bekhti_etal_2016}, taking spectra at the same positions as the absorption spectra, and using the
same weights ($w_i$) to compute the average emission.  The 64m Parkes telescope, which provided
all the data in HI4PI at these declinations, does not have a high enough gain to detect
the absorption toward the very compact emission of the extragalactic background
sources.  Some of the Galactic continuum sources are surrounded by extended continuum
emission. % (Figure \ref{fig:contin_maps}).  
For these a careful comparison
of emission and absorption will require the 
full ASKAP emission maps.  But to study the absorption in the outer Galaxy we use only the
extragalactic continuum source sample.

\subsection{Absorption Lines}

\begin{figure}[h]
\hspace{.2in}\includegraphics[width=3.5in]{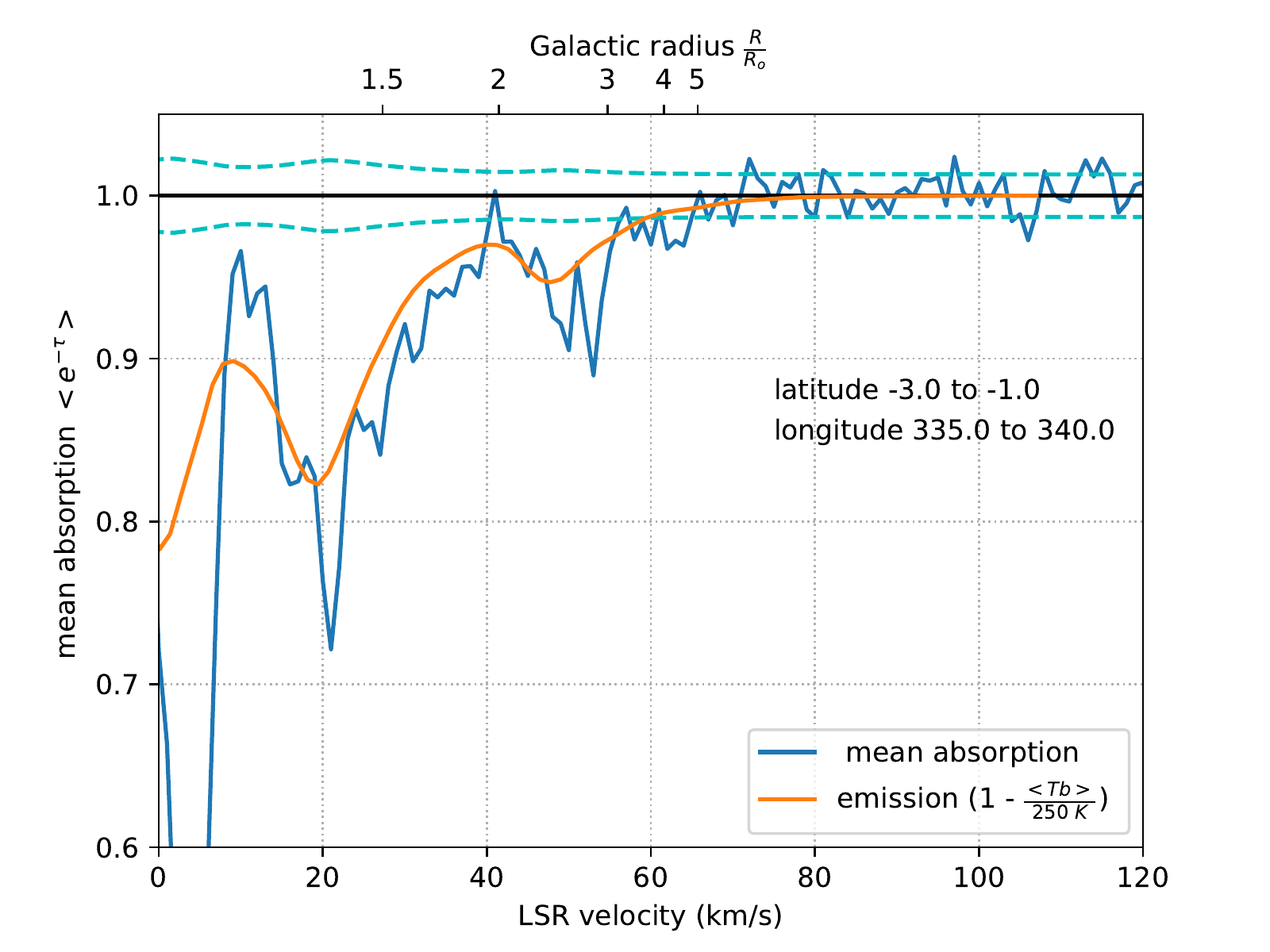}
\hspace{-.3in}\includegraphics[width=3.5in]{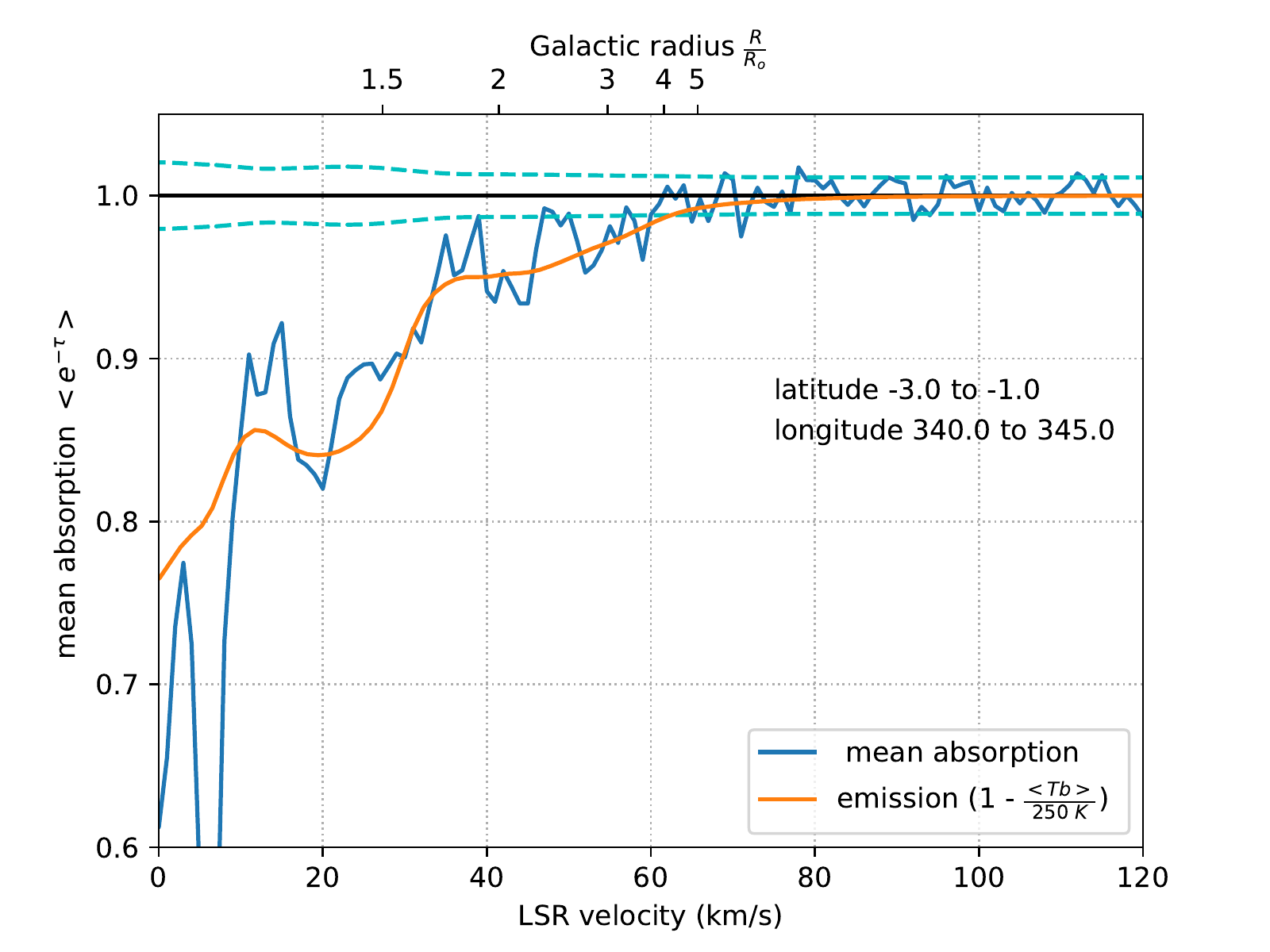}

\caption{Averages of spectra at negative latitudes $-3^o < b <  -1^o$, 
with the lower longitudes on the left ($335^o<\ell<340^o$) and the higher longitudes on
the right ($340^o<\ell<345^o$).  The weighted average absorption spectrum
is shown, and the similarly weighted average emission, $\langle  T_b \rangle$, is shown,
scaled to match the nominal absorption at $T_{s}=250$ K. If $T_{s} < 250$ K then
the absorption profile is below the scaled emission, if $T_{s} > 250$ K then
the absorption is above the emission.
The noise envelopes shown $\pm 1 \sigma$ assuming the spectra are independent.
%baseline, the upper dashed line shows the rms vs. velocity expected from the average of
%the high weight spectra, while the lower dashed curve shows the rms vs. velocity 
%expected for the low weight spectral average (equation \ref{eq:weighted_error}).  
The upper scale on the x-axis shows the kinematic galactocentric radius, $R$,
in units of the solar circle radius, $R_0$, assuming a flat rotation curve.
\label{fig:av_neglat}}
\end{figure}

\begin{figure}[h]
\hspace{.2in}\includegraphics[width=3.5in]{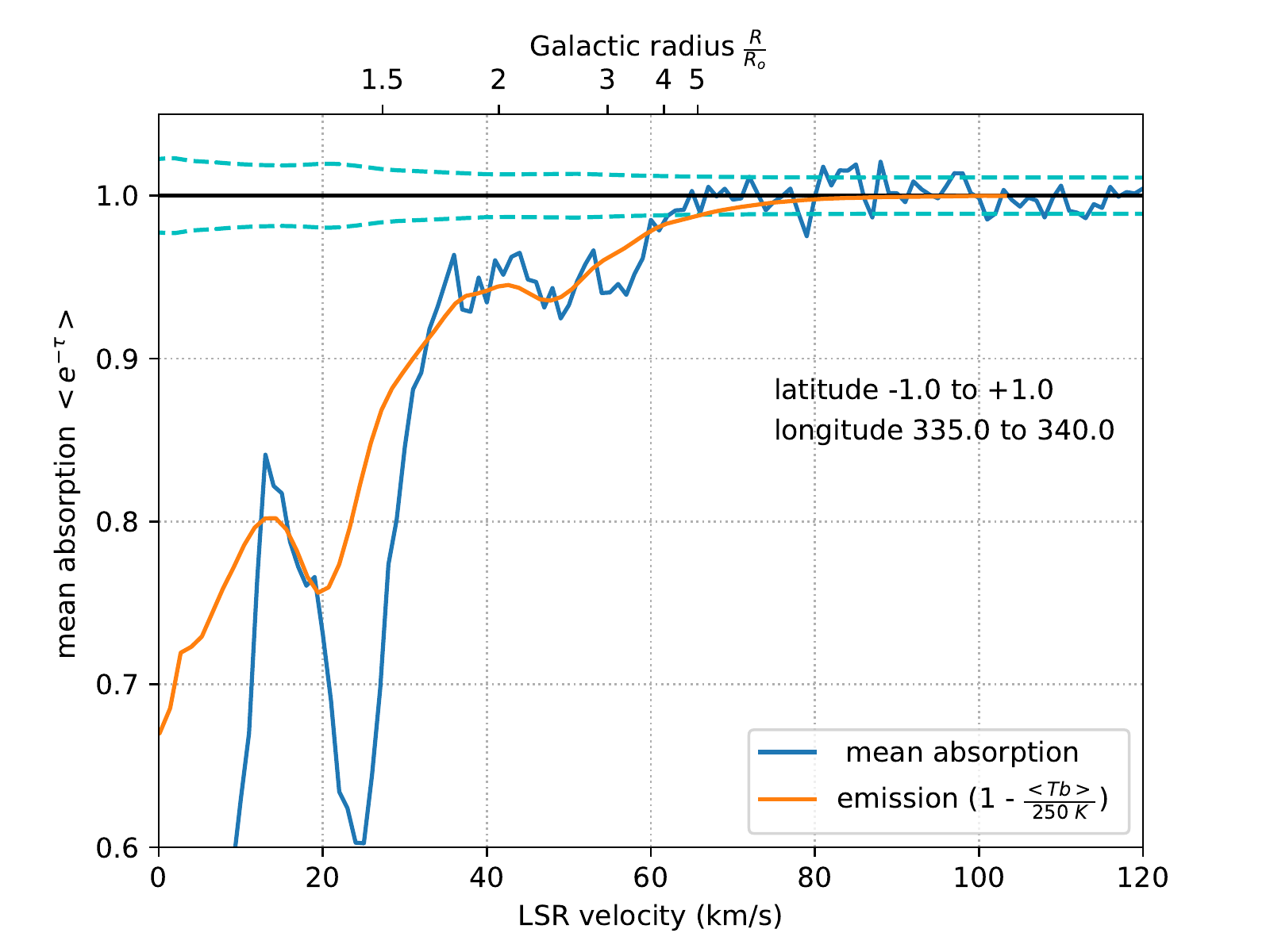}
\hspace{-.3in}\includegraphics[width=3.5in]{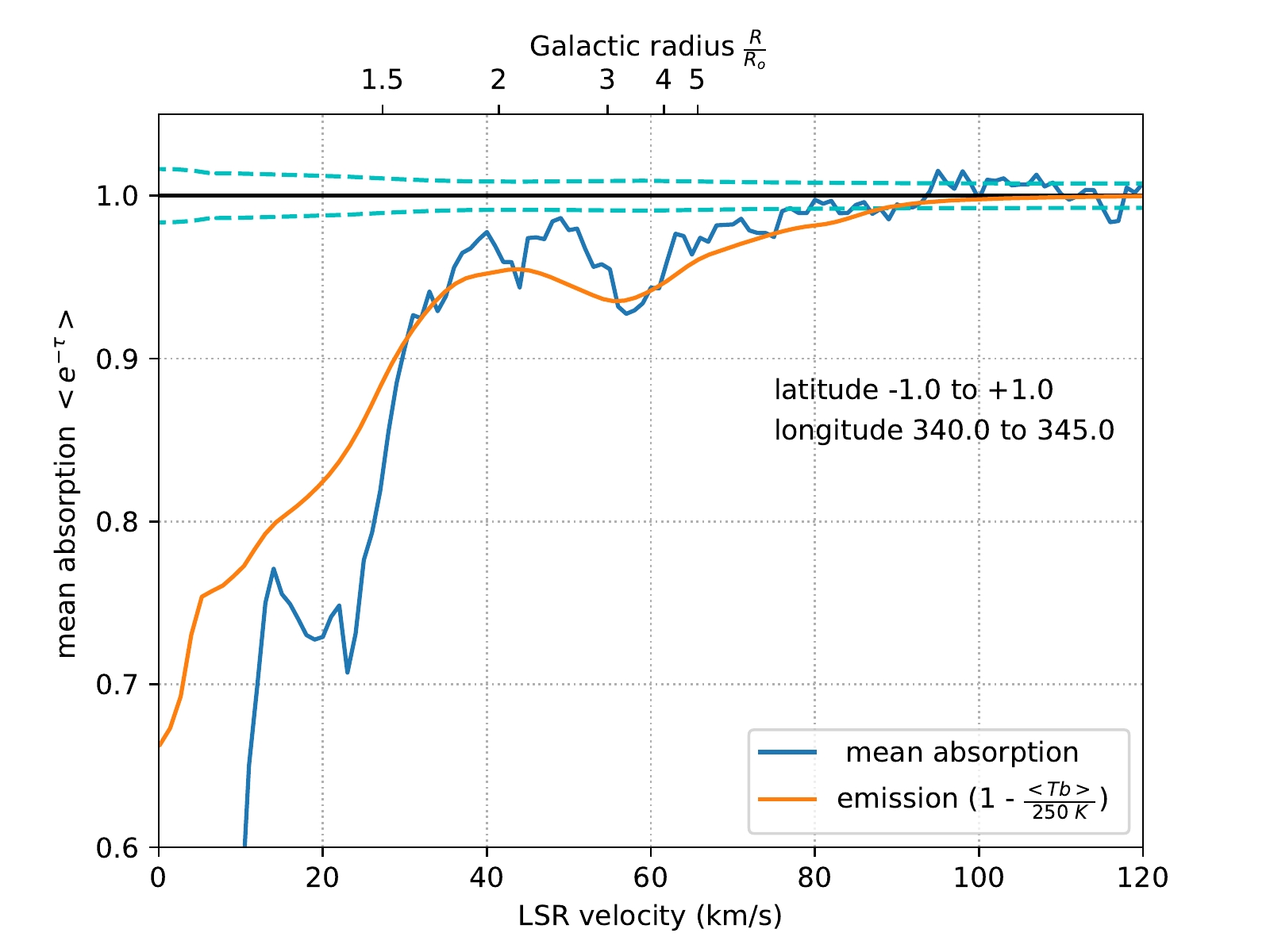}

\caption{Averages of spectra at low latitudes,  $-1^o < b <  +1^o$,
with the lower longitudes on the left ($335^o<\ell<340^o$) and the higher longitudes on
the right ($340^o<\ell<345^o$), as in Figure \ref{fig:av_neglat}.  In this case
the absorption in the Sgr-Car Arm with $v$ between 20 and 30 km s$^{-1}$ is
mostly much cooler
than the nominal $T_{s}$ = 250 K used for scaling the emission, but 250 K is typical for the very distant gas at 50$<v<$65. 
\label{fig:av_lowlat}}
\end{figure}

\begin{figure}[h]
\hspace{-.1in}\includegraphics[width=3.5in]{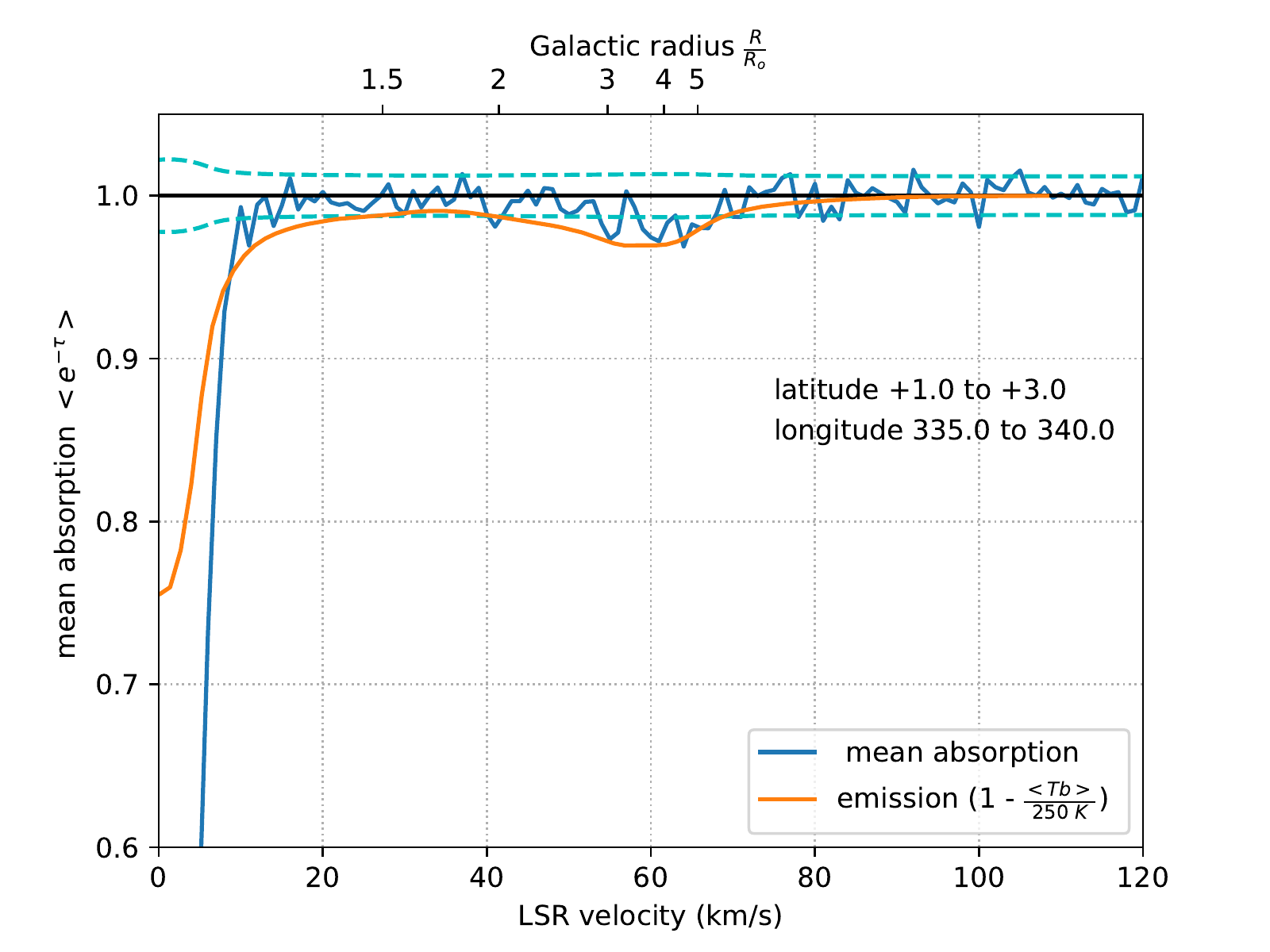}
\hspace{-.2in}\includegraphics[width=3.5in]{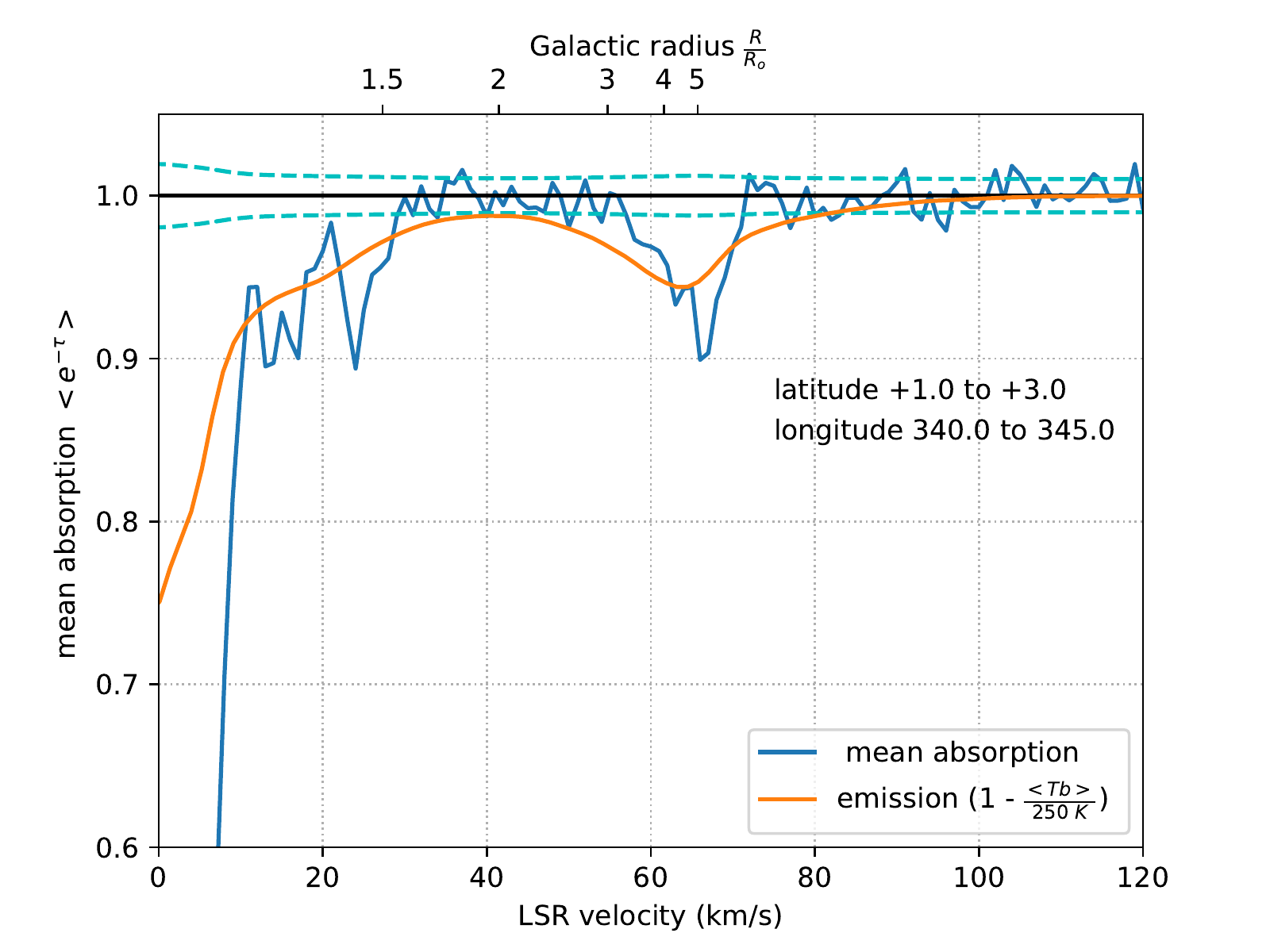}

\caption{Averages of spectra at positive latitudes, $+1^o < b < +3^o$,
with the lower longitudes on the left ($335^o<\ell<340^o$) and the higher longitudes on
the right ($340^o<\ell<345^o$), as in Figure \ref{fig:av_neglat}.
\label{fig:av_poslat}}
\end{figure}

There are not enough background sources in the GASKAP field to allow mapping of the
spatial distribution of the CNM gas that causes the absorption in the outer Galaxy.
%but not enough to draw contour diagrams of optical depth at specific velocities.
For a rough, low resolution study of the continuity of the absorbing clouds
over the field, we divide the area into six sub-fields of roughly equal areas:
two longitude ranges, $336^o < \ell < 340^o$ (low longitudes) and
$340^o < \ell < 344^o$ (high longitudes), for each of three latitude ranges:
$-3^o < b < -1^o$ (negative latitudes), $-1^o < b < +1^o$ (low latitudes),
and $+1^o < b < +3^o$ (positive latitudes).
%As a consistency check, for each area we divide the sample of spectra into two groups:
%the few spectra (4 to 12) with weights greater than 450 ($\sigma_{\tau} < 0.047$) and
%the many (15 to 42) with lower weight.  The average spectra for the two
%samples 
The stacked spectra for each region are shown in Figures
\ref{fig:av_neglat}, \ref{fig:av_lowlat}, and \ref{fig:av_poslat}.

% Denis comment: Figs. 9, 10 and 11 are hard to read (the rest are very clear). Suggest to keep same color scheme but make 1-\langle Tb>/150K as dotted lines to reduce difficulty of discerning the lines.
%You could also omit velocities greater than 100 km/s so that the interesting part is spread over a larger part of the plot.

Almost all the regions show a distinct absorption line centered at about +20 km s$^{-1}$,
with the exception of the region at positive latitude ($+1^o < b < +3^o$) and lower longitudes
(335$^o <\ell < 340^o$) (Figure \ref{fig:av_poslat}, left panel).  Two blended, narrow
lines centered at about +60 and +65 km s$^{-1}$ that are prominent in the 
right hand panel of Figure \ref{fig:av_poslat} (positive latitudes and higher longitude)
are seen 
faintly in the low latitudes at high longitudes
and very faintly in the lower longitudes at positive latitudes 
but not elsewhere.  The spectrum at low latitudes and high longitudes,
Figure \ref{fig:av_lowlat}, right hand panel, shows a feature centered on
$\sim$+58 km s$^{-1}$, 
also clearly seen on Figure \ref{fig:vscales}.
At negative latitudes and low longitudes there are two features between +45 and +55 
km s$^{-1}$, Figure \ref{fig:av_neglat}, left panel.

\subsection{Warm and Cool Phases at Large $R$}

\begin{figure}[h]
\hspace{+.7in}\includegraphics[width=5.5in]{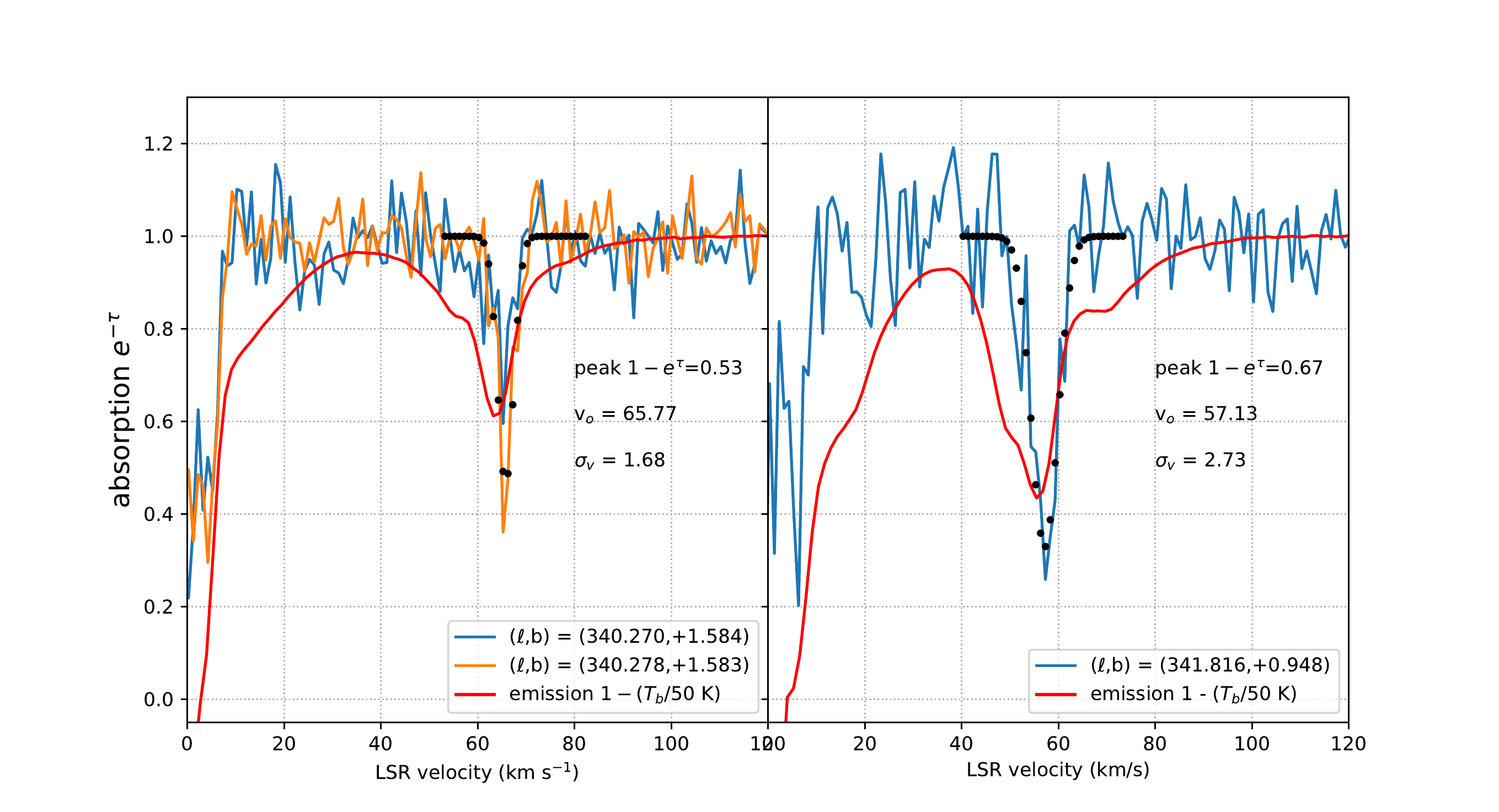}

\caption{Representative absorption lines in the far outer Galaxy.  The left panel
shows absorption spectra toward two components of a double source at
($\ell$,b) = (340.27,+1.58), the right panel shows the spectrum toward a source
at (341.82,+0.95).  Gaussian fits to the deeper line in the left panel and the
line in the right panel are shown.  The HI4PI emission spectra are plotted for 
comparison as $1 - \left(\frac{T_b}{50 {\textrm \small K}} \right)$.  Both absorption
lines go below the emission at their lowest points, suggesting that the spin
temperature is less than 50 K.
\label{fig:Two_lines}}
\end{figure}

Two examples of clouds detected on specific LoS at high positive
velocities are shown on Figure \ref{fig:Two_lines}.
Gaussian fits to $e^{-\tau}$ are shown.  For the narrower line, with
$\sigma_v \simeq 1.7$ km s$^{-1}$, 
the implied upper limit kinetic temperature is
$T_D = 340$ K, from
\begin{equation}
\frac{T_{D}}{\textrm{K}} \ = \ 121 \ \left(\frac{\sigma_v}{\textrm{km \ s}^{-1}}\right)^2
\label{eq:Tdop}
\end{equation}
where $\sigma_v$ is the fitted velocity dispersion of the 
absorption line.  Generally the line width has a turbulent contribution as well
as the Maxwellian thermal velocity width, so that $T_{s}$ is typically less
than half of $T_D$
\citep[e.g.][]{Jameson_etal_2019,Murray_etal_2018}.  The red curves on Figure 
\ref{fig:Two_lines} represent the HI4PI emission, plotted upside down and scaled by 
1/(50 K).  In both cases the absorption goes well below the emission, which
indicates that the spin temperatures of the clouds are less than 50 K.  Since the
HI4PI emission is based on the GASS Parkes survey with a beam width of 14\arcmin,
the emission-absorption comparison on a single LoS is tentative; more accurate
spin temperature analysis will be possible with the full resolution emission 
cube from ASKAP.

These CNM clouds are similar to the large, cool clouds found at
high $R$ in the second quadrant by \citet{Strasser_etal_2007}.  Most of the
Strasser et al. clouds show \textcolor{black}{mean} spin temperatures between 60
and 120 K (their Table 3) and Galactic radii $R$ in the range 12 to 27 kpc.
%In this field 
The clouds in Figure \ref{fig:Two_lines} are more distant ($R\simeq$ 27 to 42 kpc)
but otherwise consistent with the properties of the clouds in that survey.
%We may speculate that the extreme outer disk is rich in CNM clouds.

%\caption{
%Velocity integrals of the average emission and absorption in two samples of spectra toward
%different groups of 47 background sources each (high weight and low weight).  For
%each source, the emission and absorption
%spectra are integrated over 10 km s$^{-1}$ intervals, from +10 to +100 km s$^{-1}$.
%All longitudes and latitudes are included.  In each 10 km s$^{-1}$ interval, we
%construct the distribution of mean optical depth for each spectrum in the sample.
%The vertical lines indicate the standard deviations about the means plotted as points.
%The expected error on the mean, about 1/7 times the standard deviations, are indicated
%by the end caps on the heavier weighted lines.  In some cases these are smaller than the
%points themselves. Green points show the averages of absorption (1-$e^{-\tau}$), and 
%red points show the corresponding averages of the emission, $T_b$, scaled by 1/250K,
%which is a typical value of $T_{s}$ outside the solar circle.  The red and green points
%are offset in velocity for clarity, but they correspond to the same 10 km s$^{-1}$ 
%averaging intervals.  The left panel shows the results for 47 sources with intermediate
%weights (100 to 400), the right panel shows the results for 47 sources with the highest
%weights ($wt > 400$).  Weights were not applied to computation of the means and 
%standard deviations of the samples.
%\label{fig:Velavg1}}

\begin{figure}[h]
\hspace{-.05in}\includegraphics[width=3.5in]{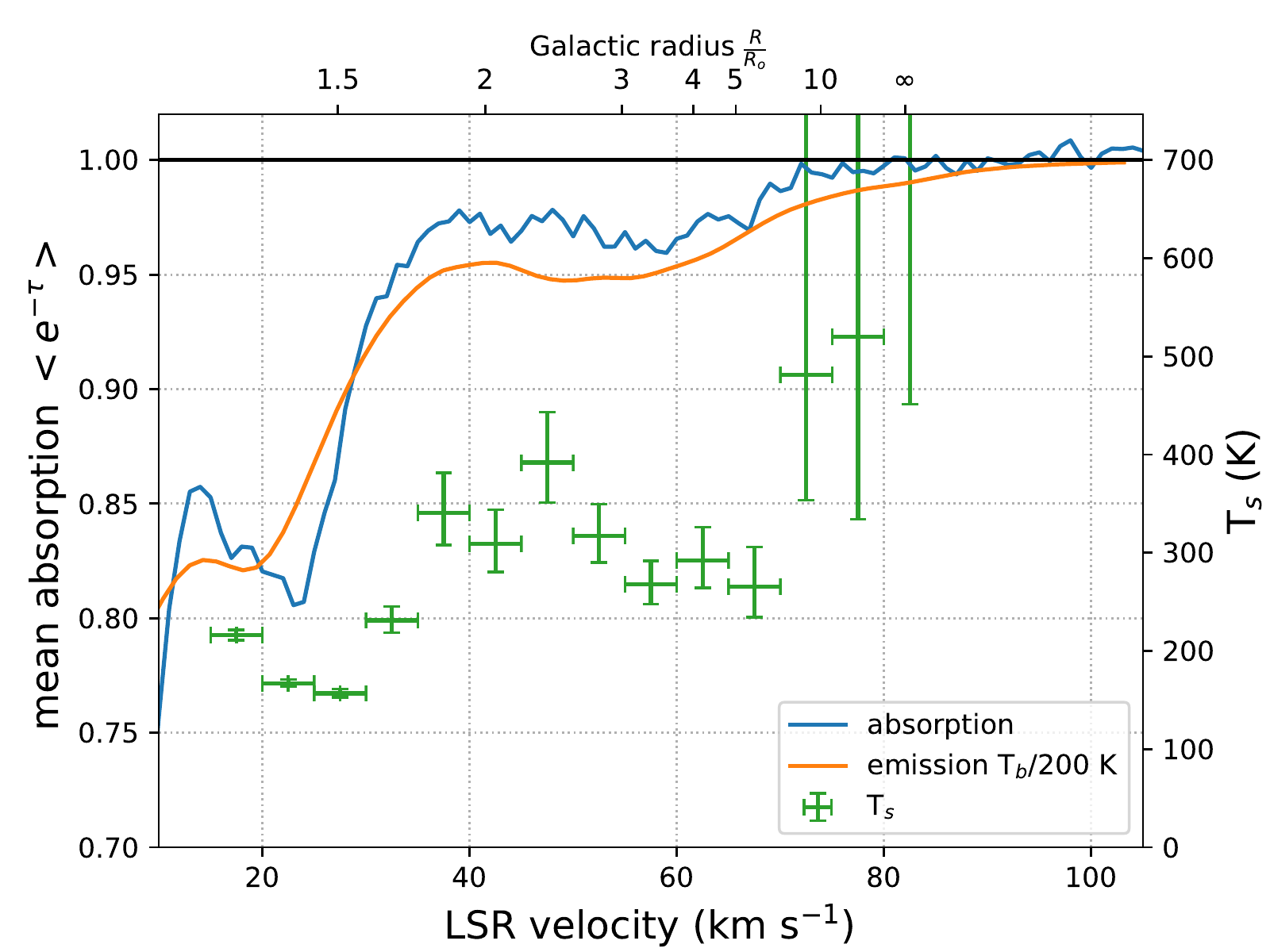}
\hspace{+.1in}\includegraphics[width=3.5in]{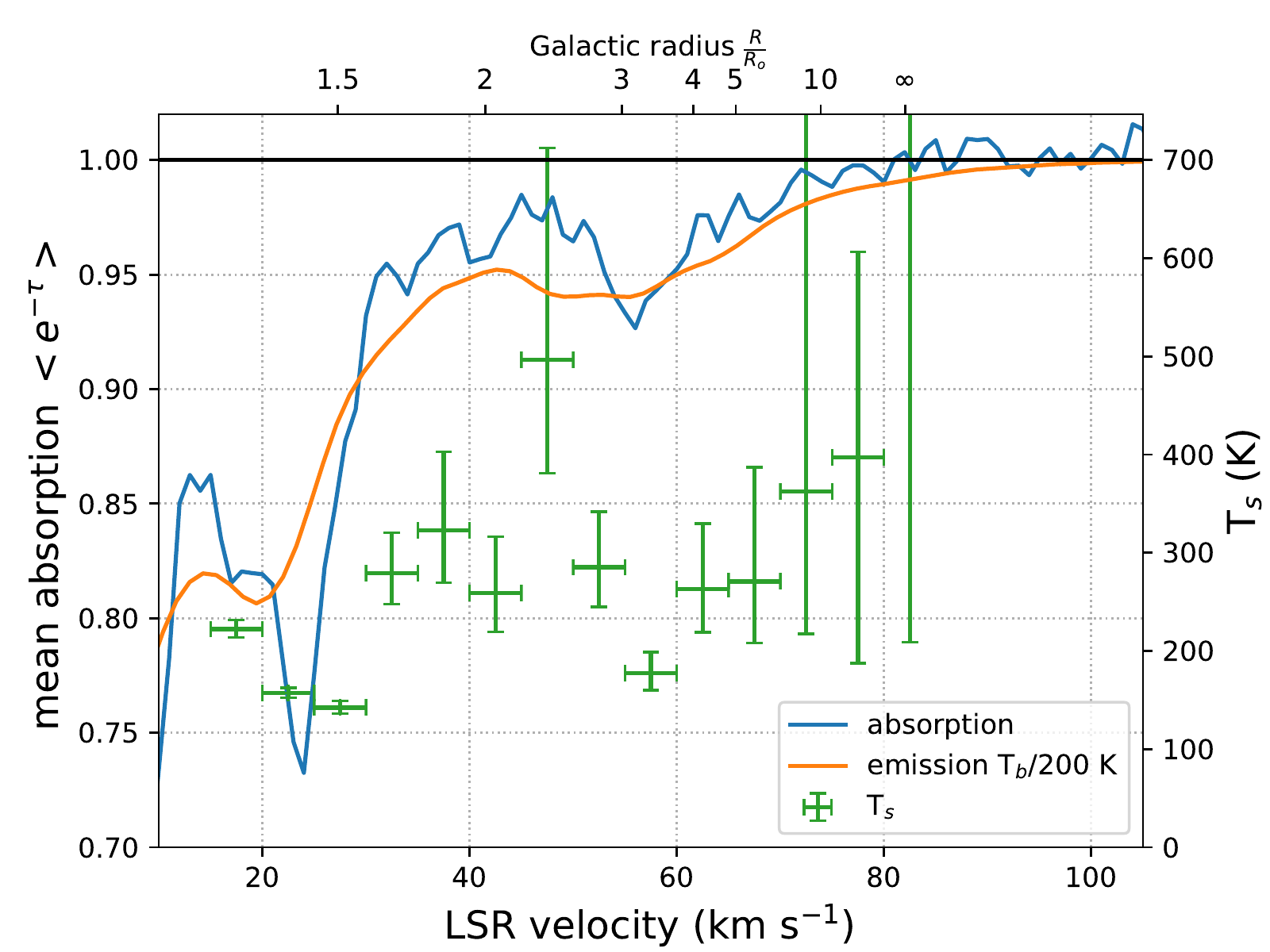}

\caption{The variation with velocity of the mean 
spin temperature, computed as the ratio of the sample mean
values of $T_b$ and $1 - e^{-\tau}$.
The corresponding values of Galactic radius, $R$ are
indicated on the upper x-axis scale in units of $R_o$, the solar circle radius.
The left panel shows overall averages of all 175 extragalactic background sources
in the PAF field.  The right panel shows the average for a subset of 
spectra with very low noise that show significant absorption lines in the 
range $55 < v < 66$ km s$^{-1}$.
\label{fig:tspin_vs_v}}
\end{figure}

To estimate the radial variation of the mean spin temperature, i.e. the ratio of
emission divided by absorption in equation \ref{eq:tspin}, we consider two 
averages of absorption and emission spectra, shown on the two panels of
Figure \ref{fig:tspin_vs_v}.  In the left panel is the weighted average of 
all 175 spectra toward extragalactic sources with noise $\sigma_{\tau} <$
0.32, giving weight $w_i >$ 10 in equation \ref{eq:wts}.
The emission spectra and the optical depth noise spectra are averaged using
the same weights, as for Figures 
\ref{fig:av_neglat}, \ref{fig:av_lowlat}, and \ref{fig:av_poslat}.
This grand average has optical depth noise $\sigma_{\tau}=0.0042$ in each
1 km s$^{-1}$ velocity channel.  Averaging channels in groups of five
reduces the noise to 0.0030.  Since the emission (red curve) is about 
0.05 * 200 = 10 K in the velocity range 40 to 70 km s$^{-1}$, and the
absorption is $\tau \sim (1 - e^{-\tau}) \sim 0.035$, \textcolor{black}{mean} spin temperatures
are in the range 260 to 310 K (one sigma).  The radiometer noise in the
weighted average emission spectrum is insignificant: the discrepency
in beam size between the emission and absorption is the main source of
uncertainty, as discussed above.  The \textcolor{black}{mean} spin temperatures computed for the
5 km s$^{-1}$ averages are shown as the green crosses on the left panel
of Figure \ref{fig:tspin_vs_v}.  For velocities between 20 and 30 km s$^{-1}$
corresponding to the Sgr-Car Arm, the \textcolor{black}{mean} spin temperatures are relatively low,
$T_{s} = 170$ and 160 K,  similar to solar neighborhood values seen at
high latitudes \citep{Heiles_Troland_2003,Murray_etal_2018}.  Beyond about 35 km s$^{-1}$,
nominal $R\simeq 15$ kpc, the \textcolor{black}{mean} spin temperatures rise to the range 250 to 400 K,
out to velocity 65 km s$^{-1}$, nominal $R=40$ kpc.  Beyond 70 km s$^{-1}$
the \textcolor{black}{mean} spin temperature rises to about 450 to 500K, but the absorption is barely
detected at these velocities, and the translation between velocity and
$R$ is compromised by random motions of a few km s$^{-1}$.
%assuming 5 km s$^{-1}$ radial inflow velocity 

Table \ref{tab:outer_temps} gives values derived from the average spectra
on the left panel of Figure \ref{fig:tspin_vs_v}.  Each row on the table 
corresponds to a velocity interval of five km s$^{-1}$, indicated in the 
first column.  The galactocentric radius, $R$ of the midpoint of the interval
is given in column 2, and the length along the LoS, $\Delta d$, corresponding
to this velocity interval is given in column 3.  The mean value of the 
absorption over the interval with error is in column 4, i.e. the 
equivalent width.  The 
mean brightness temperature of the emission average is in column 5.
The \textcolor{black}{mean} spin temperature computed with these two means is in column 6, with 
errors based on the errors in the optical depth.  Using the path length,
$\Delta d$ and the absorption and emission averages times the velocity 
interval width, we get the column density divided by path length,
i.e. the mean density of HI from the emission spectrum, column 8 on
Table \ref{tab:outer_temps}, and from the 
equivalent width we get a mean value for the density divided by 
the \textcolor{black}{mean} spin temperature in column 7 \citep[eq. 5 and 6]{Dickey_etal_2009}.
Both these quantities change by nearly two orders of magnitude from
$R\sim$10 to $R\sim$40 kpc, but their ratio, i.e. $T_{s}$, stays
remarkably constant.  The determination of $R$ for
last two rows on the table, $\Delta v = [65,70]$ and $[70,75]$ km s$^{-1}$, is 
not at all trustworthy, so values derived using distance are
not given.

\begin{table} \caption{Outer Galaxy Mean HI Spin Temperatures  \label{tab:outer_temps}}
%\hspace{-2cm}\begin{tabular}{|l|l|l|l|l|l|l|l|}
\begin{tabular}{lllllllll}
\hline
$\Delta v$ & $R$ & $\Delta d$ & $<1-e^{-\tau}>$ & $<T_b>$ & $T_{s}$ &
$<\frac{n}{T}>$ & $<n>$ \\
km s$^{-1}$ & kpc & kpc & & $\ \ $K & $\ \ $K & 10$^{-6}$ cm$^{-3}$ / K & 10$^{-3}$ cm$^{-3}$ \\
\hline
$[15,20]$&  10.5 & 0.8 & 0.169 $\pm$ 0.004 & 36.3 &  215$^{+  5}_{-   5}$ &585.68$\pm$ 14.14 & 126.13\\
$[20,25]$&  11.4 & 1.0 & 0.191 $\pm$ 0.004 & 31.8 &  167$^{+  4}_{-   3}$ &567.90$\pm$ 11.91 & 94.64\\
$[25,30]$&  12.5 & 1.2 & 0.137 $\pm$ 0.004 & 21.3 &  155$^{+  4}_{-   4}$ &343.25$\pm$  9.11 & 53.34\\
$[30,35]$&  13.7 & 1.4 & 0.059 $\pm$ 0.003 & 13.4 &  229$^{+ 14}_{-  12}$ &121.90$\pm$  6.85 & 27.89\\
$[35,40]$&  15.3 & 1.7 & 0.030 $\pm$ 0.003 & 9.9 &  328$^{+ 37}_{-  30}$ &51.06$\pm$  5.21 & 16.77\\
$[40,45]$&  17.2 & 2.2 & 0.031 $\pm$ 0.003 & 9.3 &  298$^{+ 32}_{-  26}$ &41.46$\pm$  4.02 & 12.37\\
$[45.50]$&  20 & 2.9 & 0.026 $\pm$ 0.003 & 10.4 &  396$^{+ 52}_{-  41}$ &26.80$\pm$  3.12 & 10.62\\
$[50,55]$&  23 & 3.9 & 0.033 $\pm$ 0.003 & 10.6 &  320$^{+ 33}_{-  27}$ &24.77$\pm$  2.30 &  7.92\\
$[55,60]$&  28 & 5.7 & 0.038 $\pm$ 0.003 & 10.2 &  264$^{+ 23}_{-  20}$ &19.85$\pm$  1.59 &  5.25\\
$[60,65]$&  35 & 9.0 & 0.029 $\pm$ 0.003 & 8.6 &  291$^{+ 34}_{-  27}$ & 9.61$\pm$  0.99 &  2.80\\
$[65,70]$&  $\sim$47 &$\dots$  & 0.024 $\pm$ 0.003 & 6.0 &  253$^{+ 36}_{-  28}$ &$\dots$  &$\dots$  \\
$[70,75]$&$\dots$   &$\dots$  & 0.008 $\pm$ 0.003 & 3.8 &  485$^{+274}_{- 129}$ &$\dots$  &$\dots$  \\
\hline
\end{tabular}
\end{table}

The right panel of Figure \ref{fig:tspin_vs_v} shows the result for a small
sample of spectra with very low noise ($w_i>$800 or $\sigma_{\tau}<0.035$),
which are also selected by having significant absorption lines in the range 
48 to 65 km s$^{-1}$.  Thus this is a small and biased sample, selected to
show the characteristics of the absorption at $R\simeq 25$ to 30 kpc.  At those
velocities the absorption line is prominent, and it leads to a particularly
low value for $T_{s}$ at $v=55$ to 60 km s$^{-1}$ %($R\simeq$ 25 to 30 kpc)
that gives $T_{s}=177\pm14$ K. 
In this case the error bars are given not by the noise in the absorption 
spectrum, but by the scatter in the values of $(1 - e^{-\tau})$ 
%in each 5 km s$^{-1}$ 
as the standard deviation of the sample of different spectra.
The optical depth profile is the mean of the sample, i.e. the spectra are
all given equal weight.  In spite of the small sample and selection
bias, this sample of LoS gives \textcolor{black}{mean} spin temperatures quite similar to the 
weighted average of the much larger
sample on the left panel.  On the right panel $T_{s}$ is in the range
250 to 400 K for 30$<v<70$ km s$^{-1}$, except for the high
value (500$^{+220}_{-120}$ K) for $v$ between 45 and 50 km s$^{-1}$
and the artificially low value between 55 and 60 km s$^{-1}$ due
to the sample selection criteria.

On the right panel of Figure \ref{fig:tspin_vs_v}, it is interesting
to consider the shape of the 
tail of the mean absorption spectrum at velocities between about 65 and
82 km s$^{-1}$, i.e. 40$<R<\infty$ in equation \ref{eq:v_of_R} 
%since with the spectra shifted to nominal projected velocities
%at $\ell = 340^o \ R\rightarrow \infty$ at $v=82.1$ km s$^{-1}$ for
%our chosen $\Theta_o=240$ km s$^{-1}$.  
Possibly all the gas seen beyond 65 km s$^{-1}$
is closer than $R\simeq40$ kpc, and random motions, both microscopic and
macroscopic spread out the velocity distribution by 10 to 15 km s$^{-1}$
in absorption and 25 km s$^{-1}$ in emission.  These two spectral tails
appear to be exponential in their asymptotic behavior approaching one
in the figure. For comparison,
\citet{Kalberla_Kerp_2009} find using the LAB survey data that for $R > 35$ kpc
the disk surface density of the HI, $\Sigma$, changes slope on a plot of log $\Sigma$ - log $R$,
and azimuthal symmetry breaks down \citep[e.g.][fig. 9]{Kalberla_Dedes_2008}.
At longitude 340$^o$, $R = 35$ kpc corresponds to $v\simeq 63$ km s$^{-1}$.  Thus the
last velocity interval corresponding to distinctively disk gas on Figure
\ref{fig:tspin_vs_v} is 60 to 65 km s$^{-1}$.

%###################
A value of $T_{s}$ about 300 K is consistent with the finding of
\citet[][fig. 5]{Dickey_etal_2009}, but the result here is more precise because it 
is based on a much larger sample of spectra in a much smaller area.  
Typical values of $T_{s}$ measured in extragalactic 
damped Lyman $\alpha$ systems are higher than the values measured here
\citep{Kanekar_etal_2014}, but two systems with relatively high
metallicity \citep{Kanekar_2014} show \textcolor{black}{mean} spin
temperatures of 372 K and (tentatively) 242 K.  Thus conditions in the outer
edge of the MW disk may be similar to low redshift extragalactic 21-cm absorption
systems \citep[][and references therein]{Allison_2021}.

%\input{tables/Tspin_tab.tex}

%Now compute velocity gradient and divide to get \langle kappa\rangle and mean(n)

%Table?

%\citep{Stanimirovic_etal_2014}.

% multiply by dv/dr to get n, kappa vs R

\section{Discussion \label{sec:disc}}

  Physical interpretation of \textcolor{black}{mean} spin temperature measurements in the far
outer disk of the Milky Way is ambiguous; conditions in that environment are 
very different from those in the local
ISM.  Even in the solar neighborhood, the spin temperature 
may depart from the kinetic temperature of the gas, particularly in the WNM
\citep{Liszt_2001, Davies_Cummings_1975}.  The spin temperature will
be driven to equal the kinetic temperature in low density environments by
the Wouthuysen-Field effect, i.e. 
differential radiative excitation and de-excitation of the Ly $\alpha$ transition
when the color temperature of the radiation has come into equilibrium with the
kinetic temperature \citep[][and references therein]{Seon_Kim_2020}.
Even if the excitation of the 21-cm line is in equilibrium with the neutral
hydrogen kinetic temperature in all phases, values of $T_{s}$ 
measured from comparison of the 21-cm emission and absorption
represent a harmonic mean of different thermal phases (Eq. \ref{eq:Tspin_2})
at least including the CNM and WNM, often with UNM clouds as well
\citep{Murray_etal_2018} that may be more common \textcolor{black}{in the outer disk} than they are here.
At the outer edge of the Milky Way disk, even the existence of two
equilibrium phases is in doubt, depending on the metallicity and the
trade-off between heating by UV photons vs. cosmic
rays {\citep{Bialy_Sternberg_2019}.  
%At the low densities typical
%of the outer edge of the disk, e.g. $n_H \sim 3\times10^{-4}$ cm$^{-3}$ at
%$R\sim 33$ kpc \citep{Kalberla_Dedes_2008}, heating by photoelectric
%emission of electrons from grains \citep{Draine_1978} and cooling by the inverse
%process, i.e. electron
%recombination onto positively charged grains, are expected to dominate 
%other heating and cooling processes for
%temperatures below about 8000 K \citep[][fig. 3]{Bialy_Sternberg_2019, Wolfire_etal_1995}.  
The abundance of grains is also critical \citep[][Fig. 13]{Wolfire_etal_2003}, 
since photo-electric heating dominates as long as the metallicity $Z^{\prime}>0.1$
\citep[][sec. 4]{Bialy_Sternberg_2019}.  If
the gas phase abundance gradient derived from
HII region temperatures by \citet{Wenger_etal_2019} continues as
far as $R\simeq 35$ kpc,
 the oxygen abundance (O/H) would be $< \frac{1}{10}$ of
its value at the solar circle.
It may be that the gas we see in HI emission and absorption never has time to
come to thermal equilibrium because of episodic shock heating due to converging
flows \citep{Vazquez-Semadeni_etal_2006, Gazol_Villagran_2016}. Such
events must be common in CGM gas, occuring 
at intervals not much longer than the cooling time.
Giving up on thermal equilibrium, modern simulations provide models of
the evolution of gas passing through a hot halo while condensing from
WIM to WNM to CNM \citep{Banda-Barragan_etal_2020, Nelson_etal_2020,Dutta_etal_2021}.  Such simulations indicate that cold-mode accretion can include both WNM and CNM clouds.

The connection between cool HI clouds at large $R$ like those seen in this study and the accretion of gas onto spiral
disks is considered by \citet{Bland-Hawthorn_etal_2017}, with particular attention
to the effects of the external ionizing radiation field.  Both the CNM and WNM 
fit in the category of ``cold mode" accretion in this context, in contrast to ``warm hot"
and ``hot mode" that are warmer still, and fully ionized \citep{Putman_etal_2012}.
In the models of \citet{Stewart_etal_2011} and \citet{Trapp_etal_2021},
a likely accretion pattern is for co-rotating CGM gas to migrate
inward through the outer boundary of the disk and on to the inner disk where
star-formation is active.  Simulations, like that of \citet{Nuza_etal_2019}, put this process
along with accretion by vertical infall
in context over cosmological time scales.  The CGM seen in 
faint 21-cm emission with the Green Bank Telescope around many nearby spirals
\citep{Pisano_2014, Pingel_etal_2018, Martin_etal_2019, Das_etal_2020, Sardone_etal_2021} 
is the reservoir from which the cold-mode accretion flows.  The structure, dynamics,
thermodynamics, and ionization of the CGM are not yet fully understood; they present
many compelling astrophysical questions that may ultimately tie together the
physics of the interstellar medium in spiral disks with the process
of galaxy formation 
studied through cosmological simulations and high redshift absorption line surveys.

\subsection{Future Observations}

The ASKAP Pilot survey observations were done in 2019/2020, to be followed by
pilot phase 2 in 2021, and full survey observations beginning in early 2022.
Based on the absorption line results presented here, and the emission line
cubes still being analysed, the astrophysical drivers and survey strategy 
will be finalised.  For the full survey, the GASKAP proposal
\citep{Dickey_etal_2013} plans 50 hours of integration time for each
field in a single line of PAF footprints along the Galactic plane.  This 
integration time will reduce the rms noise in every spectrum by a factor 
of 1.8 compared with the 16 hour Pilot Survey integration.  Thus
most of the spectra that currently have low weights (100$<w_i<$ 300, i.e.
$0.06 < \sigma_{\tau} < 0.1$) would move
to higher weights ($\sigma_{tau} < 0.05$), and a larger number of
new spectra would
become available at low weights.  This will more than
double the number of spectra per square degree in each sample (Figures \ref{fig:av_neglat}-\ref{fig:av_poslat}).
The current sample is too sparse to map the distribution of the absorption.
With a reduction in the noise level in all spectra
by a factor of almost two, it will be possible to grid the optical depths at each velocity into 
images of the structures that cause the absorption.  This will be very
helpful to determine the CGM cloud mass spectrum, turbulence spectrum, and
CNM filling factor.

Observing further above and below the Galactic plane is also part of the GASKAP
survey strategy.  To cover a larger latitude range, $3^o < |b| < 10^o$,
in a reasonable time, the integration time per pointing would be $\sim 10$ hours, shorter than for
this Pilot field.  Since the density of extragalactic sources is
the same at all latitudes, those fields will give numbers of spectra and
optical depth sensitivities very similar to those obtained here,
but with $\sigma_{\tau}$ higher by 26\%.  At a distance
of 35 kpc, the $\pm10$ degree range in latitude becomes a range in height above
and below the plane, $z$, of $-6 < z < +6$ kpc, enough
to map the flaring of the CNM disk,
i.e. the increase of the scale height with $R$.  Comparison with the emission
survey over this increased range of latitude
would show whether heating and partial ionization by the external UV field 
\citep{Bland-Hawthorn_etal_2017} causes a temperature gradient with $z$
in the gas in the far outer disk.  The results will help to explain how the CGM 
properties evolve as the cold-mode accretion flow brings co-rotating clouds into
the thick disk and ultimately into the thin disk.

%The kinetic temperature predicted for conditions at high $R$ by
%\citet[][Table 3]{Wolfire_etal_2003} is 

\section{Summary}

Study of HI absorption in a single pilot field observed with the ASKAP 
telescope at velocity resolution of 1 km s$^{-1}$ gives a variety of
results that demonstrate the power of the telescope and the potential
of a full survey of the Galactic plane as part of the GASKAP project.  This paper
concentrates on analysis of the spectra toward $\sim$175 extragalactic sources
to determine the distribution of CNM in the Galactic disk.
There is much more data to analyse, 
particularly using the Galactic continuum sources and their spectra.

The main topic considered here is the structure of the
Milky Way disk near longitude $\ell=340^o$ as traced by the 
21-cm absorption.  The thickness of the disk measured
near the sub-central point implies that the CNM at the far end of
the bar and in the Three Kpc Arm is very tightly confined in $z$
(near mid-plane), with scale heights $\sim$ 50 pc.  
This is the first measurement of such a narrow scale height
for the atomic medium in the inner Galaxy, in good agreement with 
molecular lines and the ultra-thin stellar disk (section \ref{sec:termv}).
Near the solar circle on the far side of the Galactic center
the scale height is $\sim$160 pc, similar to its value in the solar neighborhood.

Moving to the outer Galaxy, the disk continues to flare.  The absorption
spectra trace the CNM far outside the solar circle, to radius $R\sim$40 kpc.
Although kinematic distances become less accurate at large $R$,
the CNM follows the WNM to the edge of the disk.  The ratio of the
emission and absorption is nearly constant vs. velocity to at least
$5 R_o$.  

The number of background sources giving sensitive absorption spectra
is large enough that some structure in the outer disk is evident.
Dividing the field into six regions shows that some have much more
CNM than others at 25 $<R<$ 40 kpc.  Some lines of sight show quite deep,
narrow absorption lines at velocities of 55 to 70 km s$^{-1}$, two
examples are presented.  Comparison with emission in these directions
suggests that the CNM temperature is less than 50 K in these clouds.

The GASKAP survey will extend our knowledge of the CNM beyond the disk
and into the CGM where cold-mode accretion is underway.  It will provide
statistics of absorption lines for comparison with results from
extragalactic surveys to help quantify the thermal balance in gas
in exo-galactic environments.

\begin{acknowledgements}
The Australian SKA Pathfinder is part of the Australia Telescope National Facility which is managed by CSIRO. Operation of ASKAP is funded by the Australian Government with support from the National  Collaborative  Research  Infrastructure  Strategy.  ASKAP uses the resources of the Pawsey Supercomputing  Centre.   Establishment  of  ASKAP,  the  Murchison Radioastronomy  Observatory  and  the  Pawsey  Supercomputing Centre are initiatives of the Australian Government,  with  support  from  the  Government  of  Western  Australia  and  the  Science  and  Industry  Endowment Fund.  We acknowledge the Wajarri Yamatji people  as  the  traditional  owners  of  the  Observatory  site.  This  paper  includes  archived  data  obtained  through the  CSIRO  ASKAP  Science  Data  Archive,  CASDA.

This research has made use of of NASA's Astrophysics Data System; the SIMBAD database and VizieR catalogue access tool, CDS, Strasbourg, France
(DOI: 10.26093/cds/vizier); and matplotlib for python \citep{matplotlib}.
\end{acknowledgements}

{}

\clearpage

%\FloatBarrier

\startlongtable
\begin{deluxetable*}{|l|ll|ll|l|ll|}
\tablewidth{7in}
\tablecaption{Galactic Sources \label{tab:Galactic}}
\tablewidth{7in}
\tablehead{ \colhead{index} & \colhead{RA} &  \colhead{Dec} &
\colhead{Longitude} &
\colhead{Latitude} & $\sigma_{\tau}$ & \colhead{identification} & \colhead{structure} } 
\startdata
    3   & 16:30:25.8 & -46:02:51 & 337.474 & +1.619 & 0.472& Planetary Nebula    & point\\
   17   & 16:32:56.4 & -47:57:54 & 336.367 & -0.004 & 0.088&   WISE-HII  ATLASgal   & point\\
   20   & 16:32:57.5 & -46:40:28 & 337.315 & +0.872 & 0.126&   SNR   & mean of a double source\\
   23   & 16:33:28.9 & -47:29:53 & 336.772 & +0.247 & 0.197&   WISE-HII  ATLASgal   & complex structure\\
   26   & 16:33:58.3 & -46:50:09 & 337.314 & +0.635 & 0.198&  Planetary Nebula   & point\\
   29   & 16:34:04.7 & -47:16:30 & 337.004 & +0.324 & 0.034&   WISE-HII  ATLASgal   & point\\
   30   & 16:34:28.3 & -47:52:52 & 336.603 & -0.136 & 0.552& YSO?   & point\\
   31   & 16:34:33.9 & -47:58:15 & 336.548 & -0.208 & 0.426&   ATLASgal   & complex structure\\
   32   & 16:34:38.4 & -47:36:31 & 336.823 & +0.028 & 0.274&   ATLASgal   & complex structure\\
   33   & 16:34:39.5 & -47:36:06 & 336.830 & +0.030 & 0.133&   ATLASgal   & complex structure\\
   34   & 16:34:48.4 & -47:32:49 & 336.887 & +0.049 & 0.077&   WISE-HII  ATLASgal   & complex structure\\
   35   & 16:34:53.0 & -47:37:04 & 336.844 & -0.008 & 0.194& YSO?   & complex structure\\
   36   & 16:35:12.8 & -47:35:32 & 336.900 & -0.032 & 0.092&   WISE-HII  ATLASgal   & point\\
   37   & 16:35:32.3 & -47:31:14 & 336.990 & -0.024 & 0.060&   WISE-HII  ATLASgal   & point\\
   43   & 16:36:13.5 & -47:31:06 & 337.070 & -0.108 & 0.271& dark cloud   & point\\
   44   & 16:36:24.8 & -47:24:33 & 337.172 & -0.058 & 0.088&   WISE-HII  ATLASgal   & complex structure\\
   46   & 16:36:28.4 & -47:23:55 & 337.187 & -0.059 & 0.091&   WISE-HII   & complex structure\\
   48   & 16:36:30.9 & -47:23:36 & 337.195 & -0.061 & 0.804&   WISE-HII  ATLASgal   & complex structure\\
   50   & 16:36:42.5 & -47:31:31 & 337.120 & -0.174 & 0.008&   WISE-HII  ATLASgal   & complex structure\\
   51   & 16:36:44.1 & -47:31:59 & 337.117 & -0.182 & 0.009&   WISE-HII  ATLASgal   & complex structure\\
   52   & 16:36:47.1 & -47:31:37 & 337.127 & -0.184 & 0.018&   WISE-HII   & complex structure\\
   56   & 16:37:05.6 & -47:21:40 & 337.285 & -0.111 & 0.039&   WISE-HII  ATLASgal   & ring or shell source\\
   58   & 16:37:13.5 & -47:25:17 & 337.255 & -0.168 & 0.040&   WISE-HII  ATLASgal   & ring or shell source\\
   59   & 16:37:17.2 & -47:23:47 & 337.281 & -0.159 & 0.080&   WISE-HII  ATLASgal   & complex structure\\
   60   & 16:37:21.8 & -46:11:30 & 338.184 & +0.639 & 0.175& YSO   & point\\
   63   & 16:37:52.0 & -46:54:34 & 337.708 & +0.094 & 0.028&   WISE-HII  ATLASgal   & one component of double\\
   64   & 16:37:54.7 & -46:54:46 & 337.711 & +0.086 & 0.042&   WISE-HII  ATLASgal   & one component of double\\
   66   & 16:38:11.1 & -47:04:52 & 337.617 & -0.062 & 0.021&   WISE-HII  ATLASgal   & complex structure\\
   69   & 16:38:18.9 & -47:04:51 & 337.632 & -0.078 & 0.236&   ATLASgal   & point\\
   72   & 16:38:30.9 & -47:00:50 & 337.705 & -0.059 & 0.026&   WISE-HII  ATLASgal   & complex structure\\
   75   & 16:38:50.4 & -47:28:03 & 337.404 & -0.402 & 0.371&   WISE-HII  ATLASgal   & complex structure\\
   76   & 16:38:52.3 & -47:07:17 & 337.665 & -0.176 & 0.017&   WISE-HII  ATLASgal   & point\\
   79   & 16:39:03.4 & -46:42:27 & 337.995 & +0.077 & 0.056&   WISE-HII  ATLASgal   & point\\
   82   & 16:39:21.6 & -46:44:18 & 338.006 & +0.018 & 0.095&   WISE-HII  ATLASgal   & point\\
   83   & 16:39:32.5 & -46:55:26 & 337.889 & -0.129 & 0.285& YSO   & complex structure\\
   85   & 16:39:38.8 & -46:41:26 & 338.075 & +0.013 & 0.009&   WISE-HII  ATLASgal   & mean of a double source\\
   86   & 16:39:49.6 & -46:43:01 & 338.075 & -0.028 & 0.105&   WISE-HII  ATLASgal   & point\\
   87   & 16:39:57.5 & -46:45:10 & 338.064 & -0.069 & 0.027&   WISE-HII  ATLASgal   & point\\
   89   & 16:39:59.7 & -46:24:33 & 338.325 & +0.155 & 0.062&   ATLASgal   & point\\
   92   & 16:40:07.3 & -46:23:26 & 338.353 & +0.151 & 0.029&   ATLASgal   & complex structure\\
   93   & 16:40:07.5 & -46:25:06 & 338.333 & +0.132 & 0.028&   ATLASgal   & point\\
   96   & 16:40:10.4 & -46:24:19 & 338.348 & +0.135 & 0.196&   ATLASgal   & complex structure\\
   97   & 16:40:11.0 & -46:21:39 & 338.383 & +0.163 & 0.029& YSO   & complex structure\\
   98   & 16:40:15.4 & -45:39:03 & 338.922 & +0.625 & 0.030& YSO   & point\\
  100   & 16:40:46.2 & -46:24:36 & 338.413 & +0.054 & 0.087&   ATLASgal   & complex structure\\
  101   & 16:40:47.4 & -46:32:00 & 338.323 & -0.030 & 0.170&   SNR   & point\\
  102   & 16:40:50.3 & -46:23:24 & 338.436 & +0.059 & 0.007&   ATLASgal   & point\\
  103   & 16:40:55.2 & -45:50:56 & 338.850 & +0.406 & 0.102&   WISE-HII  ATLASgal   & point\\
  105   & 16:40:57.5 & -46:21:34 & 338.472 & +0.064 & 0.132&   SNR   & complex structure\\
  106   & 16:40:58.1 & -46:25:52 & 338.420 & +0.015 & 0.207& YSO   & complex structure\\
  107   & 16:40:58.4 & -46:27:00 & 338.406 & +0.002 & 0.037& YSO   & complex structure\\
  108   & 16:40:59.3 & -47:07:13 & 337.905 & -0.444 & 0.071&   S36 bubble & complex structure\\
  109   & 16:41:01.6 & -46:25:18 & 338.433 & +0.014 & 0.007&   ATLASgal   & complex structure\\
  110   & 16:41:03.9 & -46:22:11 & 338.477 & +0.043 & 0.068&   SNR   & mean of a double source\\
  111   & 16:41:06.3 & -46:21:32 & 338.489 & +0.045 & 0.093&   ATLASgal  SNR   & complex structure\\
  114   & 16:41:06.4 & -47:07:52 & 337.910 & -0.466 & 0.005&   WISE-HII   & complex structure\\
  113   & 16:41:08.2 & -47:06:46 & 337.928 & -0.458 & 0.015&   WISE-HII  ATLASgal   & complex structure\\
  116   & 16:41:13.2 & -46:11:31 & 338.628 & +0.140 & 0.026&   WISE-HII  ATLASgal   & ring or shell source\\
  121   & 16:41:16.1 & -45:48:53 & 338.916 & +0.383 & 0.054&   WISE-HII  ATLASgal   & point\\
  122   & 16:41:31.1 & -46:34:31 & 338.374 & -0.151 & 0.025&   WISE-HII  ATLASgal   & point\\
  124   & 16:41:36.2 & -46:17:54 & 338.592 & +0.021 & 0.035&   ATLASgal   & complex structure\\
  125   & 16:41:43.5 & -46:18:44 & 338.595 & -0.005 & 0.126&   WISE-HII  ATLASgal   & complex structure\\
  126   & 16:41:44.3 & -46:44:45 & 338.271 & -0.293 & 0.156&   WISE-HII  ATLASgal   & point\\
  128   & 16:41:51.7 & -46:35:08 & 338.405 & -0.202 & 0.005&   WISE-HII  ATLASgal   & ring or shell source\\
  131   & 16:41:54.2 & -46:34:58 & 338.412 & -0.206 & 0.018&   WISE-HII  ATLASgal   & point\\
  132   & 16:42:01.2 & -46:47:52 & 338.264 & -0.363 & 0.128&   WISE-HII   & point\\
  134   & 16:42:09.3 & -46:47:03 & 338.289 & -0.371 & 0.056&   WISE-HII  ATLASgal   & point\\
  136   & 16:42:14.1 & -46:25:25 & 338.569 & -0.144 & 0.024&   WISE-HII  ATLASgal   & one component of double\\
  137   & 16:42:19.2 & -46:34:47 & 338.461 & -0.258 & 0.093&   WISE-HII  ATLASgal   & complex structure\\
  138   & 16:42:23.9 & -46:21:04 & 338.642 & -0.118 & 0.153& IR emission lines?   & point\\
  139   & 16:42:24.0 & -46:18:01 & 338.681 & -0.084 & 0.059&   ATLASgal   & point\\
  144   & 16:42:48.6 & -45:54:06 & 339.028 & +0.124 & 0.227&  YSO  & point\\
  145   & 16:42:59.5 & -45:49:40 & 339.104 & +0.149 & 0.038&   WISE-HII  ATLASgal   & complex structure\\
  150   & 16:43:15.7 & -44:35:17 & 340.071 & +0.927 & 0.047&   WISE-HII  ATLASgal   & point\\
  158   & 16:44:14.0 & -45:31:23 & 339.476 & +0.185 & 0.071&   WISE-HII  ATLASgal   & complex structure\\
  166   & 16:44:41.0 & -45:34:32 & 339.488 & +0.091 & 0.060&   WISE-HII  ATLASgal   & complex structure\\
  168   & 16:44:49.2 & -45:59:10 & 339.193 & -0.195 & 0.017&   pulsar   & point\\
  169   & 16:45:03.0 & -45:30:34 & 339.580 & +0.086 & 0.048&   WISE-HII   & complex structure\\
  170   & 16:45:05.8 & -45:30:13 & 339.590 & +0.083 & 0.077&   WISE-HII  ATLASgal   & complex structure\\
  176   & 16:45:37.7 & -45:22:37 & 339.747 & +0.095 & 0.192&   WISE-HII  ATLASgal   & point\\
  178   & 16:45:51.1 & -45:09:52 & 339.934 & +0.204 & 0.235& YSO   & point\\
  180   & 16:46:08.8 & -46:29:01 & 338.965 & -0.693 & 0.119& SF cloud   & mean of a double source\\
  190   & 16:47:05.5 & -45:50:36 & 339.558 & -0.402 & 0.026&   WISE-HII   & complex structure\\
  193   & 16:47:36.3 & -45:45:47 & 339.678 & -0.419 & 0.268& YSO   & point\\
  196   & 16:48:05.1 & -45:05:11 & 340.249 & -0.046 & 0.058&   WISE-HII  ATLASgal   & point\\
  198   & 16:48:10.6 & -45:21:32 & 340.050 & -0.234 & 0.016&   WISE-HII  ATLASgal   & ring or shell source\\
  202   & 16:48:14.3 & -45:21:40 & 340.056 & -0.244 & 0.027&   WISE-HII  ATLASgal   & complex structure\\
  208   & 16:48:53.5 & -45:10:08 & 340.277 & -0.208 & 0.035&   ATLASgal   & complex structure\\
  212   & 16:49:14.8 & -45:36:32 & 339.980 & -0.539 & 0.131&   ATLASgal   & point\\
  215   & 16:49:29.9 & -45:17:46 & 340.248 & -0.372 & 0.091&   WISE-HII  ATLASgal   & complex structure\\
  220   & 16:51:14.4 & -46:14:21 & 339.717 & -1.208 & 0.047&   WISE-HII   & complex structure\\
  227   & 16:52:21.2 & -44:27:58 & 341.210 & -0.231 & 0.018&   WISE-HII   & ring or shell source\\
  229   & 16:52:33.1 & -43:23:46 & 342.059 & +0.420 & 0.006&   WISE-HII  ATLASgal   & complex structure\\
  234   & 16:52:52.3 & -43:54:22 & 341.702 & +0.052 & 0.021&   WISE-HII  ATLASgal   & point\\
  235   & 16:52:54.7 & -44:26:53 & 341.287 & -0.297 & 0.030&   ATLASgal   & complex structure\\
  251   & 16:54:15.3 & -45:16:58 & 340.790 & -1.009 & 0.030& YSO   & ring or shell source\\
  252   & 16:54:16.1 & -45:17:31 & 340.785 & -1.016 & 0.004&   WISE-HII  ATLASgal   & ring or shell source\\
  269   & 16:56:33.9 & -43:46:15 & 342.226 & -0.380 & 0.037&   Planetary Nebula  & point\\
  290   & 16:59:03.9 & -42:41:38 & 343.353 & -0.066 & 0.032&   WISE-HII  ATLASgal   & point\\
  296   & 16:59:19.2 & -42:34:17 & 343.478 & -0.027 & 0.040&   ATLASgal   & complex structure\\
  297   & 16:59:20.7 & -42:32:38 & 343.502 & -0.014 & 0.018&   WISE-HII  ATLASgal   & point\\
  305   & 17:00:17.1 & -43:12:46 & 343.082 & -0.562 & 0.084&   SNR   & point\\
  319   & 17:02:13.1 & -46:55:51 & 340.355 & -3.114 & 0.129& YSO   & point\\
  322   & 17:02:39.5 & -46:08:01 & 341.033 & -2.687 & 0.114& YSO   & point\\
  331   & 17:04:13.1 & -42:19:54 & 344.221 & -0.594 & 0.029&   ATLASgal   & point\\
\enddata
\end{deluxetable*}

\eject

\clearpage

%\FloatBarrier

\startlongtable
\begin{deluxetable*}{|l|ll|ll|l|l|}
\tablewidth{6.5in}
\tablecaption{Extragalactic Sources \label{tab:exgal}}
\tablewidth{6in}
\tablehead{
\colhead{index}& \colhead{RA} & \colhead{Dec} &
\colhead{Longitude} & \colhead{Latitude} &\colhead{$\sigma_{\tau}$} &\colhead{structure} }
\startdata
    0   & 16:29:14.5 & -45:41:17 & 337.593 & +2.016 & 0.435 & single component\\
    1   & 16:29:34.5 & -44:18:37 & 338.633 & +2.922 & 0.190 & single component\\
    2   & 16:30:18.3 & -44:52:32 & 338.312 & +2.440 & 0.328 & single component\\
    4   & 16:30:34.1 & -45:15:06 & 338.070 & +2.148 & 0.186 & single component\\
    5   & 16:30:40.0 & -47:00:19 & 336.805 & +0.932 & 0.074 & single component\\
    6   & 16:30:42.7 & -44:30:48 & 338.625 & +2.635 & 0.164 & single component\\
    7   & 16:30:44.7 & -44:00:14 & 339.001 & +2.980 & 0.144 & mean of a double source\\
    8   & 16:31:13.0 & -46:43:40 & 337.072 & +1.054 & 0.482 & single component\\
    9   & 16:31:29.9 & -44:28:17 & 338.752 & +2.562 & 0.465 & single component\\
   10   & 16:31:37.9 & -43:45:25 & 339.290 & +3.032 & 0.036 & single component\\
   11   & 16:32:02.0 & -46:52:53 & 337.055 & +0.847 & 0.043 & single component\\
   13   & 16:32:38.8 & -45:53:21 & 337.853 & +1.446 & 0.154 & single component\\
   14   & 16:32:39.9 & -47:34:38 & 336.620 & +0.294 & 0.268 & single component\\
   15   & 16:32:46.7 & -45:58:02 & 337.812 & +1.376 & 0.024 & single component\\
   16   & 16:32:54.3 & -43:57:00 & 339.306 & +2.733 & 0.024 & single component\\
   19   & 16:32:57.0 & -45:03:52 & 338.494 & +1.969 & 0.266 & single component\\
   21   & 16:33:05.8 & -45:09:29 & 338.443 & +1.886 & 0.017 & one component of a double\\
   24   & 16:33:47.5 & -44:09:29 & 339.261 & +2.475 & 0.206 & single component\\
   25   & 16:33:50.7 & -47:46:46 & 336.607 & +0.011 & 0.103 & complex structure\\
   27   & 16:34:00.5 & -46:19:36 & 337.693 & +0.976 & 0.386 & single component\\
   28   & 16:34:04.0 & -43:07:53 & 340.051 & +3.134 & 0.161 & one component of a double\\
   38   & 16:35:35.5 & -46:15:24 & 337.929 & +0.822 & 0.344 & single component\\
   39   & 16:35:36.7 & -43:17:37 & 340.122 & +2.817 & 0.137 & single component\\
   40   & 16:35:48.8 & -42:28:38 & 340.751 & +3.339 & 0.387 & single component\\
   41   & 16:36:04.5 & -44:57:59 & 338.940 & +1.629 & 0.100 & single component\\
   42   & 16:36:05.4 & -42:48:44 & 340.537 & +3.076 & 0.259 & single component\\
   45   & 16:36:28.0 & -46:36:47 & 337.767 & +0.470 & 0.214 & single component\\
   47   & 16:36:30.0 & -45:22:44 & 338.685 & +1.296 & 0.140 & single component\\
   49   & 16:36:38.1 & -44:45:18 & 339.163 & +1.697 & 0.123 & single component\\
   53   & 16:36:58.0 & -45:06:25 & 338.942 & +1.418 & 0.050 & single component\\
   54   & 16:37:01.6 & -43:34:02 & 340.091 & +2.443 & 0.121 & single component\\
   55   & 16:37:04.3 & -43:10:52 & 340.384 & +2.695 & 0.058 & single component\\
   57   & 16:37:13.8 & -44:23:08 & 339.508 & +1.867 & 0.133 & one component of a double\\
   61   & 16:37:25.0 & -44:10:34 & 339.686 & +1.982 & 0.123 & single component\\
   62   & 16:37:27.9 & -43:35:42 & 340.124 & +2.365 & 0.128 & complex structure\\
   65   & 16:37:57.2 & -43:01:53 & 340.603 & +2.676 & 0.104 & single component\\
   70   & 16:38:27.3 & -44:14:45 & 339.759 & +1.797 & 0.043 & single component\\
   71   & 16:38:28.6 & -45:02:19 & 339.171 & +1.265 & 0.160 & single component\\
   77   & 16:38:59.2 & -43:45:25 & 340.187 & +2.052 & 0.129 & single component\\
   78   & 16:39:00.2 & -44:50:33 & 339.379 & +1.327 & 0.128 & single component\\
   80   & 16:39:08.7 & -47:34:24 & 337.359 & -0.511 & 0.463 & single component\\
   81   & 16:39:15.3 & -46:55:04 & 337.861 & -0.088 & 0.046 & complex structure\\
   84   & 16:39:34.6 & -44:26:37 & 339.745 & +1.516 & 0.052 & single component\\
   91   & 16:40:01.5 & -44:49:14 & 339.516 & +1.206 & 0.121 & mean of a double source\\
   90   & 16:40:01.6 & -43:44:16 & 340.326 & +1.924 & 0.111 & single component\\
   94   & 16:40:07.6 & -47:02:34 & 337.866 & -0.283 & 0.033 & single component\\
   95   & 16:40:08.9 & -47:08:44 & 337.792 & -0.354 & 0.228 & single component\\
   99   & 16:40:39.5 & -43:42:35 & 340.423 & +1.858 & 0.233 & single component\\
  104   & 16:40:57.2 & -43:53:02 & 340.328 & +1.703 & 0.196 & single component\\
  115   & 16:41:13.9 & -45:12:06 & 339.372 & +0.794 & 0.216 & single component\\
  117   & 16:41:14.9 & -44:00:22 & 340.271 & +1.582 & 0.081 & one component of a double\\
  120   & 16:41:16.0 & -44:00:02 & 340.278 & +1.583 & 0.037 & one component of a double\\
  123   & 16:41:33.0 & -44:52:38 & 339.653 & +0.966 & 0.225 & one component of a double\\
  127   & 16:41:50.5 & -47:40:38 & 337.583 & -0.920 & 0.035 & single component\\
  133   & 16:42:06.8 & -44:13:33 & 340.209 & +1.320 & 0.083 & single component\\
  140   & 16:42:39.0 & -47:38:45 & 337.696 & -1.002 & 0.124 & single component\\
  141   & 16:42:45.9 & -44:00:06 & 340.454 & +1.379 & 0.053 & complex structure\\
  142   & 16:42:46.9 & -44:01:15 & 340.442 & +1.365 & 0.032 & complex structure\\
  147   & 16:43:03.7 & -46:47:23 & 338.387 & -0.492 & 0.023 & one component of a double\\
  148   & 16:43:05.5 & -44:17:38 & 340.273 & +1.143 & 0.075 & single component\\
  151   & 16:43:18.9 & -42:41:51 & 341.504 & +2.161 & 0.193 & single component\\
  152   & 16:43:29.7 & -47:10:40 & 338.143 & -0.802 & 0.125 & single component\\
  153   & 16:43:35.5 & -46:26:31 & 338.709 & -0.332 & 0.187 & single component\\
  154   & 16:43:38.2 & -45:24:14 & 339.498 & +0.342 & 0.178 & single component\\
  155   & 16:43:47.9 & -44:02:17 & 340.549 & +1.216 & 0.012 & single component\\
  160   & 16:44:22.6 & -44:36:40 & 340.183 & +0.762 & 0.035 & single component\\
  159   & 16:44:24.2 & -44:36:45 & 340.185 & +0.758 & 0.051 & complex structure\\
  161   & 16:44:25.9 & -44:19:43 & 340.403 & +0.939 & 0.063 & complex structure\\
  162   & 16:44:28.2 & -44:54:38 & 339.967 & +0.554 & 0.193 & single component\\
  163   & 16:44:30.0 & -45:41:35 & 339.379 & +0.039 & 0.145 & single component\\
  164   & 16:44:35.6 & -47:13:37 & 338.228 & -0.975 & 0.157 & single component\\
  165   & 16:44:35.7 & -47:17:23 & 338.180 & -1.017 & 0.141 & single component\\
  167   & 16:44:48.2 & -44:27:08 & 340.353 & +0.809 & 0.192 & single component\\
  171   & 16:45:10.4 & -47:32:53 & 338.048 & -1.259 & 0.134 & single component\\
  172   & 16:45:16.0 & -45:36:19 & 339.532 & -0.006 & 0.138 & one component of a double\\
  173   & 16:45:18.6 & -45:37:40 & 339.520 & -0.026 & 0.032 & single component\\
  174   & 16:45:32.9 & -47:42:34 & 337.967 & -1.412 & 0.031 & single component\\
  177   & 16:45:49.5 & -43:35:56 & 341.120 & +1.225 & 0.022 & single component\\
  179   & 16:45:52.1 & -45:18:37 & 339.825 & +0.106 & 0.249 & single component\\
  181   & 16:46:13.7 & -43:16:51 & 341.410 & +1.376 & 0.075 & single component\\
  182   & 16:46:14.0 & -44:11:53 & 340.713 & +0.780 & 0.036 & single component\\
  183   & 16:46:22.0 & -42:51:40 & 341.745 & +1.629 & 0.126 & single component\\
  184   & 16:46:34.2 & -43:39:14 & 341.166 & +1.087 & 0.053 & single component\\
  185   & 16:46:41.3 & -46:10:30 & 339.260 & -0.564 & 0.255 & one component of a double\\
  186   & 16:46:44.5 & -45:40:06 & 339.652 & -0.242 & 0.014 & single component\\
  187   & 16:47:00.0 & -46:02:31 & 339.397 & -0.519 & 0.248 & single component\\
  188   & 16:47:00.5 & -47:29:39 & 338.290 & -1.460 & 0.285 & single component\\
  189   & 16:47:01.4 & -46:49:55 & 338.797 & -1.033 & 0.053 & single component\\
  192   & 16:47:24.0 & -44:25:27 & 340.676 & +0.474 & 0.155 & single component\\
  194   & 16:47:49.5 & -43:03:13 & 341.771 & +1.301 & 0.115 & single component\\
  195   & 16:47:50.9 & -43:08:03 & 341.713 & +1.246 & 0.032 & mean of a double source\\
  201   & 16:48:12.6 & -43:02:48 & 341.822 & +1.252 & 0.152 & single component\\
  203   & 16:48:18.0 & -43:26:39 & 341.529 & +0.983 & 0.162 & single component\\
  204   & 16:48:25.9 & -46:14:33 & 339.404 & -0.838 & 0.075 & single component\\
  205   & 16:48:30.7 & -44:31:24 & 340.728 & +0.259 & 0.042 & single component\\
  206   & 16:48:40.8 & -43:33:21 & 341.488 & +0.859 & 0.198 & single component\\
  207   & 16:48:49.1 & -46:19:43 & 339.381 & -0.944 & 0.086 & single component\\
  209   & 16:48:53.8 & -47:14:34 & 338.689 & -1.542 & 0.038 & single component\\
  213   & 16:49:23.8 & -46:41:39 & 339.164 & -1.255 & 0.128 & single component\\
  214   & 16:49:28.0 & -43:14:49 & 341.816 & +0.948 & 0.103 & single component\\
  216   & 16:49:42.7 & -43:53:10 & 341.354 & +0.504 & 0.023 & single component\\
  217   & 16:50:37.0 & -46:38:42 & 339.336 & -1.385 & 0.094 & single component\\
  218   & 16:50:38.9 & -47:31:26 & 338.663 & -1.950 & 0.391 & single component\\
  219   & 16:51:06.6 & -43:08:10 & 342.093 & +0.788 & 0.036 & single component\\
  221   & 16:51:22.5 & -42:39:00 & 342.499 & +1.061 & 0.140 & complex structure\\
  222   & 16:51:23.5 & -46:50:25 & 339.271 & -1.611 & 0.160 & single component\\
  223   & 16:51:33.9 & -43:46:07 & 341.658 & +0.322 & 0.042 & single component\\
  224   & 16:52:07.7 & -47:08:21 & 339.120 & -1.898 & 0.224 & single component\\
  228   & 16:52:29.5 & -44:35:09 & 341.133 & -0.326 & 0.095 & one component of a double\\
  231   & 16:52:34.0 & -45:36:44 & 340.348 & -0.987 & 0.031 & single component\\
  232   & 16:52:40.4 & -47:28:12 & 338.922 & -2.180 & 0.312 & complex structure\\
  233   & 16:52:45.6 & -42:11:14 & 343.018 & +1.158 & 0.146 & single component\\
  236   & 16:52:54.3 & -46:37:58 & 339.596 & -1.680 & 0.234 & single component\\
  237   & 16:53:00.2 & -46:44:00 & 339.529 & -1.756 & 0.038 & single component\\
  238   & 16:53:03.3 & -46:26:17 & 339.763 & -1.577 & 0.095 & single component\\
  240   & 16:53:18.1 & -45:19:21 & 340.654 & -0.904 & 0.115 & single component\\
  241   & 16:53:23.9 & -42:45:34 & 342.650 & +0.704 & 0.030 & single component\\
  242   & 16:53:26.1 & -42:03:14 & 343.201 & +1.145 & 0.030 & single component\\
  244   & 16:53:34.5 & -42:30:40 & 342.863 & +0.836 & 0.024 & single component\\
  245   & 16:53:39.7 & -42:17:05 & 343.049 & +0.967 & 0.085 & single component\\
  246   & 16:53:59.7 & -42:20:19 & 343.046 & +0.885 & 0.098 & single component\\
  247   & 16:54:04.1 & -43:05:09 & 342.475 & +0.403 & 0.032 & one component of a double\\
  249   & 16:54:10.7 & -43:59:05 & 341.790 & -0.180 & 0.175 & single component\\
  250   & 16:54:13.8 & -47:34:37 & 339.005 & -2.451 & 0.380 & single component\\
  254   & 16:54:29.9 & -46:51:30 & 339.593 & -2.034 & 0.014 & single component\\
  255   & 16:54:34.7 & -45:24:34 & 340.728 & -1.133 & 0.156 & single component\\
  256   & 16:54:39.8 & -47:03:49 & 339.451 & -2.185 & 0.093 & single component\\
  257   & 16:54:53.7 & -45:34:16 & 340.637 & -1.278 & 0.022 & single component\\
  259   & 16:55:05.6 & -47:05:25 & 339.477 & -2.259 & 0.124 & single component\\
  260   & 16:55:14.3 & -45:33:41 & 340.683 & -1.318 & 0.305 & single component\\
  261   & 16:55:36.3 & -43:48:44 & 342.085 & -0.271 & 0.050 & single component\\
  262   & 16:55:40.1 & -46:45:25 & 339.798 & -2.126 & 0.177 & mean of a double source\\
  264   & 16:56:01.8 & -42:20:03 & 343.286 & +0.595 & 0.047 & single component\\
  265   & 16:56:14.9 & -47:29:15 & 339.289 & -2.660 & 0.057 & single component\\
  266   & 16:56:19.9 & -42:43:09 & 343.020 & +0.311 & 0.201 & single component\\
  267   & 16:56:25.2 & -43:17:05 & 342.589 & -0.056 & 0.040 & mean of a double source\\
  270   & 16:56:36.7 & -44:37:16 & 341.568 & -0.918 & 0.132 & one component of a double\\
  272   & 16:56:58.2 & -42:10:11 & 343.523 & +0.562 & 0.111 & complex structure\\
  274   & 16:57:05.6 & -44:04:02 & 342.054 & -0.639 & 0.067 & one component of a double\\
  276   & 16:57:20.7 & -45:18:22 & 341.113 & -1.447 & 0.078 & single component\\
  277   & 16:57:27.2 & -44:16:30 & 341.932 & -0.819 & 0.182 & single component\\
  278   & 16:57:53.2 & -44:02:50 & 342.159 & -0.739 & 0.017 & mean of a double source\\
  280   & 16:57:56.4 & -42:47:49 & 343.144 & +0.032 & 0.096 & complex structure\\
  282   & 16:58:09.4 & -45:19:01 & 341.193 & -1.565 & 0.174 & single component\\
  283   & 16:58:15.0 & -43:30:12 & 342.625 & -0.452 & 0.063 & single component\\
  284   & 16:58:17.4 & -47:30:51 & 339.483 & -2.947 & 0.087 & one component of a double\\
  286   & 16:58:28.5 & -42:43:38 & 343.259 & -0.002 & 0.105 & single component\\
  287   & 16:58:31.7 & -47:00:07 & 339.911 & -2.661 & 0.130 & single component\\
  288   & 16:58:55.5 & -43:54:17 & 342.386 & -0.797 & 0.105 & single component\\
  289   & 16:59:01.1 & -43:24:51 & 342.782 & -0.506 & 0.019 & single component\\
  292   & 16:59:07.0 & -42:53:03 & 343.209 & -0.192 & 0.053 & single component\\
  294   & 16:59:08.0 & -46:19:42 & 340.505 & -2.326 & 0.258 & single component\\
  293   & 16:59:08.4 & -44:39:21 & 341.820 & -1.292 & 0.036 & single component\\
  295   & 16:59:15.1 & -47:25:35 & 339.653 & -3.020 & 0.261 & single component\\
  299   & 16:59:49.2 & -44:41:48 & 341.863 & -1.412 & 0.106 & single component\\
  300   & 16:59:54.3 & -46:55:01 & 340.123 & -2.793 & 0.274 & single component\\
  301   & 17:00:03.1 & -42:28:58 & 343.631 & -0.079 & 0.204 & single component\\
  302   & 17:00:04.8 & -42:12:28 & 343.850 & +0.087 & 0.058 & single component\\
  304   & 17:00:16.4 & -42:29:41 & 343.646 & -0.118 & 0.091 & single component\\
  306   & 17:00:30.6 & -42:34:49 & 343.606 & -0.205 & 0.080 & mean of a double source\\
  308   & 17:00:34.2 & -46:57:51 & 340.156 & -2.912 & 0.067 & single component\\
  309   & 17:00:36.5 & -46:44:34 & 340.334 & -2.781 & 0.218 & mean of a double source\\
  310   & 17:01:05.9 & -42:52:28 & 343.440 & -0.472 & 0.074 & single component\\
  311   & 17:01:08.0 & -43:52:07 & 342.660 & -1.087 & 0.113 & one component of a double\\
  313   & 17:01:13.2 & -44:05:07 & 342.499 & -1.233 & 0.106 & single component\\
  314   & 17:01:23.1 & -42:46:22 & 343.553 & -0.450 & 0.105 & one component of a double\\
  316   & 17:01:26.3 & -46:39:58 & 340.482 & -2.846 & 0.134 & single component\\
  317   & 17:01:54.0 & -43:28:48 & 343.052 & -0.958 & 0.353 & single component\\
  318   & 17:02:05.7 & -47:02:28 & 340.254 & -3.164 & 0.075 & single component\\
  320   & 17:02:21.4 & -46:40:07 & 340.577 & -2.972 & 0.173 & single component\\
  321   & 17:02:36.0 & -44:55:47 & 341.981 & -1.944 & 0.024 & single component\\
  323   & 17:02:47.5 & -44:07:07 & 342.645 & -1.476 & 0.013 & single component\\
  324   & 17:02:50.9 & -45:15:51 & 341.743 & -2.183 & 0.212 & single component\\
  325   & 17:03:03.7 & -46:11:30 & 341.030 & -2.778 & 0.265 & single component\\
  326   & 17:03:21.3 & -45:50:53 & 341.334 & -2.609 & 0.054 & single component\\
  327   & 17:03:23.8 & -44:22:32 & 342.507 & -1.719 & 0.201 & single component\\
  328   & 17:03:50.6 & -45:38:21 & 341.552 & -2.550 & 0.136 & single component\\
  329   & 17:03:54.1 & -43:26:06 & 343.309 & -1.219 & 0.321 & one component of a double\\
  330   & 17:04:10.6 & -43:00:01 & 343.685 & -0.994 & 0.248 & single component\\
  332   & 17:04:36.0 & -45:13:15 & 341.965 & -2.402 & 0.402 & single component\\
  333   & 17:04:39.3 & -45:29:29 & 341.755 & -2.573 & 0.378 & single component\\
  334   & 17:04:51.9 & -45:17:20 & 341.939 & -2.480 & 0.256 & single component\\
  335   & 17:04:53.4 & -46:24:00 & 341.056 & -3.155 & 0.405 & one component of a double\\
  336   & 17:05:07.6 & -45:31:06 & 341.784 & -2.655 & 0.063 & mean of a double source\\
  338   & 17:05:42.4 & -45:41:30 & 341.707 & -2.841 & 0.083 & complex structure\\
  340   & 17:06:20.0 & -45:55:18 & 341.589 & -3.067 & 0.294 & single component\\
  341   & 17:06:22.0 & -46:00:18 & 341.526 & -3.121 & 0.072 & single component\\
  342   & 17:06:30.2 & -44:08:29 & 343.031 & -2.020 & 0.261 & single component\\
  343   & 17:06:57.9 & -45:37:09 & 341.897 & -2.973 & 0.078 & complex structure\\
  346   & 17:07:48.6 & -46:00:42 & 341.671 & -3.326 & 0.492 & single component\\
\enddata
\end{deluxetable*}
\eject

\end{document}